\documentclass[longauth]{aa}

\usepackage{graphicx}
\usepackage{txfonts}
\usepackage{lipsum}
\usepackage{subcaption}         
\usepackage{lscape}             
\usepackage{placeins}           
\usepackage{xcolor}
\usepackage{natbib}
\bibpunct{(}{)}{;}{a}{}{,} 
\usepackage[referable]{threeparttablex}
\usepackage{longtable}
\usepackage{multirow}
\usepackage{amsmath}
\usepackage{hyperref}

\usepackage{dcolumn}

\newcolumntype{d}[1]{D{.}{.}{#1}}

\newcommand{\MPG}{MPG~2.2m}
\newcommand{\dP}{du~Pont}

\newcommand{\Ha}{H$\alpha$}
\newcommand{\pha}{PHANGS-H$\alpha$}
\newcommand{\Rc}{$R_\mathrm{c}$}
\newcommand{\Hii}{\ion{H}{ii}}
\newcommand{\Nii}{[\ion{N}{ii}]}
\newcommand{\fNii}{$\mathcal{F}_{[\mathrm{N}\,\mathrm{II}]}$}
\newcommand{\niiha}{[\ion{N}{ii}]$\lambda 6583/\mathrm{H}\alpha$}
\newcommand{\lognii}{$\log_{10}($[\ion{N}{ii}]$\lambda 6583/\mathrm{H}\alpha)$}
\newcommand{\logmass}{$\log_{10}(\mathrm{M_*}/\mathrm{M}_{\odot})$}
\newcommand{\gbp}{$G_\mathrm{BP}$}
\newcommand{\grp}{$G_\mathrm{RP}$}
\newcommand{\gcol}{$G_\mathrm{BP}-G_\mathrm{RP}$}

\newcommand{\sex}{\texttt{SExtractor}}
\newcommand{\scamp}{\texttt{SCAMP}}
\newcommand{\swarp}{\texttt{SWarp}}

\newcommand{\dra}{$\Delta\alpha$}
\newcommand{\ddec}{$\Delta\delta$}

\newcommand{\ew}{EW$_{\mathrm{H}\alpha}$}
\newcommand{\ewhan}{EW$_{\mathrm{H}\alpha + [\mathrm{N}\,\mathrm{II}]}$}
\newcommand{\ewph}{EW$_{\mathrm{ph}}$}

\begin{document}



\title{The \pha\ survey}

\subtitle{Ground-based narrow-band imaging of nearby star-forming galaxies}


\newcommand{\GEMINI}{\label{GEMINI}Gemini Observatory/NSF’s NOIRLab, 950 N. Cherry Avenue, Tucson, AZ, 85719, USA
}
\newcommand{\ASCL}{\label{ASCL}Astrophysics Source Code Librar
y, Michigan Technological University, 1400 Townsend Drive, Houghton, MI 49931}
\newcommand{\OSU}{\label{OSU}Department of Astronomy, The Ohio State University, 140 West 18th Avenue, Columbus, Ohio 43210, USA}
\newcommand{\Alberta}{\label{Alberta}Department of Physics, University of Alberta, Edmonton, AB T6G 2E1, Canada}
\newcommand{\ANU}{\label{ANU}Research School of Astronomy and Astrophysics, Australian National University, Canberra, ACT 2611, Australia}
\newcommand{\IPARCOS}{\label{IPARCOS}Instituto de F\'{\i}sica de Part\'{\i}culas y del Cosmos, Universidad Complutense de Madrid, E-28040 Madrid, Spain}
\newcommand{\IPAC}{\label{IPAC}Caltech-IPAC, 1200 E. California Blvd. Pasadena, CA 91125, USA}
\newcommand{\Carnegie}{\label{Carnegie}Observatories of the Carnegie Institution for Science, 813 Santa Barbara Street, Pasadena, CA 91101, USA}
\newcommand{\CCAPP}{\label{CCAPP}Center for Cosmology and Astroparticle Physics, 191 West Woodruff Avenue, Columbus, OH 43210, USA}
\newcommand{\CfA}{\label{CfA}Harvard-Smithsonian Center for Astrophysics, 60 Garden Street, Cambridge, MA 02138, USA}
\newcommand{\CITEVA}{\label{CITEVA}Centro de Astronomía (CITEVA), Universidad de Antofagasta, Avenida Angamos 601, Antofagasta, Chile}
\newcommand{\CNRS}{\label{CNRS}CNRS, IRAP, 9 Av. du Colonel Roche, BP 44346, F-31028 Toulouse cedex 4, France}
\newcommand{\ESO}{\label{ESO}European Southern Observatory, Karl-Schwarzschild Stra{\ss}e 2, D-85748 Garching bei M\"{u}nchen, Germany}
\newcommand{\ESOChile}{\label{ESOChile}European Southern Observatory, Alonso de C\'{o}rdova 3107, Casilla 19, Santiago, Chile}
\newcommand{\Heidelberg}{\label{Heidelberg}Astronomisches Rechen-Institut, Zentrum f\"{u}r Astronomie der Universit\"{a}t Heidelberg, M\"{o}nchhofstra\ss e 12-14, D-69120 Heidelberg, Germany}
\newcommand{\ICRAR}{\label{ICRAR}International Centre for Radio Astronomy Research, University of Western Australia, 35 Stirling Highway, Crawley, WA 6009, Australia}
\newcommand{\IRAM}{\label{IRAM}Institut de Radioastronomie Millim\'{e}trique (IRAM), 300 Rue de la Piscine, F-38406 Saint Martin d'H\`{e}res, France}
\newcommand{\IRAP}{\label{IRAP}CNRS, IRAP, 9 Av. du Colonel Roche, BP 44346, F-31028 Toulouse cedex 4, France}
\newcommand{\UPS}{\label{UPS}Universit\'{e} de Toulouse, UPS-OMP, IRAP, F-31028 Toulouse cedex 4, France}
\newcommand{\ITA}{\label{ITA}Universit\"{a}t Heidelberg, Zentrum f\"{u}r Astronomie, Institut f\"{u}r Theoretische Astrophysik, Albert-Ueberle-Str 2, D-69120 Heidelberg, Germany}
\newcommand{\IWR}{\label{IWR}Universit\"{a}t Heidelberg, Interdisziplin\"{a}res Zentrum f\"{u}r Wissenschaftliches Rechnen, Im Neuenheimer Feld 205, D-69120 Heidelberg, Germany}
\newcommand{\JHU}{\label{JHU}Department of Physics and Astronomy, The Johns Hopkins University, Baltimore, MD 21218, USA}
\newcommand{\Leiden}{\label{Leiden}Leiden Observatory, Leiden University, P.O. Box 9513, 2300 RA Leiden, The Netherlands}
\newcommand{\Maryland}{\label{Maryland}Department of Astronomy, University of Maryland, College Park, MD 20742, USA}
\newcommand{\MPE}{\label{MPE}Max-Planck-Institut f\"{u}r extraterrestrische Physik, Giessenbachstra{\ss}e 1, D-85748 Garching, Germany}
\newcommand{\MPIA}{\label{MPIA}Max-Planck-Institut f\"{u}r Astronomie, K\"{o}nigstuhl 17, D-69117, Heidelberg, Germany}
\newcommand{\Nagoya}{\label{Nagoya}Department of Physics, Nagoya University, Furo-cho, Chikusa-ku, Nagoya, Aichi 464-8602, Japan}
\newcommand{\NRAO}{\label{NRAO}National Radio Astronomy Observatory, 520 Edgemont Road, Charlottesville, VA 22903-2475, USA}
\newcommand{\OAN}{\label{OAN}Observatorio Astron\'{o}mico Nacional (IGN), C/Alfonso XII, 3, E-28014 Madrid, Spain}
\newcommand{\ObsParis}{\label{ObsParis}Sorbonne Universit\'{e}, Observatoire de Paris, Universit\'{e} PSL, CNRS, LERMA, F-75014, Paris, France}
\newcommand{\Ox}{\label{Ox}Sub-department of Astrophysics, Department of Physics, University of Oxford, Keble Road, Oxford OX1 3RH, UK}
\newcommand{\JBCA}{\label{JBCA}UK ALMA Regional Centre Node, Jodrell Bank Centre for Astrophysics, Department of Physics and Astronomy, The University of Manchester, Oxford Road, Manchester M13 9PL, UK}
\newcommand{\Princeton}{\label{Princeton}Department of Astrophysical Sciences, Princeton University, Princeton, NJ 08544 USA}
\newcommand{\UToledo}{\label{UToledo}Ritter Astrophysical Research Center, University of Toledo, Toledo, OH, 43606}
\newcommand{\Toulouse}{\label{Toulouse}Universit\'{e} de Toulouse, UPS-OMP, IRAP, F-31028 Toulouse cedex 4, France}
\newcommand{\UBonn}{\label{UBonn}Argelander-Institut f\"ur Astronomie, Universit\"at Bonn, Auf dem H\"ugel 71, 53121 Bonn, Germany}
\newcommand{\UChile}{\label{UChile}Departamento de Astronom\'{i}a, Universidad de Chile, Camino del Observatorio 1515, Las Condes, Santiago, Chile}
\newcommand{\UCM}{\label{UCM}Departamento de F\'{\i}sica de la Tierra y Astrof\'{\i}sica, Universidad Complutense de Madrid, E-28040 Madrid, Spain}
\newcommand{\UCSD}{\label{UCSD}Center for Astrophysics and Space Sciences, Department of Physics,  University of California,\\ San Diego, 9500 Gilman Drive, La Jolla, CA 92093, USA}
\newcommand{\ULyon}{\label{ULyon}Univ Lyon, Univ Lyon 1, ENS de Lyon, CNRS, Centre de Recherche Astrophysique de Lyon UMR5574,\\ F-69230 Saint-Genis-Laval, France}
\newcommand{\UMass}{\label{UMass}University of Massachusetts—Amherst, 710 N. Pleasant Street, Amherst, MA 01003, USA}
\newcommand{\UniCA}{\label{UniCA}Université Côte d'Azur, Observatoire de la Côte d'Azur, CNRS, Laboratoire Lagrange, 06000, Nice, France}
\newcommand{\UWyoming}{\label{UWyoming}Department of Physics and Astronomy, University of Wyoming, Laramie, WY 82071, USA}
\newcommand{\LAM}{\label{LAM}
Aix Marseille Univ, CNRS, CNES, LAM (Laboratoire d’Astrophysique de Marseille),  F-13388 Marseille,
France}
\newcommand{\UHawaii}{\label{UHawaii}Institute for Astronomy, University of Hawaii, 2680 Woodlawn Drive, Honolulu, HI 96822, USA}
\newcommand{\UGent}{\label{UGent}Sterrenkundig Observatorium, Universiteit Gent, Krijgslaan 281 S9, B-9000 Gent, Belgium}
\newcommand{\IPARC}{\label{IPARC}Instituto de F\'{\i}sica de Part\'{\i}culas y del Cosmos IPARCOS, Facultad de Ciencias F\'{\i}sicas, Universidad Complutense de Madrid, E-28040, Spain}
\newcommand{\STScI}{\label{STScI}Space Telescope Science Institute, 3700 San Martin Drive, Baltimore, MD 21218, USA}
\newcommand{\McMaster}{\label{McMaster}Department of Physics and Astronomy, McMaster University, Hamilton, ON L8S 4M1, Canada}
\newcommand{\INAF}{\label{INAF}INAF -- Osservatorio Astrofisico di Arcetri, Largo E. Fermi 5, I-50157, Firenze, Italy}
\newcommand{\UniSQ}{\label{UniSQ}Centre for Astrophysics, University of Southern Queensland, Toowoomba, QLD 4350, Australia}
\newcommand{\UA}{\label{UA}Centro de Astronomía (CITEVA), Universidad de Antofagasta, Avenida Angamos 601, Antofagasta, Chile}
\newcommand{\LERMA}{\label{LERMA}Observatoire de Paris, PSL Research University, CNRS, Sorbonne Universit\'es, 75014 Paris}
\newcommand{\SAIMSU}{\label{SAIMSU}Sternberg Astronomical Institute, Lomonosov Moscow State University, Universitetsky pr. 13, 119234 Moscow, Russia}
\newcommand{\Rad}{\label{Rad}Elizabeth S. and Richard M. Cashin Fellow at the Radcliffe Institute for Advanced Studies at Harvard University, 10 Garden Street, Cambridge, MA 02138, U.S.A.}
\newcommand{\unam}{\label{unam}Instituto de Astronom\'{\i}a, Universidad Nacional Aut\'onoma de M\'exico, Ap. 70-264, 04510 CDMX, Mexico}
\newcommand{\COOL}{\label{COOL}Cosmic Origins Of Life (COOL) Research DAO, \href{https://coolresearch.io}{https://coolresearch.io}}
\newcommand{\TKU}{\label{TKU}Department of Physics, Tamkang University, No.151, Yingzhuan Road, Tamsui District, New Taipei City 251301, Taiwan}
\newcommand{\UCT}{\label{UCT}Department of Astronomy, University of Cape Town, Rondebosch 7701, Cape Town, South Africa}
\newcommand{\AIP}{\label{AIP}Leibniz-Institut for Astrophysik Potsdam (AIP), An der Sternwarte 16, 14482 Potsdam, Germany}
\newcommand{\Zagreb}{\label{Zagreb}Department of Physics, Faculty of Science, University of Zagreb, Bijeni\v{c}ka 32, 10 000 Zagreb, Croatia}
\newcommand{\ARK}{\label{ARK}Department of Physics, University of Arkansas, 226 Physics Building, 825 West Dickson Street, Fayetteville, AR 72701, USA}
\newcommand{\UConn}{\label{UConn}University of Connecticut, Department of Physics, 196A  Auditorium Road, Unit 3046, Storrs, CT, 06269}

\author{
    Alessandro~Razza\inst{\ref{UChile}} \and
    Guillermo~A.~Blanc\inst{\ref{Carnegie}, \ref{UChile}} \and
    Brent~Groves\inst{\ref{ICRAR}} \and
    Enrico~Congiu\inst{\ref{ESOChile}} \and
    Justus~Neumann\inst{\ref{MPIA}} \and
    Hsi-An~Pan\inst{\ref{TKU}} \and
    I-Ting~Ho\inst{\ref{MPIA}} \and
    Ashley~T.~Barnes\inst{\ref{ESO}} \and
    Francesco~Belfiore\inst{\ref{INAF}} \and
    M\'ed\'eric~Boquien\inst{\ref{UniCA}} \and
    Charlie~Burton\inst{\ref{Alberta}} \and
    M\'elanie~Chevance\inst{\ref{ITA},\ref{COOL}} \and
    Oleg~Egorov\inst{\ref{Heidelberg}} \and
    Eric~Emsellem\inst{\ref{ESO},\ref{ULyon}} \and
    Chris~Faesi\inst{\ref{UConn}} \and
    Simon~C.~O.~Glover\inst{\ref{ITA}} \and
    Kathryn~Grasha\inst{\ref{ANU}} \and
    Ralf~S.~Klessen\inst{\ref{ITA},\ref{IWR},\ref{CfA},\ref{Rad}} \and
    Kathryn~Kreckel\inst{\ref{Heidelberg}} \and
    Adam~K.~Leroy\inst{\ref{OSU}} \and
    Rebecca~McElroy \inst{\ref{UniSQ}} \and
    Ismael~Pessa\inst{\ref{AIP}} \and 
    Eva~Schinnerer\inst{\ref{MPIA}} \and
    Neven~Tomi\v{c}i\'{c}\inst{\ref{Zagreb}} \and   
    Amirnezam~Amiri\inst{\ref{ARK}} \and
    Gagandeep~S.~Anand\inst{\ref{STScI}} \and
    Yixian~Cao\inst{\ref{MPE}} \and
    Daniel~A.~Dale\inst{\ref{UWyoming}} \and
    Simthembile~Dlamini\inst{\ref{UCT}} \and
    Jing~Li\inst{\ref{Heidelberg}} \and
    J.~Eduardo~M\'endez-Delgado\inst{\ref{unam}} \and
    Eric~J.~Murphy\inst{\ref{NRAO}} \and
    Debosmita~Pathak\inst{\ref{OSU}} \and
    Miguel~Querejeta\inst{\ref{OAN}} \and
    Lise~Ramambason\inst{\ref{ITA}} \and
    Erik~Rosolowsky\inst{\ref{Alberta}} \and
    Fabian~Scheuermann\inst{\ref{Heidelberg}} \and
    Leonardo~\'Ubeda\inst{\ref{STScI}} \and
    Thomas~G.~Williams\inst{\ref{JBCA}}   
    % ...
    }
    
    \institute{
    \UChile{} \and
    \Carnegie{} \and
    \ICRAR{} \and
    \ESOChile{} \and
    \MPIA{} \and
    \TKU{} \and
    \ESO{} \and
    \INAF{} \and
    \UniCA{} \and
    \Alberta{} \and
    \ITA{} \and
    \COOL{} \and
    \Heidelberg{} \and
    \ULyon{} \and
    \UConn{} \and
    \ANU{} \and
    \IWR{} \and
    \CfA{} \and
    \Rad{} \and
    \OSU{} \and
    \UniSQ{} \and
    \AIP{} \and
    \Zagreb{} \and 
    \ARK{} \and
    \STScI{} \and
    \MPE{} \and
    \UWyoming{} \and
    \UCT{} \and
    \unam{} \and
    \NRAO{} \and
    \OAN{} \and
    \JBCA{}
    }

\date{Received XXX, 2025; accepted YYY, 2025}


\abstract{
    We present \pha, a narrow-band imaging survey that maps H$\alpha$ emission over a sample of 65 nearby massive star-forming galaxies. The data were obtained using the MPG-ESO 2.2-meter telescope at La Silla and the du Pont 2.5-meter telescope at Las Campanas Observatory, in the framework of the multi-wavelength cloud-scale (50--100~pc) resolution mapping of molecular gas and star formation conducted by the Physics at High Angular resolution in Nearby GalaxieS (PHANGS) collaboration. \pha\ complements the already published PHANGS-ALMA, PHANGS-MUSE, PHANGS-HST, and PHANGS-JWST surveys, providing an anchor point for the photometric and astrometric calibration of these datasets, as well as samples of \Hii\ regions, and star formation rate maps for the bulk of the PHANGS sample. We present observations, data processing, and calibration of the \pha\ dataset, as well as the procedures used to derive emission-line fluxes from narrow-band imaging. A subset of galaxies with available spectroscopic Ha mapping from the PHANGS-MUSE survey allows for a detailed comparison with the narrow-band photometry presented here. This informs a series of best practices for the processing of narrow-band H$\alpha$ imaging that we apply to the full dataset.
}


\titlerunning{\pha}
\authorrunning{\pha\ team}

\maketitle
\nolinenumbers

\section{Introduction}
\label{sec:intro}

The \Ha\ nebular emission line at 6563~\AA\ is typically the brightest Balmer and rest-frame optical emission line in galaxies.
It arises from hydrogen recombination in the ionized interstellar medium (ISM), which is primarily ionized by stars more massive than $\sim$15 M$_\odot$ with lifespans shorter than $\sim$10 Myr. These massive stars create discrete \Hii\ regions and contribute to the ionization of diffuse ionized gas (DIG). Owing to their short lifetimes, the \Ha\ emission effectively traces recent star formation, measuring a nearly instantaneous star formation rate \citep[SFR, e.g.][]{Kennicutt_1998}.

In recent years, integral-field spectroscopy (IFS) surveys such as CALIFA \citep[][]{Sanchez_2012}, SAMI \citep[][]{Croom_2012}, VENGA \citep[][]{Blanc_2013}, and MaNGA \citep[][]{Bundy_2015} have extensively mapped galaxies in the nearby universe. These surveys have produced spatially resolved maps of nebular emission in galaxies at $\sim 1$ kpc scale spatial resolution. Recently, the \mbox{PHANGS-MUSE} ESO large program \citep{Emsellem_2022} used the Multi Unit Spectroscopic Explorer (MUSE) instrument at the ESO Very Large Telescope (VLT) to deliver spatially resolved spectroscopy for a sample of 19 nearby galaxies. This data covers a plethora of nebular emission lines at the `cloud-scale' (50--100 pc) resolution needed to isolate individual ionized nebulae in these systems. Reaching cloud-scale resolution is essential to directly link star formation to its local environment and to resolve the impact of stellar feedback on individual star-forming regions.

The PHANGS-MUSE data can be used to measure the SFR, traced by H$\alpha$, at high physical resolution, but the limited $\sim1\arcmin$ FoV of MUSE makes it expensive to collect data for a large sample. On the other hand, SFR maps from all-sky UV \citep[GALEX,][]{Martin_2005} and IR \citep[WISE,][]{Wright_2010} surveys are compiled in the z0MGS atlas for $\sim$15,750 local galaxies \citep{Leroy_2019}. However, these maps are limited by coarser resolutions of $7.5\arcsec$ and $15\arcsec$ ($\sim$0.7 and $\sim$1.5~kpc at $D=20$~Mpc, respectively) and trace star formation on longer timescales ($\lesssim 100$~Myr), reflecting emission from a broader range of stellar masses than the short-lived massive stars responsible for \Ha\ emission.

Narrow-band imaging offers an observationally more efficient approach to map \Ha\ nebular emission across large samples of local star-forming galaxies. Compared to IFS, narrow-band imaging suffers from various limitations, including limited spectral coverage and lower spectral resolution (filter widths of $\sim$ 10--100 \AA), as well as more subtle complications such as the challenges of modelling and subtracting the underlying stellar continuum contribution and the contribution of neighbouring emission lines within the narrow-band filter pass-band. Nevertheless, narrow-band imaging remains an affordable technique that surpasses integral field unit (IFU) spectroscopy in certain aspects: by offering much larger fields of view, it enables more accurate astrometric and photometric calibration and allows efficient mapping of very nearby galaxies at high spatial resolution.

Various surveys have exploited the narrow-band imaging approach over the last few decades, spanning from Milky Way (MW) studies of stars emitting in \Ha\ \citep[e.g. IPHAS,][]{Drew_2005} to extra-galactic studies of star formation in very large sample of galaxies \citep[e.g. the ALFALFA \Ha\ Survey, 1555 galaxies between $\sim$20 and $\sim$100 Mpc,][]{vanSistine_2016}. Relevant studies on nearby galaxies include the SIRTF Nearby Galaxies Survey \citep[SINGS, 75 galaxies at $<$ 30 Mpc,][]{Kennicutt_2003}, the \Ha\ Galaxy Survey \citep[\Ha GS, 334 galaxies at $<$ 45 Mpc,][]{James_2004}, the Survey for Ionization in Neutral-Gas Galaxies \citep[SINGG, 468 galaxies with $\sim$75\% of the sample at $\lesssim$  30 Mpc,][]{Meurer_2006}, the 11 Mpc Ultraviolet and \Ha\ Survey \citep[11HUGS, 258 galaxies at $<$ 11 Mpc,][]{Lee_2007,Kennicutt_2008}, and the Javalambre Photometric Local Universe Survey \citep[J-PLUS, 805 galaxies at $<$ 75 Mpc][]{Vilella_2021}. Other surveys have targeted a sample of 140 irregular galaxies in the local universe \citep{Hunter&Elmegreen_2004}, or mapped extraplanar diffuse ionized gas (DIG) in 74 edge-on spiral galaxies \citep{Rossa&Dettmar_2003}.

In this work, we present the PHANGS-H$\alpha$ survey, a narrow-band imaging campaign focused on producing wide-field H$\alpha$ emission line maps covering the full extent of the disks of 65 nearby massive star-forming galaxies. The survey is conducted as part of the Physics at High Angular resolution in Nearby GalaxieS (PHANGS\footnote{\url{https://sites.google.com/view/phangs/home}}) collaboration, which coordinates a systematic multi-wavelength mapping of the interstellar medium (ISM) and the stellar component of a representative sample of nearby star-forming galaxies, reaching cloud-scale spatial resolution.

PHANGS-H$\alpha$ supplements already published datasets such as PHANGS-ALMA \citep[][which provides the parent sample for this work]{Leroy_2021}, PHANGS-MUSE \citep{Emsellem_2022}, PHANGS-HST \citep{Lee_2022} and PHANGS-JWST \citep{Williams_2024}. These datasets probe different phases of the ISM (molecular, ionized, and atomic via dust and PAH emission), and different stages of the star formation process (e.g. embedded and visible young clusters, HII regions, and supernova remnants). 

The 65 galaxies mapped by \pha , when supplemented with previous surveys like SINGS, SINGG, and 11HUGS, provide complete H$\alpha$ coverage of the full extent of the disks of the 90 targets in the PHANGS project parent sample (i.e.\ PHANGS-ALMA). In the context of the PHANGS multi-wavelength project, the \pha\ survey role is fundamental (in combination with PHANGS-ALMA CO maps) to trace the molecular gas-star formation lifecycle \citep{Schinnerer_2019,Chevance_2020,Pan_2022,Kim_2022,Romanelli_2025}, and to measure the SFR from \Ha\ + IR traces \citep{Belfiore_2023,Sun_2023,Querejeta_2024,Querejeta_2025,Leroy_2025}. In addition, the ground-based broad-band imaging taken as part of \pha\ also provides a common anchoring point for the photometric and astrometric calibration of the multi-wavelength PHANGS datasets. Furthermore, the availability of both integral-field spectroscopy from PHANGS-MUSE and narrow-band H$\alpha$ imaging from \pha\ on the same targets allows us to directly evaluate the performance of the methodologies used to extract emission-line fluxes from narrow-band images, and define a new set of best practices for processing this type of dataset.

We start by presenting the \pha\ survey in Section~\ref{sec:sample} where the sample selection is introduced in the context of the PHANGS-ALMA parent sample. The observational campaign for the survey is described in Section~\ref{sec:imaging}. The methods used to reduce and calibrate the narrow- and broad-band images with a custom data reduction pipeline are detailed in Section~\ref{sec:reduction}. The procedure used to derive the \Ha\ emission-line flux maps from the calibrated images is explained in Section~\ref{sec:final_calib}. A thorough validation of these methodologies is presented in Section~\ref{sec:muse_comp} where we compare our measurements to the spectroscopically derived \Ha\ line flux maps from PHANGS-MUSE. Finally, conclusions are given in Section~\ref{sec:conclusions}.

\section{The \pha\ sample selection}
\label{sec:sample}

\subsection{The PHANGS parent sample and surveys}
\label{subsec:parent_sample}

The PHANGS surveys share a common parent sample of nearby ($D\lesssim 20$ Mpc) massive star-forming galaxies ($\log(M_*/M_\odot) > 9.75$) spanning the SFR-$M_*$ main sequence ($\log(\mathrm{sSFR}/\mathrm{yr}^{-1})>-11$) that are also moderately inclined ($i < 75^{\circ}$) and observable from the southern hemisphere ($-75^{\circ}<\delta < +25^{\circ}$). These criteria were applied to select targets for the PHANGS-ALMA large program \citep{Leroy_2021}, which produced CO (2-1) maps of 90 galaxies at $\sim 1\arcsec$ resolution using the Atacama Large Millimeter/submillimeter Array (ALMA). The sample selection and distribution of target properties are thoroughly discussed by \cite{Leroy_2021}. The goal was to create a representative sample of massive disk galaxies across the star forming main sequence that are sufficiently nearby, so clouds of 50-100~pc scales could be resolved at the $\sim 1\arcsec$ resolution of the adopted ALMA array configuration, and of seeing-limited optical ground-based instruments like MUSE.

PHANGS-ALMA serves as the parent sample for all the main PHANGS surveys. PHANGS-HST \citep{Lee_2022} produced high-resolution (0.08\arcsec) broad-band UV-optical imaging for 38 galaxies using the WFC3 and ACS instruments on the Hubble Space Telescope (HST). The survey has mapped and catalogued over $\sim$100,000 young stellar clusters and stellar associations \citep{Maschmann_2024}. The PHANGS-JWST Treasury Survey \citep{Lee_2023, Williams_2024} has obtained eight near and mid-IR imaging bands from 2 to 21~$\mu$m for 19 galaxies using the NIRCam and MIRI instruments on the James Webb Space Telescope (JWST), mapping the embedded young cluster population and the emission from dust and Polycyclic Aromatic Hydrocarbons (PAHs) at high spatial resolution in the ISM of these targets. As mentioned in Section \ref{sec:intro}, the same sample of 19 PHANGS galaxies were mapped using the MUSE integral-field-spectrograph on the ESO VLT as part of the PHANGS-MUSE large program \citep{Emsellem_2022}. The MUSE data delivers seeing-limited ($\sim$1\arcsec) integral-field spectroscopic mapping over the 4750--9350 \AA\ range at low spectral resolution ($R\sim2500$) for these targets.

\subsection{\pha\ sample}
\label{subsec:halpha_sample}

The PHANGS-H$\alpha$ sample selection aims to overcome the limited sample size of the PHANGS-MUSE survey and yield seeing-limited ($\sim$1\arcsec) SFR tracer maps that complement the CO maps available for the full PHANGS-ALMA parent sample. The \pha\ sample consists of 63 galaxies from this parent sample that do not have available high quality narrow-band H$\alpha$ imaging from the SINGS \citep{Kennicutt_2003}, SINGG \citep{Meurer_2006}, and 11HUGS \citep{Lee_2007} surveys. Two additional targets (IC 1993 and NGC 1326), observed with \dP , are not currently in the ALMA sample.

\begin{figure}[h!]
    \centering
    \includegraphics[width=0.5\textwidth]{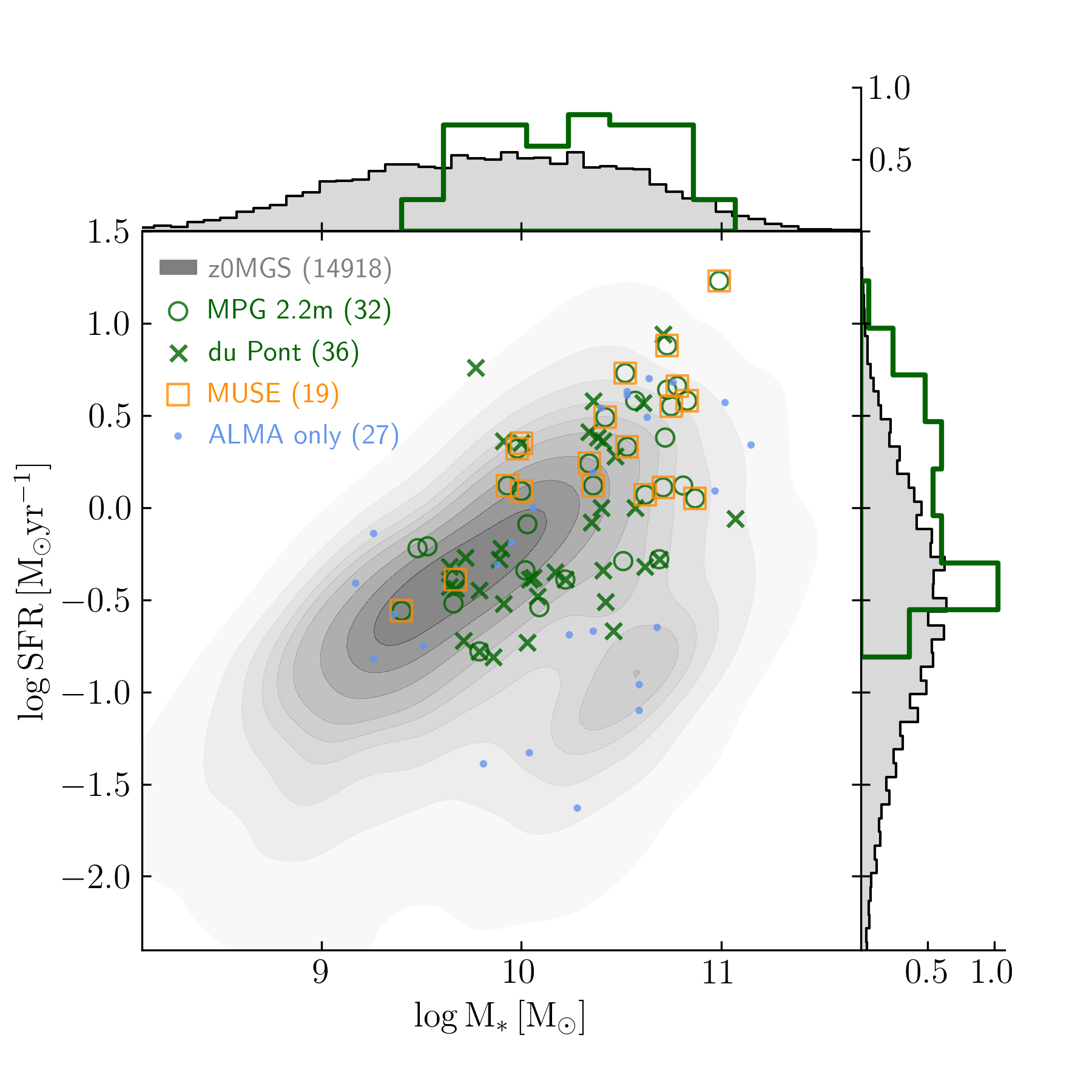}
    \caption{SFR-M$_*$ plot showing the distribution of the \pha\ galaxy sample. Green circles indicate those observed with WFI on the MPG 2.2m telescope and green crosses the galaxies observed with the Direct CCD camera on the du Pont telescope (see Section~\ref{sec:imaging} for the observations and instruments description). There are 3 targets observed with both instruments. The galaxies overlapping with the PHANGS-MUSE survey are highlighted by orange squares. All the remaining galaxies from PHANGS-ALMA sample are marked with blue dots. The grey contours in the background show the \citet{Leroy_2019} atlas of 14918 local ($D\leq 50$ Mpc) galaxies observed with GALEX and WISE. In the top and right plots, 1D distributions of the SFRs and stellar masses show that \pha\ sample is on average more massive and star forming than z0MGS. The central plot shows that \pha\ galaxies are star-forming main sequence and green valley galaxies. The SFR and M$_*$ values are from \citet{Leroy_2021} and are listed in this paper in Table~\ref{tab:sample}.}
    \label{fig:SFMS}
\end{figure}

Figure~\ref{fig:SFMS} presents the distribution of the \pha\ sample in the M$_*$--SFR plane. Galaxies are shown differentiated by the instruments used to observe them and the sub-sample in common with PHANGS-MUSE is highlighted. In the background, we show the distribution of a large sample (14918) of z0MGS galaxies at $D\leq 50$ Mpc \citep{Leroy_2019}, for which SFR and M$_*$ measurements are available. Figure \ref{fig:SFMS} clearly shows that \pha\ galaxies are representative of massive star-forming main sequence galaxies in the local universe. The complete target list of \pha\ is presented in Table~\ref{tab:sample}.

\section{Observations}
\label{sec:imaging}

Imaging for the \pha\ galaxies was collected from October 2016 through April 2019 during 7 observing runs with the MPG-ESO\footnote{The telescope is owned by the Max Plank Institute for Astronomy and operated by the European Southern Observatory (ESO).} 2.2-meter telescope (hereafter \MPG ) at ESO's La Silla Observatory, and 6 observing runs with the Ir\'{e}n\'{e}e du~Pont 2.5-meter telescope (hereafter \dP ) at Las Campanas Observatory (LCO\footnote{LCO is owned and operated by The Observatories of the Carnegie Institution for Science.}). In total, 32 galaxies were observed with the Wide Field Imager \citep[WFI,][]{Baade_1999} at the \MPG\ telescope and 36 with the Direct CCD Camera (hereafter DirectCCD) at the \dP, with three targets observed with both instruments.

As a general scheme, galaxies with larger angular sizes (R$_{25}\gtrsim 3\arcmin$) were targeted with WFI, which offers a wider FoV, while the remaining galaxies were observed with DirectCCD. Each target was observed with an \Ha\ narrow-band (NB) and an $R$ broad-band (BB) filter to subtract the underlying stellar continuum contribution to the NB filter (see Section~\ref{sec:final_calib}). As an initial strategy, we acquired three 300~s frames with the BB filter and six 445~s frames with the NB filter for each galaxy. For galaxies with a bright nucleus, we acquired additional short exposures with the BB filter to recover potentially saturated pixels. We applied dithering between individual exposures to remove gaps between CCDs (WFI) and bad pixels (in WFI and DirectCCD). Standard de-trending calibration frames were taken, including bias frames, twilight flats, and dome flats for each filter.

When possible, we acquired more or longer exposures with the goal of increasing the depth of the images and replacing observations with bad image quality, i.e.\! with PSF Full Width Half Maximum (FWHM) $>\,1.5\arcsec$, or acquired in non-optimal conditions (thin or tick clouds). The image quality constraint was required to match the ALMA data resolution and to avoid problems with the PSF-matching process before continuum-subtraction (Section~\ref{sec:final_calib}). The PSF FWHM for each individual exposure spans from $0.62\arcsec$ to $1.46\arcsec$ for the WFI and from $0.66\arcsec$ to $1.41\arcsec$ for the DirectCCD images.

For most observations, the atmospheric conditions were non-photometric. While initially spectrophotometric standard star fields were observed as part of the survey, we later chose to use Gaia DR2 \citep{Gaia_DR2} stars to perform the absolute photometric calibration of the images (see Section~\ref{subsec:photometry}). Using foreground stars in the science images makes the flux calibration process less sensitive to the absence of photometric conditions.

\subsection{MPG 2.2m/WFI observations}
\label{subsec:2p2}

A total of 23 nights of observation were collected from October 2016 to February 2019 with WFI at the MPG 2.2m telescope. WFI consists of a mosaic of 8 2k$\times$4k pixels CCDs with a scale of 0.238$\arcsec/$pixel. Each CCD covers an area of $8.12\arcmin\times 16.25\arcmin$ with vertical and horizontal gaps between CCDs of 23$\arcsec$ and 14$\arcsec$, respectively, contributing to a total FoV for the camera of $\sim 34\arcmin\times 33\arcmin$. We estimated each CCD gain and readout noise by means of the Janesick's method \citep{Janesick_2001}, finding gain values of $\sim$1.5--2.1 e$^-$/ADU and readout noise of $\sim$4.3--6.9 e$^-$ \footnote{Values from the observing date 2016/10/07.}.

\begin{figure}[h!]
    \centering
    \includegraphics[width=0.5\textwidth]{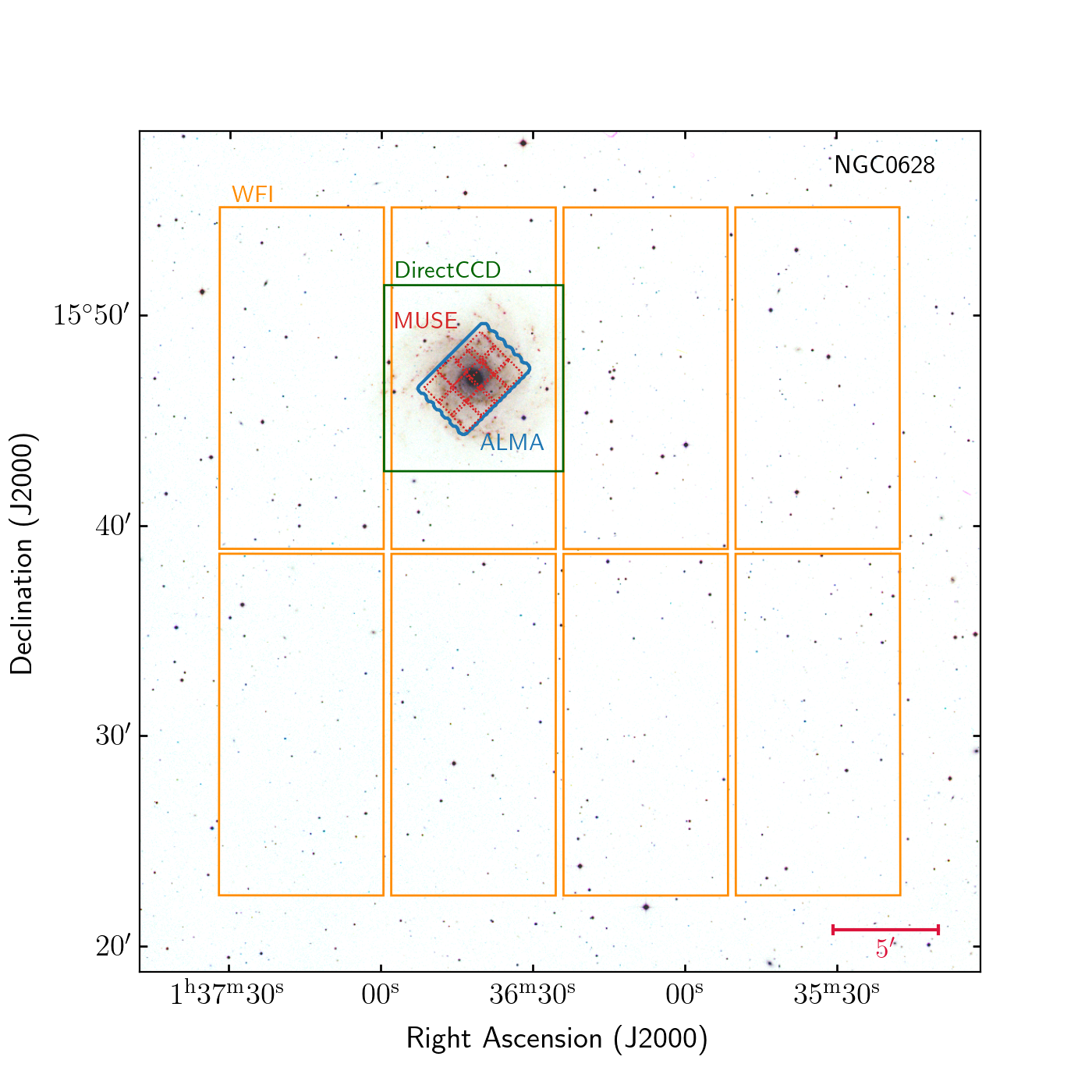}
    \caption{RGB coloured image of NGC 0628 from DSS2 blue, red and IR images. Overlaid are FoVs of WFI (orange), DirectCCD (green), MUSE (red) and ALMA (blue). ALMA and MUSE footprints showed in the plot represent PHANGS-ALMA and PHANGS-MUSE observations of the same galaxy, originating from several beams and pointings.}
    \label{fig:FoV}
\end{figure}

In Figure~\ref{fig:FoV}, we compare the WFI FoV (orange) with that of DirectCCD (Section~\ref{subsec:dupont}, green), on top of a composite colour image of NGC0628 from the DSS2 survey (blue, red and IR images). The PHANGS-ALMA and PHANGS-MUSE coverage of the same galaxy are shown for comparison in blue and red respectively. To achieve sufficiently extended spatial coverage, PHANGS-MUSE required a mosaic of 12 individual pointings for this galaxy. It should be pointed out that NGC0628 is shown centred in WFI CCD\#2 in Figure~\ref{fig:FoV}, as an indication of the strategy adopted for all the galaxies observed with this instrument. This strategy is motivated by the need of recovering a sufficient number of foreground stars outside the galaxy for calibration purposes, while avoiding the low quality CCD\#1 (high number of bad pixels) and CCD\#4 (high bias levels). During the observational campaign, the number of working CCDs in WFI went down. While we have a few galaxies observed with the full mosaic, we could use 6 adjacent CCDs for the majority of the targets.

\begin{figure*}[h!]
    \centering
    \includegraphics[width=\textwidth]{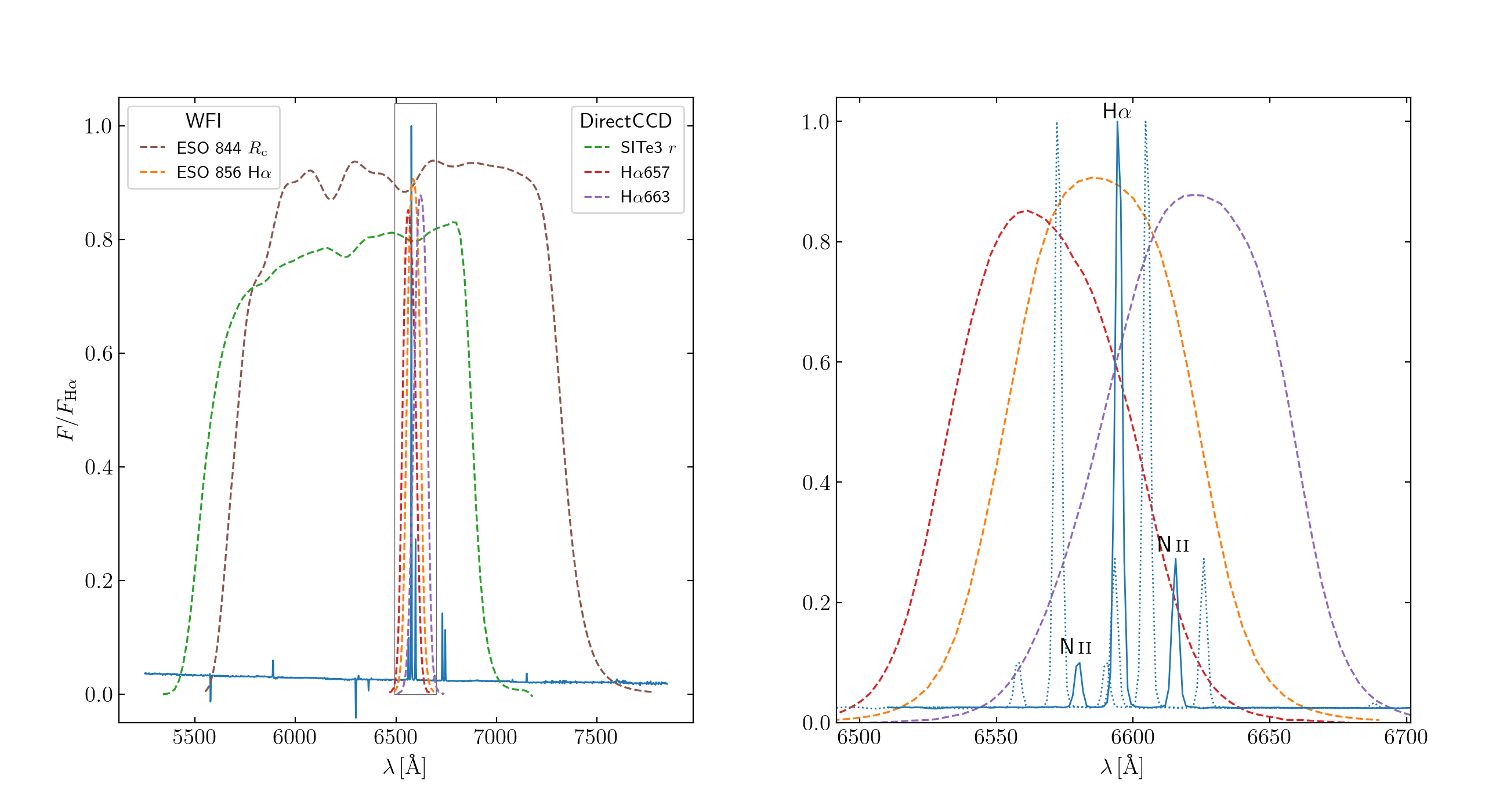}
    \caption{\textit{Left}: Set of WFI and DirectCCD broad- and narrow-band filter transmission curves overlaid on an NGC0628 \Hii\ region spectrum from PHANGS-MUSE data. The \emph{ESO $R_\mathrm{c}$ 844} (brown dashed line) and the SITe3 $r$ (green dashed lines) broad-band filter curves are clearly seen. \textit{Right}: Inset of the left-hand side plot in the wavelength range of the \Ha\ emission line and the \Nii\ doublet. The ESO narrow-band filter for WFI (orange dashed line) and the two \Ha 657 (red dashed line) and \Ha 663 (magenta dashed line) filters employed in the DirectCCD observations are shown. The solid blue line represents the spectrum shifted at the recessional velocity of 1460 km s$^{-1}$. Below this threshold, galaxies were observed with the DirectCCD filter \Ha 657, whereas the filter \Ha 663 was used for galaxies with higher receding velocities. The dotted blue lines show the closest and the most distant galaxy within the \dP\ sample.}
    \label{fig:Filters}
\end{figure*}

The broad- and narrow-band filters used in WFI are \emph{ESO $R_\mathrm{c}$ 844} (hereafter \Rc\ filter) and \emph{ESO \Ha\ 856} (hereafter \Ha\ filter), respectively. All the filters employed in the observational campaign are shown in the left-hand side of Figure~\ref{fig:Filters} where the filter transmission curves are plotted on top of a normalized PHANGS-MUSE spectra of a NGC0628 \Hii\ region. A zoom on \Ha\ and the two \Nii\ nebular lines is shown in the right panel of the figure, where the observed spectrum is plotted as a solid blue line redshifted for the line-of-sight velocity of 1460 km s$^{-1}$, which is the threshold chosen to differentiate the DirectCCD targets to be observed with the bluer or the redder \Ha\ filter (see Section~\ref{subsec:dupont}).  Shown as dotted blue lines are spectra of the same \Hii\ region shifted at the receding velocities of the closest (NGC 4424, 447 km s$^{-1}$) and the most distant (NGC 1317, 1931 km s$^{-1}$) DirectCCD target. The filter characteristics, specifically the reference wavelength and the filter width are reported in Table~\ref{tab:filters} for the WFI BB and NB filters.

\begin{table*}[h!]
\caption{\label{tab:filters}\pha\ Filters}
\centering
\begin{tabular}{@{\extracolsep{4pt}}cccccccc}
\hline\hline\\[-6pt]
 &  & \multicolumn{3}{c}{Broad-band filter} & \multicolumn{3}{c}{Narrow-band filter}\\[2pt]
\cline{3-5} \cline{6-8}\\[-6pt]
Telescope & Instrument & Name & $\lambda_\mathrm{p}$\tablefootmark{\scriptsize{a}} & Width\tablefootmark{\scriptsize{b}} & Name & $\lambda_\mathrm{p}$\tablefootmark{\scriptsize{a}} & Width\tablefootmark{\scriptsize{b}}\\[4pt]
 &  &  & (\AA ) & (\AA ) &  & (\AA ) & (\AA )\\[2pt]
\hline
& & & & & & & \\[-6pt]
MPG 2.2m & WFI & ESO $R_\mathrm{c}$ 844 & 6515 & 1523 & ESO \Ha\ 856 & 6588 & 77\\
\multirow{2}*{du Pont} & \multirow{2}*{DirectCCD} & \multirow{2}*{SITe3 $r$} & \multirow{2}*{6228} & \multirow{2}*{1257} & H$\alpha$657 & 6566 & 75\\
 & & & & & H$\alpha$663 & 6621 & 78\\[3pt]
\hline
\end{tabular}
\tablefoot{This table pivot wavelength values are used to convert \pha\ map units from $\mu$Jy to ergs s$^{-1}$ cm$^{-2}$ \AA $^{-1}$ for the surface brightness profile comparisons in Section~\ref{subsec:compare_profile}.\\
\tablefoottext{a}{Pivot wavelength calculated from the filter transmission curves and the formula from \citet{Tokunaga_2005}.}\\
\tablefoottext{b}{Filter width computed as $\int T_\lambda\,\mathrm{d}\lambda / \mathrm{max}(T_\lambda )$, where $T_\lambda$ is the given filter transmission curve.}
}
\end{table*}

\subsection{du Pont/DirectCCD observations}
\label{subsec:dupont}

Imaging with the DirectCCD camera at the \dP\ telescope was conducted over the period between January 2018 and April 2019 with a total of 22 full nights of observations. The DirectCCD is a $8.85\arcmin\times 8.85\arcmin$ FoV camera with a single CCD detector (SITe2K) with a scale of $0.259\arcsec$ per pixel. We observed our targets in fast readout mode, which gives an inverse-gain of 1.54 e$^-/$ADU and a readout noise of 6.5 e$^-$. Given the limited FoV compared with that of WFI (see Figure~\ref{fig:FoV}) we focused these observations on the targets with smaller angular diameters. The limited FoV posed a problem concerning the number of foreground stars available for astrometric and photometric calibration for some galaxies in fields with a low foreground stellar density (see Section~\ref{sec:reduction}).

The set of filters used with the DirectCCD comprises a Gunn $r$ LCO BB filter (hereafter $r$ filter) and two shifted LCO NB filters, namely \Ha 657 and \Ha 663, to account for the different galaxy redshifts. The criteria to use the bluer or redder \Ha\ filter by the galaxy receding velocity is demonstrated in the right-hand plot of Figure~\ref{fig:Filters}. Given the lack of a measured filter curve for the DirectCCD $r$ filter, we adopt that of the identical SITe3 Camera $r$ filter at the LCO Swope telescope instead, for which extensive characterization is available from the Carnegie Supernova Project \citep[CSP,][]{Stritzinger_2011,Krisciunas_2017}. The filters used for the DirectCCD observations and their characteristics are described in Table~\ref{tab:filters}. 


\section{Data reduction}
\label{sec:reduction}

The \pha\ data reduction process uses a set of pipelines designed to obtain high quality data through rigorous CCD processing and accurate flux and astrometric calibrations. An accurate astrometric solution is necessary to assess the spatial distribution of \Ha\ emission relative to the CO emission maps from the PHANGS-ALMA parent sample. In this context, previous works have shown the importance of high astrometric precision of both \Ha\ and CO emission to measure the spatial offset between them \citep{Schinnerer_2019, Pan_2022} and accurately infer cloud and feedback timescales \citep{Chevance_2020, Kim_2022}.

Likewise, the flux calibration of the BB and NB images is required to be accurate enough so that the flux uncertainties reach the level of systematic uncertainty associated with the recipes adopted to estimate the \Ha\ line fluxes from the NB images (see Section~\ref{sec:final_calib}). In addition, calibrated fluxes of the BB images (along with their astrometric solutions) have been used as a reference for calibrating other PHANGS datasets \citep[e.g.][]{Emsellem_2022}.

In the following subsections, we describe in detail each of the processing steps used to reduce and calibrate the BB and NB images. Initially, each DirectCCD image and each CCD of the WFI mosaic is processed separately (Section~\ref{subsec:processing}). The detrended CCD images are subsequently projected onto a common coordinate system derived from an astrometric solution (Section~\ref{subsec:astrometry}), to then be photometrically calibrated against Gaia DR2 stars present in each exposure (Section~\ref{subsec:photometry}). Individually, each of the astrometrically and photometrically calibrated exposures are background subtracted (Section~\ref{subsec:bkg_subtr}) before being combined to produce a pair of BB and NB final images for each galaxy in the survey (Section~\ref{subsec:weight_comb}).

\subsection{CCD processing}
\label{subsec:processing}

We use the \texttt{python} \texttt{ccdproc} package \citep{Craig_2016} to implement the data reduction steps, such as bias and overscan subtraction, flat-fielding, cosmic rays removal, bad pixels, and columns masking into our CCD processing pipeline. In addition to these standard CCD detrending steps, the pipeline finds a rough astrometric solution for each CCD to enable a more precise image registration as explained in Section~\ref{subsec:astrometry}. Other major products of the CCD processing pipeline are the standard deviation and mask images, created for each processed CCD in each exposure, to account for the flux error and to flag bad pixels.  What follows is a description of the main processes in the order they are executed in the CCD processing pipeline with a primary focus on the non-standard procedures adopted.

\medskip
 
\noindent\textit{Rough astrometric solution\ ---}
For all the CCDs, the World Coordinate System (WCS) keywords are written in the headers to attach right ascension (RA) and declination (DEC) coordinates to each pixel position. The galaxy centres are taken as reference point for the transformation of the pixel coordinates onto the celestial sphere. A transformation matrix is then used to account for the instrument pixel scale to obtain an initial astrometric solution. The solution for this CCD is then propagated to the other CCDs while accounting for the gaps. This procedure provides a rough initial WCS which significantly simplifies the more rigorous WCS computation described in Section~\ref{subsec:astrometry}.

\medskip

\noindent\textit{Cosmic rays rejection and pixel masking\ ---}
Cosmic rays are identified with the L.A. Cosmic technique as described in \citet{vanDokkum_2001} and implemented in \texttt{ccdproc} via the \texttt{astroscrappy} package \citep{McCully_2018}. The routine creates the initial mask for each CCD by marking all the pixels with detected cosmic rays. Particular attention is spent on fine tuning the L.A. Cosmic routine to reduce the number of large areas with false positives in the galaxy centres. At this stage, pixels with values greater than 40000 Analogue-Digital Unit (ADU) are also masked for being saturated or in the CCD non-linear regime\footnote{Arbitrary value based on a linear limit for a CCD of $\sim$ 70\% of its typical full well capacity ($\sim 60000$ ADU).}. This mask image associated to each CCD is further updated in later stages of the data reduction.

\medskip

\noindent\textit{Overscan- and bias-subtraction\ ---}
Both the WFI and DirectCCD cameras are cryogenically cooled, therefore the dark current is negligible and the overscan essentially measures the bias level. By using the overscan regions, we are also safe against small image-to-image variations in the bias over the course of a single night\footnote{We noticed variations in the bias level between 180 and 220 ADU over the course of the observational campaign with WFI. Three CCDs (\#4, \#7 and \#8) experienced high jumps for few nights, as well.}. We average the data in the overscan region, present at the right edge of the exposure frames for both WFI and DirectCCD, computing the mean along the x direction for each row, and the results are fitted with a high-order polynomial along the y direction. The high order polynomial serves to capture variations as a function of position in the chip, as we noticed towards the edges of the WFI images. Typical overscan levels are of $\sim 200$ ADU for WFI and $\sim 540$ ADU\footnote{We observed variations of 450--640 ADU during the observation campaign.} for DirectCCD. A row-wise subtraction of the fitted overscan region is made for each CCD and for each science, bias and flat-field image. The overscan-subtracted bias exposures, taken in the same night, are then combined together in a masterbias\footnote{The masterbias is anyhow needed to remove remaining column-wise variations of the bias level within the CCD.}. A per CCD (master)bias-subtraction is performed on the science and flat-field images. 

\medskip

\noindent\textit{Gain scaling and error propagation\ ---}
After the bias-subtraction, standard deviation frames are created. By assuming Poisson statistics and neglecting the dark current, the error for each pixel at coordinates $(x, y)$ is initially
\begin{equation}
    \sigma (x, y) = \sqrt{\mathcal{G}\times p(x,y) + \sigma_\mathrm{RN}^2},
    \label{eq:error_prop}
\end{equation}
where $\mathcal{G}$ is the inverse-gain, $p(x,y)$ is the bias-subtracted pixel value in ADUs and $\sigma_\mathrm{RN}$ is the readout noise in electrons. A per pixel error propagation is then realised throughout the CCD processing and the rest of the data reduction.

Owing to a lack of precise gain and read out measurements for WFI detectors, we employed Janesick's method \citep{Janesick_2001} to compute a first guess of inverse-gain and readout values for each CCD from a pair of bias and a pair of flat-field exposures. The gain corrected masterflats (see below) of each CCD are then scaled to each other by computing the ratios between the average counts in 500 pixels taken from adjacent corners to estimate an ``empirical" gain correction between WFI CCDs. This procedure assumes that the flat-field image is uniform in the centre with equal counts in contiguous areas. Although not accurate, the inverse-gain values found with this method satisfy the relative scaling we are interested in, since after the reprojection of the CCDs onto the common WCS (Section~\ref{subsec:astrometry}), the WFI reconstructed mosaic is treated as an individual image for the photometric calibration (Section~\ref{subsec:photometry}).

\medskip

\noindent\textit{Flatfielding and normalization\ ---}
When available, we use the twilight flat exposures acquired in the same night of the science exposures to correct them for the point-to-point CCD sensitivity, or those taken during the same observing run, assuming no significant variations over the course of a few nights. We paid attention to take exposures short enough to keep the flat-field images counts below the 40k ADU, therefore within the linear regime\footnote{Simultaneously, we paid attention to keep them long enough with the \dP\ to avoid a shadow effect of the telescope shutter. Above 12 seconds the flux is affected by less than $0.5$\%.}. As for the masterbias images, we create a masterflat for each CCD by combining all the flat-field images from the same night. The masterflats normalization is different for the two instruments: we divide by the median value of the masterflat counts for the DirectCCD, whereas we divide by the median of the median values of each of the WFI 8 CCDs as common normalization. Each science exposure is divided by the respective masterflat and by the exposure time to normalize it to units of electrons s$^{-1}$ pixel$^{-1}$. The standard deviation images of each CCD are treated in the same way and the errors are propagated. 

\medskip

\noindent\textit{Masking artifacts\ ---}
The masterflats are used to update the mask frames with a \texttt{ccdproc} routine based on \texttt{IRAF} \texttt{ccdmask} task \citep{Tody_1986} where isolated or grouped pixels with values deviating from the median of neighbouring pixels are flagged. With this method, the majority of camera hot spots, electron traps and bad pixel columns are identified and included in the mask. However, a few artifacts are still present, especially in WFI images. For example, there are bad pixel columns and cosmic rays that escaped the automatic detection routines, satellite and asteroid tracks, and ghosts due to cross-talk between the amplifiers. To mask all the remaining defects not removed when combining the dithered images, we add manually masked regions created with DS9.

\subsection{Astrometry}
\label{subsec:astrometry}

\begin{figure*}[ht!]
    \centering
    \includegraphics[width=0.49\textwidth]{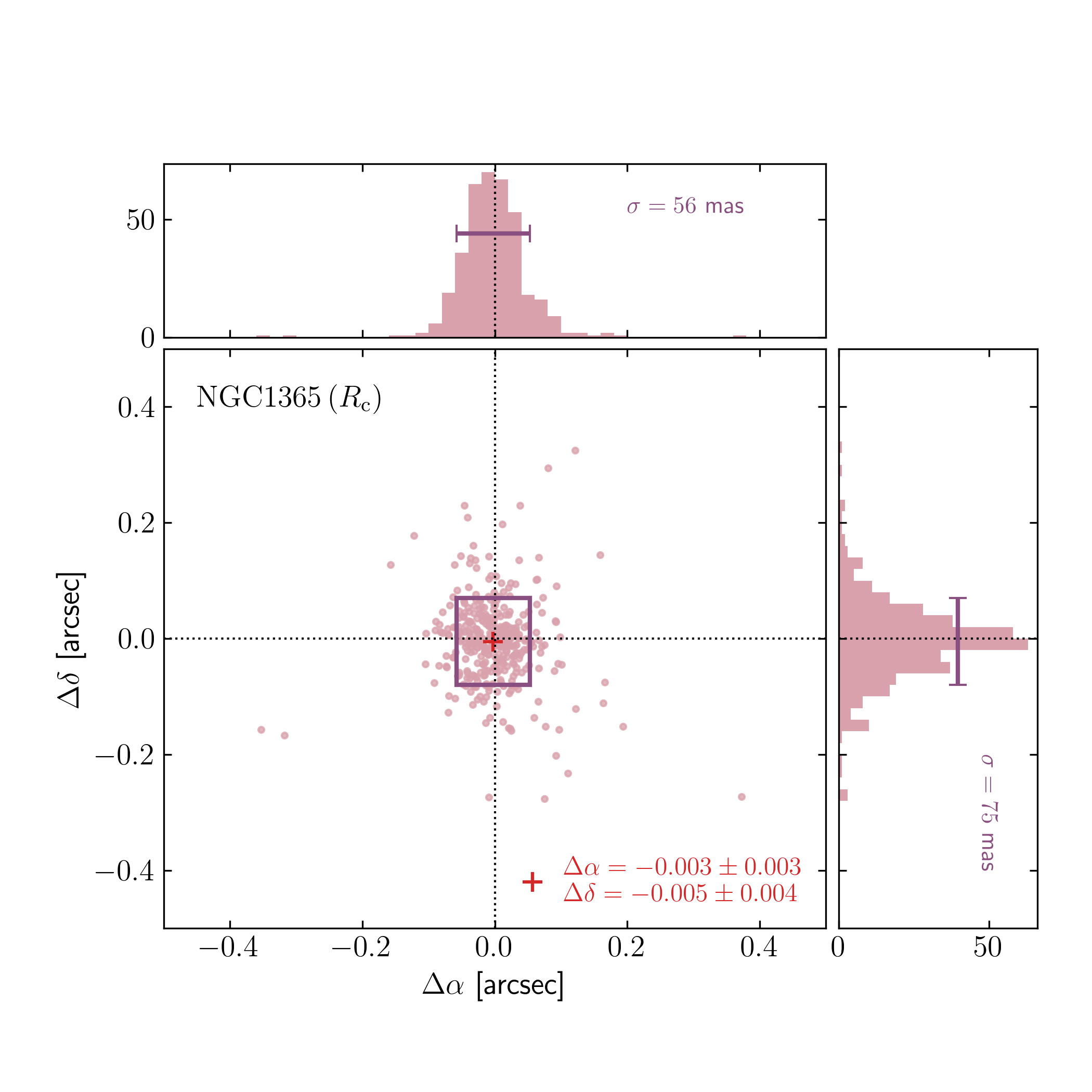}
    \includegraphics[width=0.49\textwidth]{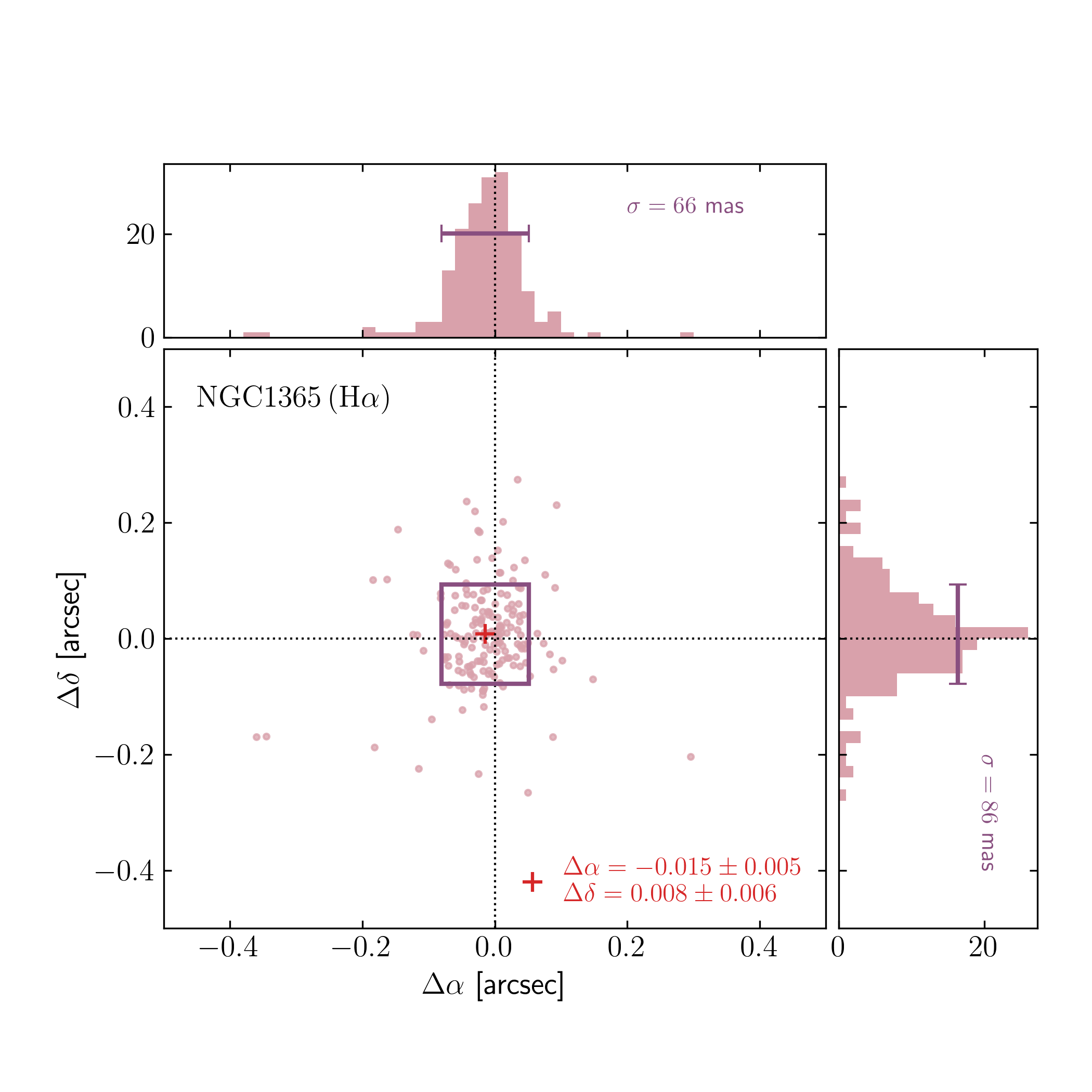}
    \caption{In the central scatter plots, RA and DEC offsets (\dra\ and \ddec , respectively) between stars centre positions computed from \pha\ final combined images for NGC 1365 and their positions from matched stars in Gaia DR2 catalogue are shown for (\emph{left}) WFI \Rc\ filter and (\emph{right}) WFI \Ha\ filter images. The red crosses in the plot centres show the mean, whereas the rectangle sides (violet) the 2$\sigma$ of \dra\ and \ddec\ distributions. Mean and standard deviation values are reported in red in the bottom-right corner of the central plot. The plots above and to the right of each central scatter plot are the 1D histograms of \dra\ and \ddec\ distributions, respectively. A violet bar representing the 2$\sigma$ of the distribution is shown in each histogram. This plots do not reflect the final astrometric solution that comes from applying a simple translation of the WCS coordinates in the opposite direction of the found mean offsets.} 
    \label{fig:astro_ngc1365}
\end{figure*}

The rough WCS solution described in Section~\ref{subsec:processing} provides the initial conditions needed to achieve an accurate astrometric calibration by means of three different steps where: (1) the correct WCS solution is found for each CCD by using Gaia DR1\footnote{At the time we performed the survey data reduction, v2.6.2 of SCAMP was used, which only included GAIA DR1. After the completion of the data reduction, Gaia DR2 became available. We thus computed and corrected the residual astrometry offset of the images against this later version (see astrometric calibration step (3) described in this section).} \citep{Gaia_mission, Gaia_DR1} stars as reference, (2) all the CCDs/exposures belonging to the same galaxy and the same filter observation are projected onto a common WCS and (3) a residual shift toward Gaia DR2 stars is applied to the final broad- and narrow-band images (produced as described in Section~\ref{subsec:weight_comb}).

The astrometric calibration steps (1) and (2) are based on a wrapper around three pieces of software: \sex\footnote{\href{https://www.astromatic.net/software/sextractor/}{https://www.astromatic.net/software/sextractor/}} \citep{Bertin_1996} to build catalogues of sources extracted from the input CCD images, \scamp\footnote{\href{https://www.astromatic.net/software/scamp/}{https://www.astromatic.net/software/scamp/}} \citep{Bertin_2006} to perform a cross-identification between these sources and Gaia DR1 stars to then compute the astrometric solutions to be read by \swarp\footnote{\href{https://www.astromatic.net/software/swarp/}{https://www.astromatic.net/software/swarp/}} \citep{Bertin_2002} that resamples and co-adds all the CCDs (from all the exposures) belonging to the same galaxy into their common astrometric projection. This latter projection is then used, another time by \swarp , as target WCS to reproject each individual science, error and mask frame. For the reprojection, we employ the \swarp\ built-in interpolation function LANCZOS3, as it gives the best compromise between flux conservation and artifact generation around saturation trails and bad pixel columns\footnote{\url{https://raw.githubusercontent.com/astromatic/swarp/legacy_doc/prevdoc/swarp.pdf}, page 18.}. The same interpolation function is employed for the mask frames to guarantee the same area is flagged after the reprojection. At the end of this per-frame process, all the science, error and mask frames belonging to the same galaxy share the same WCS solution, one for each set of images from the BB and NB filter observations.

To accomplish the step (3) of the astrometric registration, which consists in a final refinement of the WCS solution found in the previous steps, we develop a set of \texttt{python} scripts for the pipeline to detect stars in each galaxy field and compare their positions with those of Gaia DR2 stars. The final images from combining all the exposures (Section~\ref{subsec:weight_comb}) are used in order to maximize the star detection. By assuming a circular PSF, stars are extracted and their centre positions are computed with the IRAFFIND-like algorithm from \texttt{photutils} package \citep{Bradley_2021}\footnote{We noticed that the employed \texttt{python}-based IRAFFIND algorithm rejected a few cases more of blended stars compared to its DAOPHOT counterpart.}. These centre positions are then one-to-one matched with Gaia DR2 stars $\leqslant 0.5\arcsec$ distant to measure per-galaxy and per-filter mean RA and DEC offsets (\dra\ and \ddec). The mean of each image offsets is then used to further refine the astrometric solution by applying a simple translation in the opposite direction.

\begin{figure*}[ht!]
    \centering
    \includegraphics[width=1\linewidth]{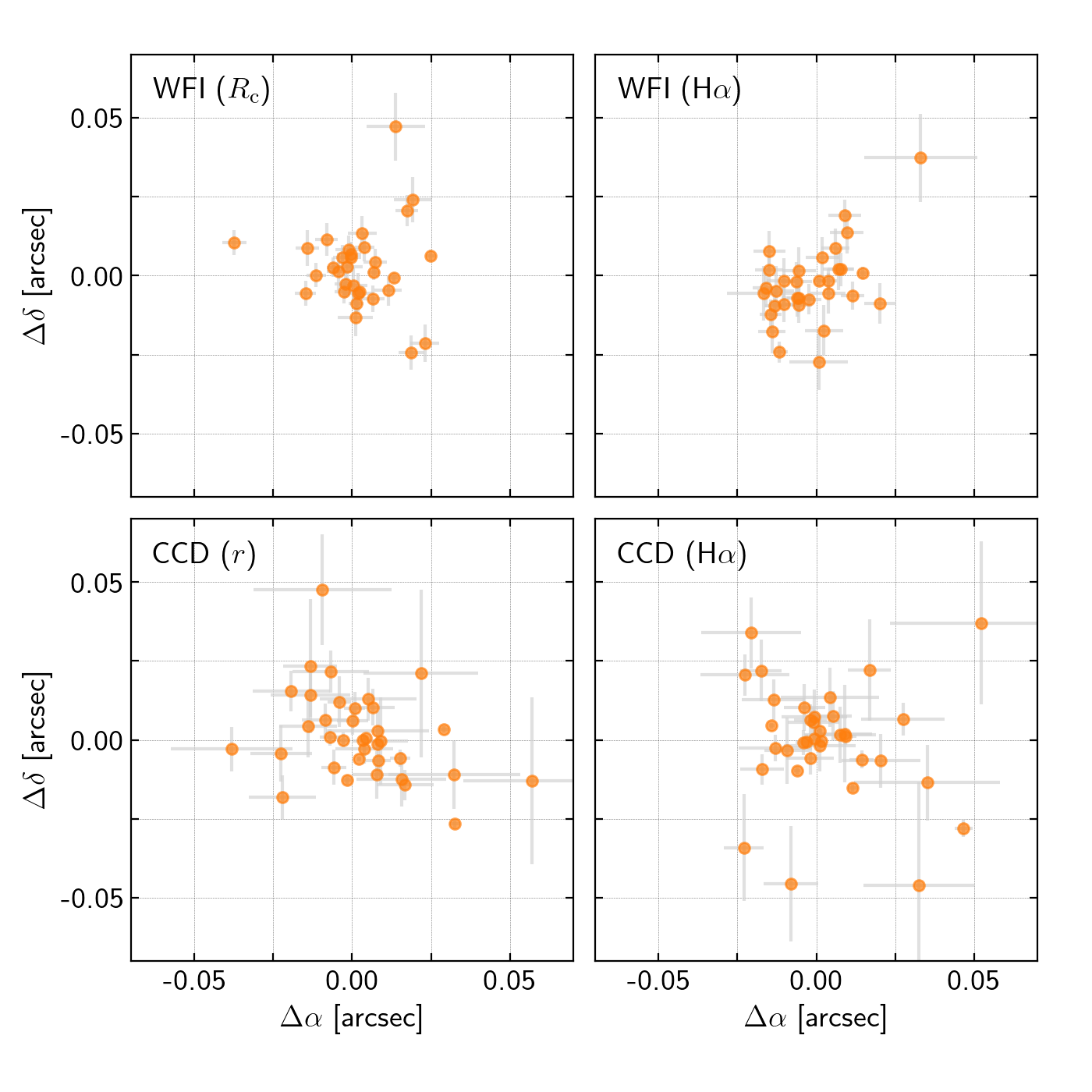}
    \caption{Mean values of the \dra\ and \ddec\ offsets computed for each galaxy of \pha\ survey as in the example of Figure~\ref{fig:astro_ngc1365}. Each point represents the resulting mean value for each galaxy observed with (\textit{top left}) WFI BB, (\textit{top left}) WFI NB,(\textit{bottom left}) CCD BB and (\textit{bottom right}) CCD NB filters, whereas the error bars show the errors of the mean in each coordinate.}
    \label{fig:all_gals}
\end{figure*}

Unlike the previous astrometric registration steps (2) and (3), which rely on \sex , \scamp\ and \swarp\ for a thorough WCS registration and projection, the simple coordinate translation of step (3) allows the use of our own custom-made pipeline. This approach permits a greater flexibility in star selection criteria, ultimately improving the astrometric solution precision. Before cross-matching, Gaia DR2 foreground stars are indeed photometrically selected by their colour and those within an ellipse with a semimajor axis 1.2 times larger than the galaxy R$_{25}$ and the semiminor defined by the inclination (values from Table~\ref{tab:sample}) are rejected. Both criteria to select the stars, i.e.\ the stellar colour selection and the position inside/outside the galaxy ellipse are consistent with those used for the photometric calibration described in Section~\ref{subsec:photometry}. In particular, selecting stars by their position with respect to the galaxy gives rise to a more accurate astrometric solution since we noticed that stars within the galaxy ellipse show systematically larger \dra\ and \ddec\ offsets compared to those outside. This is likely owing to the background galaxy flux causing the star centre to be poorly determined. On the other hand, we did not find any trend of the astrometry accuracy with the star position in the WFI mosaic or with the star proper motion as measured in the Gaia catalogue. Therefore we do not apply any cut for star location (apart from those inside the galaxy ellipse) or for its proper motion.
 
In Figure~\ref{fig:astro_ngc1365}, we show an example of offset calculations for NGC 1365. The \dra\ and \ddec\ offsets between each star detected in the final WFI \Rc\ (left) and  \Ha\ (right) images and its matched position in Gaia DR2 catalogue (after applying the selection criteria described above) are shown in the central plots. The 1D distributions for \dra\ and \ddec\ are shown above and to the right, respectively. The central red cross in each plot of Figure~\ref{fig:astro_ngc1365} represents the mean of the offset distribution with its value reported in the bottom-right corner. The dispersion, shown as a rectangle in the 2D plots and as bars in the 1D histograms, is calculated as the standard deviation of the offsets and provides an estimate of the uncertainty of the astrometric solution.

While Figure~\ref{fig:astro_ngc1365} exemplifies the method used to assess the accuracy and precision of the astrometric solution for each galaxy, Figure~\ref{fig:all_gals} presents the results for the full sample. Each point in Figure~\ref{fig:all_gals} represents the \dra\ and \ddec\ mean offset calculated for each galaxy with the error bar representing the offset uncertainty. Results are divided in four plots, one for each instrument and filter. As can be seen, the vast majority of the \dra\ and \ddec\ offsets falls within 25 mas with moderately larger scatter and errors for the CCD observations owing to the smaller number of stars used to find the astrometric solutions.

As mentioned, the resulting mean offsets are used to apply the WCS coordinates translation in order to centre at zero the astrometric solution for each galaxy and filter. After this correction, we repeat the procedure to estimate the astrometric solution accuracy against Gaia DR2 finding residuals in the mean of the offset distributions of 1--2 mas for 89\% of WFI and 85\% of DirectCCD galaxies. 
The dispersions remain unchanged after the final corrections and extend over an interval of 50--100 mas for WFI and 10--100 mas for DirectCCD for the 95\% of the galaxies in both cases. The median values of the astrometric uncertainties, thus taking into account the number of stars used to compute the dispersions, are $\sim$6 and $\sim$8 mas respectively for WFI \Rc\ and \Ha , and $\sim$10 and $\sim$12 mas respectively for CCD $r$ and \Ha\ images.

As a final remark, the WCS solutions for the BB final images have been used as reference to project the NB image before performing the continuum subtraction (Section~\ref{subsec:psf_match}) and to anchor the WCS of other datasets such as PHANGS-MUSE.

\subsection{Flux calibration}
\label{subsec:photometry}

The photometric calibration is achieved for the BB images by finding the average zero point (ZP) of all the stars in each frame to match the instrumental flux to the Gaia magnitudes. To achieve this, photometric relationships from \citet{Evans_2018} are used to transform Gaia passbands ($G$, \gbp\ and \grp ) to the \Rc\ Johnson-Cousin filter for WFI and Sloan $r$ filter for DirectCCD. The reference stars in the Gaia DR2 catalogue are selected with a colour cut chosen to minimize the scatter in the ZPs. 
Subsequently, the NB images are calibrated from the transformation of Gaia bands into our BBs by fitting a BB$-$NB vs.\ \gcol\ relation, from a set of spectrophotometric standard stars (SPSSs) from the Gaia SPSS project \citep{Pancino_2021}. All the individual frames are calibrated separately to correctly account for the variation in observing conditions.

\subsubsection{Broad-band images}
\label{subsec:broad}


We start by selecting stars in our images that are suitable for accurate aperture photometry and have counterparts in the Gaia catalogue with reliable photometric measurements. The star selection is realized by first extracting the star-like sources outside the galaxy disk (1.2 times larger than R$_{25}$, as in Section~\ref{subsec:astrometry}) with the IRAFFIND routine assuming a FWHM of 4 pixels for the PSF (corresponding to the mean image quality of our observations) and imposing a lower limit for the source extraction of 5$\sigma$ above the background. 
The sources are then matched with Gaia DR2 stars at a distance of 0.5\arcsec\ for which Gaia photometry is available. Differently from the stars selection adopted for the astrometric calibration, we also require that the stars are isolated from masked pixels for a correct aperture photometry, which results in the rejection of bright stars, as pixels above the CCD linear regime (40000 ADU) are masked by the pipeline (see Section~\ref{subsec:processing}). This requirement, combined with the imposed 5$\sigma$ detection threshold, effectively defines a bright and faint limit for the stars to be used for the calibration.  

\begin{figure*}[ht!]
    \centering
    \includegraphics[width=\textwidth]{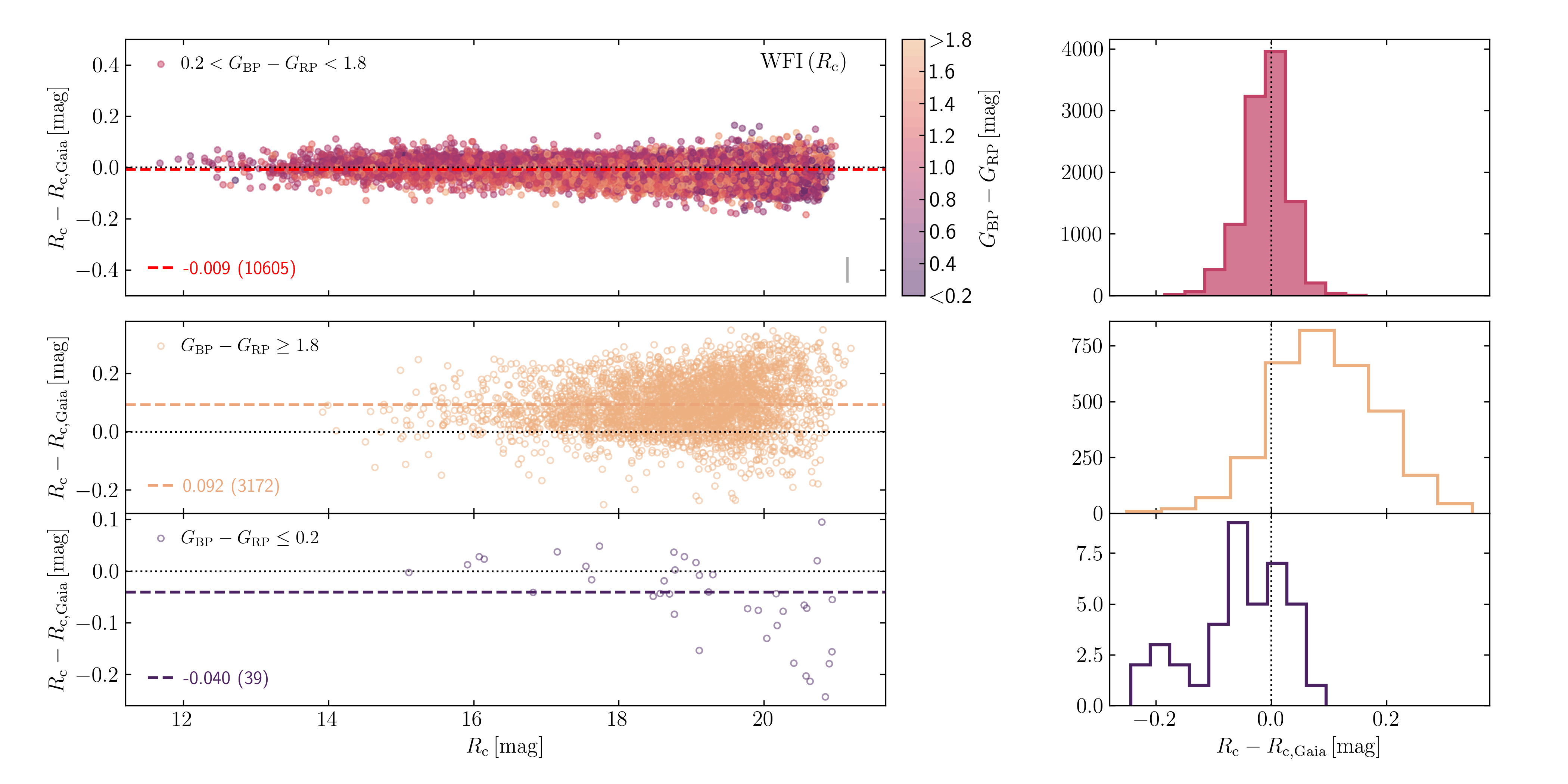}
    \caption{\emph{Left panels}: scatter plots where each point represents a star detected in any of the final galaxy images observed with the WFI, after combining the exposures. For each star, the difference between the calibrated \Rc\ photometry and the Gaia passbands transformations to the Johnson-Cousins \Rc\ ($R_\mathrm{c}^\mathrm{Gaia}$) is plotted against \Rc . Each point is colour-coded by Gaia star colour \gcol\ within the range from Equation~\ref{eq:Gaia_transf_color_cut}. Yellow and magenta open circles are used to indicate, respectively, the stars with Gaia colour $\geq 1.8$ (centre-left panel) and $\leq 0.2$ (bottom-left panel). The dashed horizontal lines indicate the mean of \Rc $-R_\mathrm{c}^\mathrm{Gaia}$ values among stars with the adopted colour cut (red dashed line on the top-left panel) and for stars outside the colour range (yellow and magenta dashed lines for stars redder than 1.8 mag and bluer than 0.2 mag, respectively). The number of stars used for the mean computations is given in parenthesis. In the bottom-right corner of the top-left plot, the average photometric error is displayed. \emph{Right panels}: 1D Gaia colour distributions of the set of stars within and outside our adopted colour cut, plotted in the left panels.}
    \label{fig:photo_BB}
\end{figure*}

Additional selection criteria must be implemented for the Gaia-matched stars to fulfil the requirements of \citet{Evans_2018} to apply their transformations from Gaia $G$, \gbp\ and \grp\ passbands to Johnson-Cousins \Rc\ and SDSS $r$ photometric systems (see the transformation equations in Appendix~\ref{app:photocal}). 
We require Gaia DR2 stars to satisfy the same conditions of brightness, colour and uncertainty in Gaia passbands photometry described by \citet{Evans_2018}\footnote{See also the documentation (Table 5.5, 5.6, and 5.9) in  \url{https://gea.esac.esa.int/archive/documentation/GDR2/Data_processing/chap_cu5pho/sec_cu5pho_calibr/ssec_cu5pho_PhotTransf.html}.}, with few exceptions: we include stars with high uncertainty in Gaia passbands photometry to increase the statistics\footnote{As we noticed, these are stars with BB magnitudes larger than $\sim$19--20 that increase the ZP computation precision without modifying its accuracy.} and a stricter condition in the colour cut is imposed as follows 
\begin{equation}
    0.2<G_\mathrm{BP}-G_\mathrm{RP}<1.8. 
\label{eq:Gaia_transf_color_cut}
\end{equation}
The use of this colour range is empirically justified by the photometric residuals shown in Figure~\ref{fig:photo_BB} (see the discussion below). In addition, inspection of the transformation relations presented by \citet{Evans_2018} indicates that the polynomial fits are most tightly constrained within this interval, while larger colour terms and increased scatter appear toward the boundaries of the calibration range. Restricting the sample to the domain in Equation~\ref{eq:Gaia_transf_color_cut}, therefore, minimizes residual colour dependencies in the adopted transformations.

Once the stars are selected by means of all the mentioned criteria, we compute their instrumental fluxes from the aperture photometry with local background subtraction. The apertures are chosen in pixels to be 3 times larger than the FWHM adopted to detect the stars in the images and the background annuli are defined with inner and outer radii of 20 and 30 pixels, respectively. Provided the associated error map, the aperture photometry error is estimated as a combination of the error of the pixels in the aperture and the error in the background. The zero-points for each star are then calculated and combined into a single zero-point for the image, following the procedure outlined in Appendix~\ref{app:photocal}.

In Figure~\ref{fig:photo_BB}, we present the photometric residuals measured after applying our photometric calibration procedure to the final exposure-combined images (Section~\ref{subsec:weight_comb}) of all the 32 galaxies observed with the WFI in \Rc\ band.
The difference between the calibrated \Rc\ and the Gaia passbands transformed to the Johnson-Cousins \Rc\ ($R_\mathrm{c}^\mathrm{Gaia}$) is plotted against \Rc\ for each star in all the 32 images. Each point representing a star is colour-coded by the Gaia colour \gcol. The lower and upper colour limits we adopt are chosen to reduce a bias in our ZP computations. While the low limit of 0.2 already corresponds to the limit of \citet{Evans_2018} (and has a minor impact on WFI calibration for the limited number of stars bluer than 0.2), the upper limit removes a large number of stars in a colour range where the transformations are more poorly constrained.

As we can see, there is no preferential colour distribution within the colour cut from Equation~\ref{eq:Gaia_transf_color_cut}. On the other hand, the stars with \gcol $\,\geq 1.8$ (yellow open circles) and \gcol $\,\leq 0.2$ (magenta open circles) tend to exhibit lower and higher values of $R_\mathrm{c}^\mathrm{Gaia}$, respectively. This is particularly evident for the large number of stars redder than 1.8 where the mean value of the \Rc $-R_\mathrm{c}^\mathrm{Gaia}$ differences represents a one order of magnitude larger offset (yellow dashed line) than for the stars within our adopted colour cut (red dashed line). This result confirms that the adopted BB calibration procedure yields consistent zero points across the sample. We do not show this post-facto photometric accuracy check for the galaxies observed with the DirectCCD $r$ band or with the narrow-band filters, as their results look similar to those in Figure~\ref{fig:photo_BB}.

\subsubsection{Narrow-band images}
\label{subsec:narrow}

\begin{figure*}[ht!]
    \centering
    \includegraphics[width=0.49\textwidth]{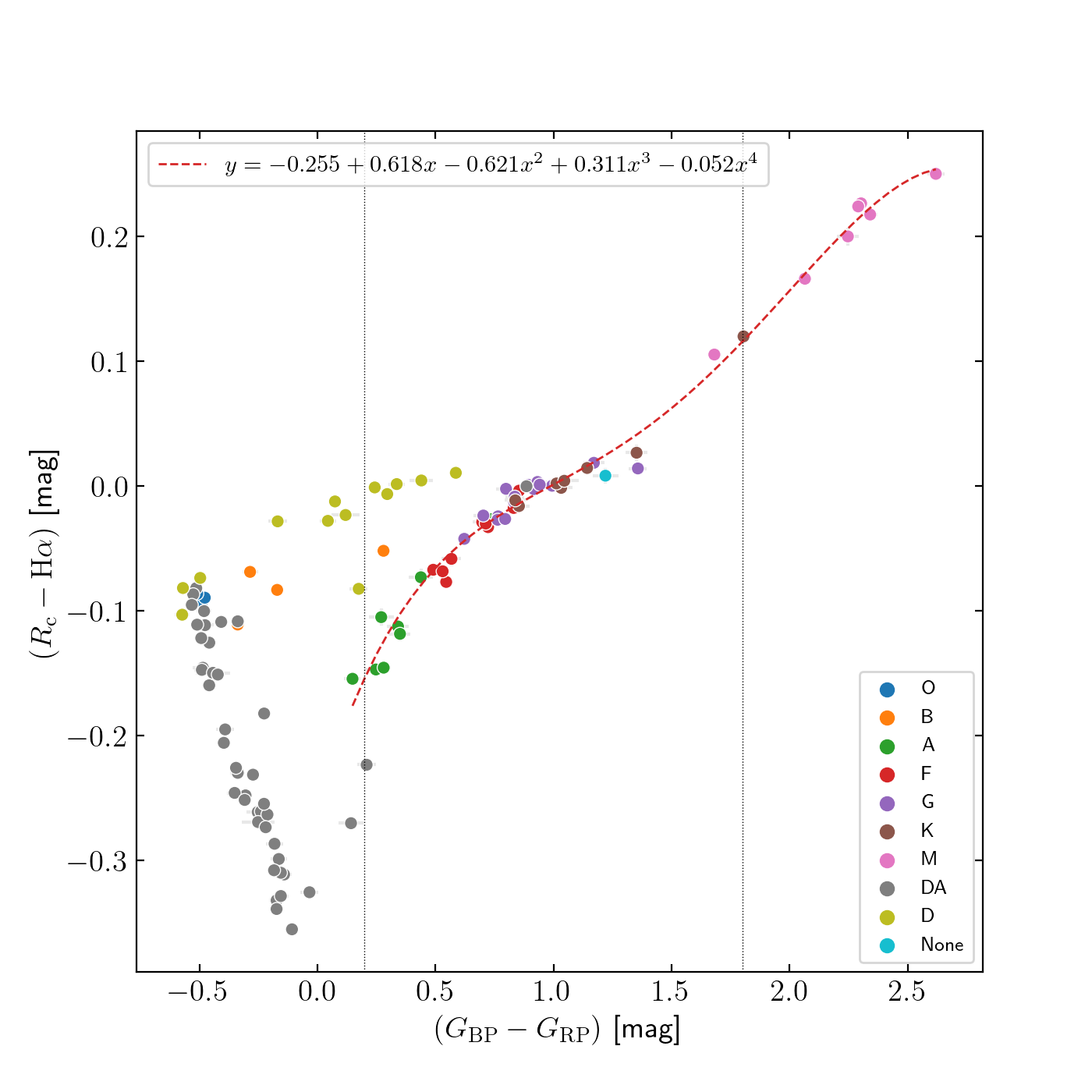}\\
    \includegraphics[width=0.49\textwidth]{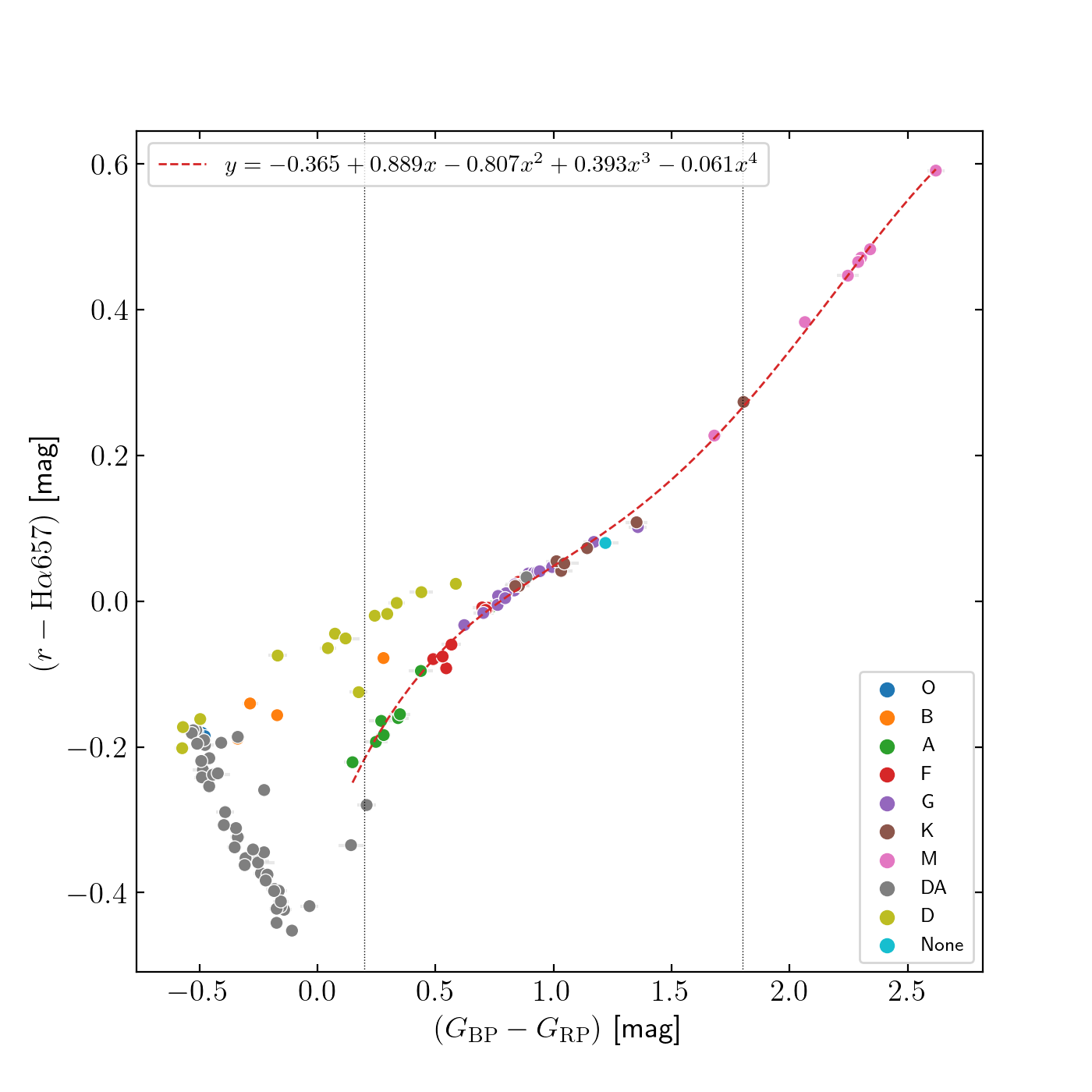}
    \includegraphics[width=0.49\textwidth]{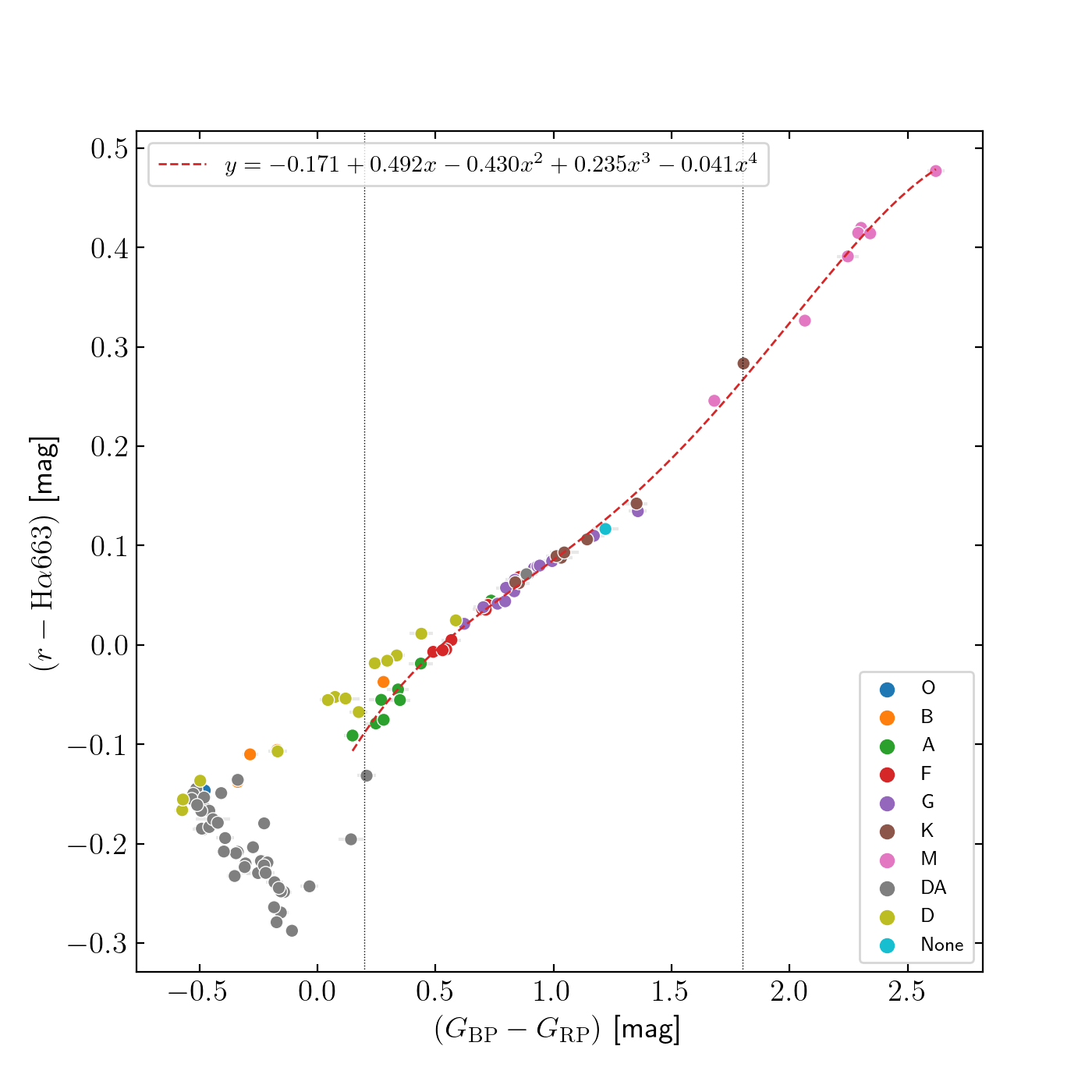}
    \caption{Colour-colour relations are shown for (\emph{top}) $R_\mathrm{c}-$\Ha , (\emph{bottom-left}) $r-$\Ha 657 and (\emph{bottom-right}) $r-$\Ha 663 against \gcol . To fit these relations, a grid of 112 SPSSs from Gaia spectrophotometric standard star survey \citep{Pancino_2021} is used. White Dwarf (types D and DA) and OB stars are excluded from the fit. The two vertical dashed lines, representing the Gaia colour limits used for the photometric calibration, are drawn to highlight the range of validity of the fit, although the whole colour range of the Gaia SPSSs is used for fitting, to better constrain the redder part of the colour limit.}
    \label{fig:photo_SPSS}
\end{figure*}

To calibrate the NB images, we adopt the same star selection criteria as in Section~\ref{subsec:broad} for both the catalogue of sources extracted from each individual exposure and the Gaia DR2 catalogue. After calculating the Gaia-predicted BB magnitudes of the selected stars using the \citet{Evans_2018} transformations (Appendix~\ref{app:photocal}), we determine their NB magnitudes by fitting a colour-colour relation for $R_\mathrm{c}-$\Ha , $r-$\Ha 657 and $r-$\Ha 663 against \gcol. To fit a ``universal" colour-colour relation, i.e.\! to be used to calibrate all the NB exposures against the BB ones, we employ a grid of 112 SPSSs from Gaia spectrophotometric standard star survey \citep{Pancino_2021}, for which calibrated spectra and \gcol\ colours are available. The synthetic magnitudes for the \pha\ filters, used to compute the BB$-$NB colours, are obtained from the SPSS spectra.

The colour-colour plots in Figure~\ref{fig:photo_SPSS} show the relations we derive by fitting a 4th-order polynomial. This was done after removing the White Dwarfs (D and DA types) and OB main sequence stars that do not share the same trend as the main sequence stars from type A to M, which make up the bulk of the population of foreground stars in our images. From these fits we find the following colour-colour relations:
\begin{equation}
\begin{aligned}
    R_\mathrm{c} - \mathrm{H}\alpha &= - 0.255 + 0.618x - 0.621x^2\\ 
    & + 0.311x^3 - 0.052x^4\\ 
    r - \mathrm{H}\alpha 657 &= - 0.365 + 0.889x - 0.807x^2\\ 
    & + 0.393x^3 - 0.061x^4\\
    r - \mathrm{H}\alpha 663 &= - 0.171 + 0.492x - 0.430x^2\\ 
    & + 0.235x^3 - 0.041x^4
\end{aligned}
\label{eq:CC_relations}
\end{equation}
with $x=$ \gcol.

The Gaia-based NB magnitudes, resulting from combining the \citet{Evans_2018} transformations and the Equation~\ref{eq:CC_relations} fitted relations, are compared with the aperture photometry to find a ZP for each star (Equation~\ref{eq:zp_formula}). A final average among all the ZPs, weighted by the ZP errors, is computed for each frame as detailed in Appendix~\ref{app:photocal} for both the BB and NB photometric calibrations.

\subsection{Background subtraction}
\label{subsec:bkg_subtr}

\begin{figure*}[ht!]
    \centering
    \includegraphics[width=\textwidth]{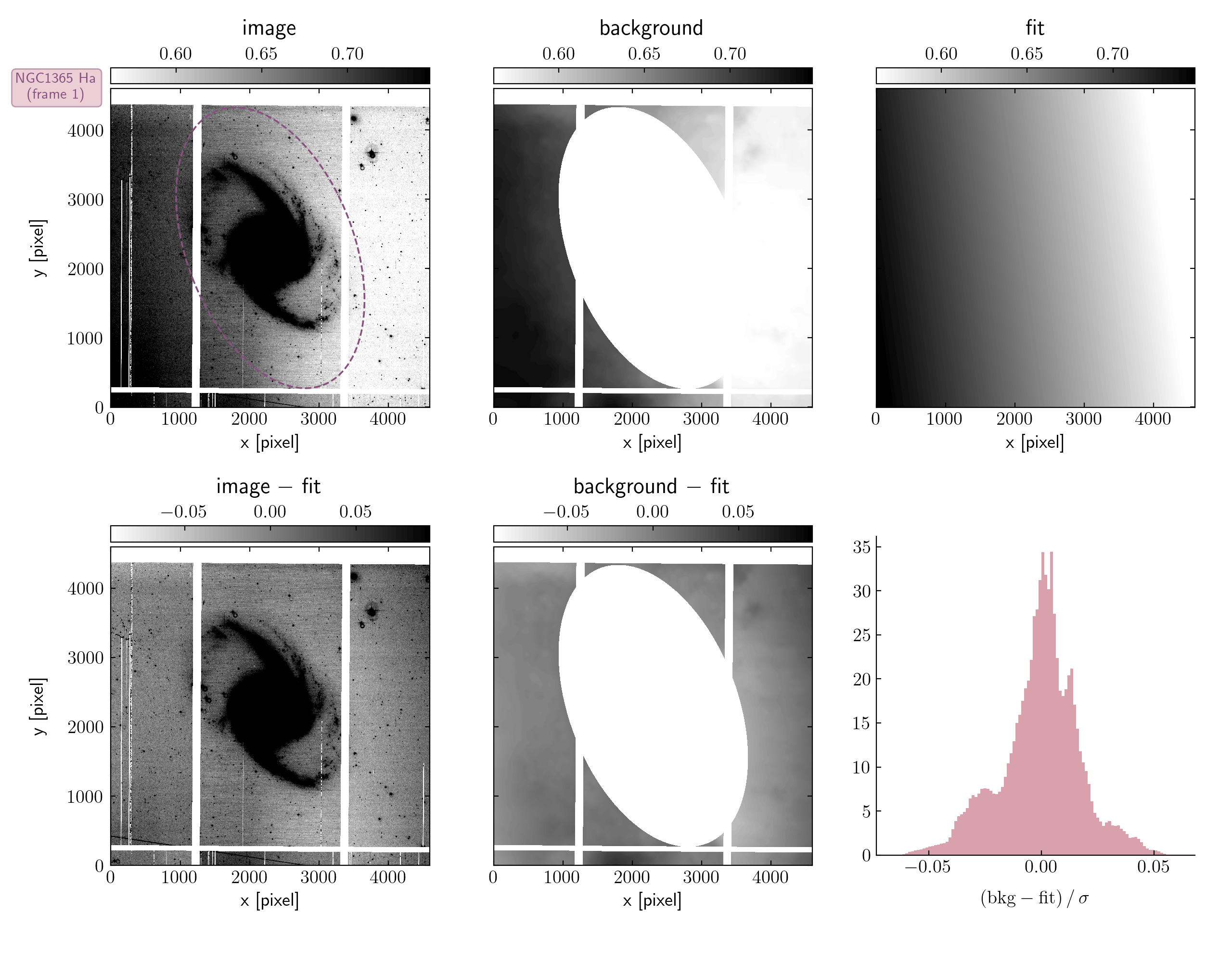}
    \caption{Illustration of the background subtraction procedure for a 450-second exposure of NGC 1365 taken with the WFI \Ha\ filter. A smaller cutout is used to show the procedure effectiveness. \textit{Top panels}: Plots presenting the fitting method where, starting from the science frame (\emph{image}), the SExtractor-like background of the unmasked pixels is generated (\emph{background}), and used to fit a first order 2D surface (\emph{fit}). For comparison, the same gray scale limits are used for all the three plots. \textit{Bottom panels}: Result plots showing the background-subtracted frame (\emph{image}$-$\emph{fit}), the SExtractor-like background residual (\emph{background}$-$\emph{fit}), and a histogram of the background residuals normalized by the image noise from the error map of the same frame. Both the background-subtracted and background residuals plots have the same scale limits as those on top but with the range centred to zero. Although other sources like stars are not shown in white in the plots as the CCD gaps and the galaxy ellipse, they are masked in the procedure.}
    \label{fig:bkg}
\end{figure*}

Once the science exposures are astrometrically and photometrically calibrated, the background needs to be subtracted from each of them before being combined together. A per-frame background subtraction is required to account for the observing conditions (e.g. gradients induced by moon illumination, structure in airglow emission, etc.) that may change from one exposure to another producing a different background level. To compute the background, the bad pixel mask produced by the pipeline (Section~\ref{subsec:processing}) is used along with a "source" mask that includes all pixels brighter than a 2-$\sigma$ threshold above the image (sigma-clipped) median. An area 1.4 times larger than the galaxy ellipse defined by the semimajor (R$_{25}$) axes, the ratio between the major and the minor axes and position angle from \citet{Lang_2020} (see Table~\ref{tab:sample}) is also masked. The factor 1.4 is chosen to sufficiently mask any galactic light while still preserving enough sky within the FoV to determine the background. This is especially important in the smaller FoV of the \dP\ observations.

After masking bad pixels and sources, a "SExtractor-like" background computation\footnote{\url{https://sextractor.readthedocs.io/en/latest/Background.html}} is implemented using the \texttt{photutils} routine to create a smoothed image of the background-only (unmasked) pixels. A tilted plane is then fitted to this masked smooth background using a Levenberg–Marquardt least-squares algorithm \citep{LMA_1978}. This fitted background plane captures the large-scale background variations outside the galaxy ellipse while recovering the background level inside. We noticed that fitting the SExtractor-like background image reduces the background rms compared with fitting directly the science image.

In Figure~\ref{fig:bkg}, we illustrate our procedure to measure, fit and subtract the background from an individual science frame. As an example, a 450-second exposure of NGC 1365 taken with the \Ha\ filter is used to demonstrate the effectiveness of the procedure, even when limited sky area is available around the galaxy ellipse. The top panels show the procedure steps where a masked exposure (\emph{image}) is used to compute the SExtractor-like background (\emph{background}) that is in turn employed to find the first order fitting surface (\emph{fit}). The procedure successfully recovers the background gradient present in the image. The bottom plots show the background-subtracted exposure (\emph{image}$-$\emph{fit}), and the background residual image (\emph{background}$-$\emph{fit}), both obtained after subtracting the 2D plane surface fit, along with a histogram of all the pixels in the background residual image normalized by the exposure error map. The histogram provides an estimate of both the background subtraction accuracy and the goodness of the 2D surface fit.

We noticed a few cases where small residuals in a CCD-to-CCD variation in the background level yield larger variations in the final images. For the exposure shown in Figure~\ref{fig:bkg}, we measure an average difference of 0.019 $\mu$Jy between background areas in the upper-right and lower-left corners, which is negligible compared to the background RMS in the same areas (0.102 $\mu$Jy). In a future release, a per-CCD plane fitting routine will be explored to mitigate more significant cases of CCD-to-CCD background variation, mostly found in large apparent-size galaxies from the forthcoming release of the \pha\ extended sample (not discussed in this paper).

\subsection{Weighted combination}
\label{subsec:weight_comb}

The last step of the \pha\ data reduction combines all the calibrated, background-subtracted exposures of the same galaxy, observed with the same filter. Each science exposure has an associated bad pixel and error map which are respectively used to select the valid pixels and to weight them by means of their inverse variance. In each position $(x,y)$, the pixel $p$ and its associated error $\sigma$ are calculated from the $N$ exposures as
\begin{equation}
\begin{aligned}
    p(x,y) &= \frac{\sum_{i=1}^Np_i(x,y)/\sigma_i^2(x,y)}{\sum_{i=1}^N1/\sigma_i^2(x,y)}\\
    \sigma (x,y) &= \frac{1}{(\sum_{i=1}^N1/\sigma_i^2(x,y))^{1/2}},
\end{aligned}\label{eq:inv_var_weight}
\end{equation}
where the pixel from the $i$-th exposure is weighted by $1/\sigma_i^2$. This inverse-variance weighted combination is of particular importance to account for the lower signal-to-noise of the short exposures (taken to recover the saturated pixels in the galaxy core) compared to the long ones. In conclusion, the process yields a final image in the BB and one in the NB filter for each galaxy, with an associated standard deviation map.

\section{H$\alpha$ emission line maps}
\label{sec:final_calib}

The primary goal of our survey is to produce calibrated \Ha\ emission-line flux maps. Starting from the calibrated and WCS-aligned BB and NB images, this requires a few key processes: (1) the subtraction of the underlying stellar continuum contribution to the narrow-band, leaving only the nebular emission component, (2) the correction for the loss of flux resulting from the relative position between the \Ha\ + \Nii\ lines and the filter transmission curve (see Figure~\ref{fig:Filters}), and (3) the correction for the \Nii $\lambda\lambda 6548,6583$ doublet contribution to the total nebular flux measured within the narrow-band filter.

The BB images are used as a proxy for the stellar continuum to be subtracted from the total (\Ha\ + \Nii\ + stellar continuum) flux in the NB images. This process requires the PSF of the NB and BB images to be matched as described in Section~\ref{subsec:psf_match}. The stellar continuum subtraction is detailed in Section~\ref{subsec:continuum_sub}. The methods used to  remove the \Nii\ contribution from the continuum-subtracted \Ha\ + \Nii\ images and to correct for the transmission loss in order to recover \Ha\ emission line fluxes are both presented in Section~\ref{subsec:N2_cont}. Results are summarized in Section~\ref{subsec:ha_flux_results}.

\subsection{PSF matching}
\label{subsec:psf_match}

To perform the PSF matching we measure the average empirical PSFs of the BB and NB calibrated images and then convolve the image with the best resolution to match the one with the broader PSF. In this process we adopt simple Gaussian 2D models for the PSFs and the convolution kernels. While a more complex PSF characterization is possible, we found this approach to be sufficient for our purposes. For each galaxy we select all stars from the GAIA DR2 catalogue that are outside the galaxy ellipse (described in Section~\ref{subsec:bkg_subtr}), more than 5 pixels away from the image edges, and faint enough to ensure they are not saturated (GAIA $G_\mathrm{RP} > 16$). The PSF of each image is then determined from these stars using the IDL adaptation of the DAOPHOT package \texttt{getpsf}\footnote{\url{https://asd.gsfc.nasa.gov/archive/idlastro/ftp/pro/idlphot/getpsf.pro}}. 

With the Gaussian FWHMs of the two empirical PSFs determined, we then convolve the image with the best resolution to match the image with the worse resolution using a Gaussian kernel. The image with the worse resolution is left unconvolved. On average the differences between the NB and BB PSF FWHMs are small, although in some cases they are up to 50\%, demonstrating the need for PSF matching.

Once convolved, we confirm that the images are astrometrically aligned and matched in pixels. During the weighted combination of the frames (described in Section~\ref{subsec:weight_comb}), the NB and BB are placed onto the same astrometrically-defined grids. However, even if highly accurate, the astrometric solutions for the BB and NB images are independently computed, as described in Section~\ref{subsec:astrometry}. Even slight offsets between the images leave asymmetric positive through negative residuals around bright point sources when the BB images are subtracted from the NB images. To confirm the astrometry between the images we visually inspect the residuals around unsaturated point sources in simple NB$-$BB images. If clear systematic residuals are observed, we determine the mean offset between the source centroids and shift the images (always keeping one of the two WCS solutions as reference). In all cases, this applied offset is smaller than the astrometric uncertainty described in Section~\ref{subsec:astrometry}.

\subsection{Continuum subtraction}
\label{subsec:continuum_sub}

Once the NB and BB images are correctly aligned and PSF-matched, we remove the stellar continuum contribution from the NB image by subtracting the BB from the NB image. For a flat and featureless continuum spectrum, and in the absence of nebular emission, the BB image provides a perfect estimate for the continuum in the NB. In areas with a high \Ha\ + \Nii\  emission equivalent width (\ewhan), such as \Hii\ regions, the nebular contribution to the BB filter can be significant. This leads to an over-subtraction of the continuum, resulting in an underestimation of the \Ha\ + \Nii\ flux.

To remove the nebular contribution from the BB flux we perform an iterative process. First, we subtract our continuum estimate (initially the BB) from the NB image by scaling the BB to the NB image using a common set of stars. Next, from this continuum-subtracted NB image, we derive maps of the \Ha\ + \Nii\ emission-line flux and \ewhan, calculated as the line flux divided by the continuum estimate. We then determine the relative contribution of the emission lines to the BB image as \mbox{\ewhan $/$(\ewhan $+\Delta_\mathrm{BB}$),} where $\Delta_\mathrm{BB}$ represents the width of the BB filter. Using this quantity, we remove the estimated nebular contamination from the BB image to obtain a new continuum estimate. This process is repeated iteratively until two consecutive continuum estimates differ by less than 1\% (see Appendix~\ref{app:remove_em_line} for further details).

In scaling the BB to the NB image, no colour correction is accounted for, i.e.~no variation in the continuum slope is assumed between the BB and the NB. Assuming a constant NB$/$BB ratio can introduce systematic uncertainties if significant stellar population variations are present within the galaxy. Another associated, but unavoidable issue, is that with narrow-band imaging we are unable to measure the depth of the Balmer absorption feature present in the stellar continuum underneath the nebular emission line \citep{Blanc_2009}. The \Ha\ absorption equivalent width (hereafter \ewph) can have values in the 2-7 \AA\ range \citep[see Figure 1 in][]{Groves_2012} depending on the mean stellar ages of the underlying stellar population. This means the absorption effect is minimal at high \Ha\ emission equivalent width (hereafter \ew ), but can be quite significant for low \ew areas, such as those found in galaxy centres or in DIG-dominated regions. We further discuss these sources of systematic uncertainty in Section \ref{sec:muse_comp}.

\subsection{\Nii\ contamination and filter transmission}
\label{subsec:N2_cont}

After the subtraction of the stellar continuum, the residual flux in the NB image includes a contribution from the \Nii $\lambda\lambda 6548,6583$ doublet lines (see Figure~\ref{fig:Filters}). Without spectra, the \Nii/\Ha\ ratio across each galaxy is unknown. To obtain the \Ha -only fluxes, we estimate the intrinsic \Nii$/$\Ha\ flux ratio for each galaxy using the N2-based mass-metallicity calibration of \citet{Kewley_2008}, and the N2-metallicity relation of \citet{Pettini_2004}, where N2 is defined as N2$\equiv$\lognii . To do so, we adopt the stellar masses listed in Table \ref{tab:sample}. To include the differences in filter transmission across the three lines, we then model the emission lines as simple Gaussian profiles, whose amplitudes are scaled by means of the N2 value and the theoretical \Nii$\lambda 6548/$\Nii$\lambda 6583$ ratio of 0.34 \citep[e.g.][]{Condon_1934,Galavis_1997}. Weighting by the filter transmission curve allows us to estimate the \Nii\ lines contribution to the total NB flux. A detailed explanation of this procedure is provided in Appendix~\ref{app:f_Nii}.

While a single value of the \Nii/\Ha\ ratio for a galaxy is simplistic, a fully accurate \Nii\ contamination removal requires further assumptions about the variation of the ratio due to radial metallicity gradients, differing ionization mechanisms (e.g.~supernovae, AGN), or the inclusion of diffuse ionized gas \citep[see e.g.][]{Belfiore_2022}. An exploration of how our simplistic correction compares with real values from our associated PHANGS-MUSE emission-line maps is discussed in Section~\ref{sec:muse_comp}.

The different line-of-sight velocities of the targets imply that different fractions of the \Ha\ and \Nii\ line fluxes are lost due to the shape of the NB filter curves. These transmission losses can be corrected as the radial velocity of each galaxy and the wavelength response of the three different NB filters used in our survey are known. We apply a single correction that takes into account the \Nii\ contamination and these transmission losses. First we calculate the effective \Nii\ contribution:
\begin{equation}
    \mathcal{F}_{[\mathrm{N}\,\mathrm{II}]} = \frac{\mathrm{F}_{[\mathrm{N}\,\mathrm{II}]\lambda 6548} + \mathrm{F}_{[\mathrm{N}\,\mathrm{II}]\lambda 6583}}{\mathrm{F}_{\mathrm{H}\alpha\lambda 6562} + \mathrm{F}_{[\mathrm{N}\,\mathrm{II}]\lambda 6548} + \mathrm{F}_{[\mathrm{N}\,\mathrm{II}]\lambda 6583}},
    \label{eq:NII_frac_formula}
\end{equation}
where $\rm{F_L}$ ($\rm{L}=\rm{H}\alpha\lambda 6562,[\mathrm{N}\,\mathrm{II}]\lambda 6548,[\mathrm{N}\,\mathrm{II}]\lambda 6583$) is the line flux weighted for the NB filter transmission curve, corrected for the radial velocity of each galaxy\footnote{We do not take into account smaller wavelength shifts arising from other motions such as galaxy rotation.}, and where \fNii\ is calculated from the adopted intrinsic \Nii/\Ha\ line ratio (see Appendix~\ref{app:f_Nii}). 

From Equation~\ref{eq:NII_frac_formula}, it is straightforward to show that to correct simultaneously for the \Nii\ contamination and the filter transmission we can use the following formula:
\begin{equation}
    f_{\mathrm{H}\alpha}=f_{\mathrm{H}\alpha + [\mathrm{N}\,\mathrm{II}]}\cdot\frac{(1-\mathcal{F}_{[\mathrm{N}\,\mathrm{II}]})}{T_{\mathrm{H}\alpha}},
    \label{eq:halpha_corr}
\end{equation}
where $f_{\mathrm{H}\alpha}$ is the corrected \Ha -only flux, $f_{\mathrm{H}\alpha + [\mathrm{N}\,\mathrm{II}]}$ represents the continuum subtracted flux observed through the NB filter\footnote{$f_{\mathrm{H}\alpha + [\mathrm{N}\,\mathrm{II}]} \equiv \mathrm{F}_{\mathrm{H}\alpha\lambda 6562} + \mathrm{F}_{[\mathrm{N}\,\mathrm{II}]\lambda 6548} + \mathrm{F}_{[\mathrm{N}\,\mathrm{II}]\lambda 6583}$} and $T_{\mathrm{H}\alpha}$ is the effective filter transmission for the \Ha\ line, i.e.\! a transmission correction that accounts for the flux loss owing to both the NB filter shape and the galaxy radial velocity (see the computation in Appendix~\ref{app:f_Nii}). The \fNii\ and $T_{\mathrm{H}\alpha}$ values found for each galaxy can be found in columns (11) and (12) of Table~\ref{tab:results}.

\subsection{\pha\ imaging products and properties}
\label{subsec:ha_flux_results}

For the 65 \pha\ targets, the final products consist of 68 sets of three images\footnote{Including those for the 3 galaxies observed with both telescopes.}: the calibrated BB and NB images, together with the \Ha\ continuum-subtracted and \Nii -corrected emission line flux maps (hereafter \Ha SUB) as obtained by following the flux correction described in Section~\ref{subsec:N2_cont} (Equation~\ref{eq:halpha_corr}).

To characterize these products, we derive the main observational parameters, which are summarized in Table~\ref{tab:results}. For both BB and NB images, four columns list the filter name, the total integration time after combining the individual exposures, the limiting surface brightness and the image resolution. The surface brightness limit is defined as the 3$\sigma$ detection threshold above the background noise, measured in $\sim 2$ arcsec$^2$ emission-free apertures (35 pixels for WFI and 30 pixels for DirectCCD, chosen to be slightly larger than the image quality limit imposed in the observational campaign). Clean background maps, were generated by masking sources in the final BB and NB images, using the same procedure described in Section~\ref{subsec:bkg_subtr} for fitting the background of the individual exposure. Following the convention in photometry, the surface brightness limit is given in Table~\ref{tab:results} in magnitude per square arcsecond. The image resolution, calculated from the image PSF (see Section~\ref{subsec:psf_match}), is given in both apparent size (arcsec) and physical dimensions (parsec), with galaxy distances taken from Table~\ref{tab:sample}.

In Table~\ref{tab:results}, a third group of four columns report key quantities derived from (and for) each of the \Ha SUB image: the surface brightness limit, the \fNii\ factor, the effective filter transmission $T_{\mathrm{H}\alpha}$, and the integrated \Ha\ luminosity. The surface brightness detection limits are estimated from the \Ha SUB images using the same procedure applied to the BB and NB images, but expressed in flux units to match the convention for emission-line maps. The reported \fNii\ factors and effective filter transmissions $T_{\mathrm{H}\alpha}$ are determined individually for each galaxy, as detailed in Appendix~\ref{app:f_Nii}. As explained, these quantities are then used to remove the \Nii\ contamination from the continuum-subtracted NB fluxes, resulting in the science-ready \Ha SUB flux maps. 

The \Ha\ luminosity is measured by integrating the \Ha SUB fluxes in 1 kpc-wide concentric annuli, deprojected onto the galaxy disk using the inclinations from Table~\ref{tab:sample}, and corrected for Galactic foreground extinction. No correction for the galaxy intrinsic dust extinction is applied, which is common practice in photometric \Ha\ studies when spectroscopic constraints (e.g.\! Balmer decrement) are unavailable. After masking the point sources, the total luminosity is then the sum of all the annuli with measurable positive flux. The point-source masks yield to a minor underestimation of the total flux which is estimated (by filling the missing pixels with a per-annulus mean value) to be in the order of 0.5\%--1.5\%, with a few galaxies around 3\%--8\% difference. This annular flux measurement follows the methodology used in Section~\ref{subsec:compare_profile} to derive the \Ha\ surface brightness profiles.

In Figures \ref{fig:wfi_img_collection} and \ref{fig:ccd_img_collection}, we present a gallery of the final \Ha SUB images for the \MPG/WFI and \dP/DirectCCD sub-samples, respectively. 
For visualization, we adopted non-linear intensity scales with different colormaps for bright and faint regions, in order to highlight the \Ha\ emission structure and any low-level residuals in the background.
Apart from being low-level features, some of the background artifacts appear far from the region of interest.

These 68 final images presented here, obtained through the \pha\ pipeline (Section~\ref{sec:reduction}) and the flux corrections (\Nii\ removal and continuum subtraction; Section~\ref{sec:final_calib}), constitute the PHANGS collaboration internal release version 2.3. This release, together with earlier versions, has been the reference dataset for all PHANGS publications to date using \pha\ narrow-band imaging data \citep[][Pan et al.\! 2025 in preparation]{Schinnerer_2019,Chevance_2020,Chevance_2022,Querejeta_2021,Querejeta_2024,Querejeta_2025,Pan_2022,Kim_2022,Sun_2023,Stuber_2023,Romanelli_2025}. It also provides the foundation for the ongoing effort to extend the sample with $\sim$ 10 galaxies from ESO archival data, reduced with the \pha\ pipeline \citep[e.g. NGC 253 presented by][]{Congiu_2025}, and new \MPG /WFI observations of 68 targets, yielding a total number of nearly 150 galaxies.

\section{Comparison to PHANGS-MUSE \Ha\ fluxes}
\label{sec:muse_comp}

In this section we compare our \Ha\ narrow-band flux maps (i.e.\! the \Ha SUB images) against spectroscopically constructed \Ha\ maps from the PHANGS-MUSE survey. We want to test whether the narrow-band \Ha SUB images are consistent with ``true" \Ha\ emission-line maps to understand the systematic uncertainties introduced by the standard assumptions made in the process of converting the narrow-band and broad-band fluxes to \Ha\ emission-line fluxes. We also explore further corrections to the data processing that optimize the fidelity with which \Ha\ narrow-band emission line fluxes can be recovered, particularly in \Hii\ regions. 

\subsection{Potential biases and correction strategy}
\label{subsec:biases}

The first step is to consider the systematic effects that may bias the recovery of \Ha\ fluxes and to outline the strategy we adopt to address them.
We are particularly interested in the biases introduced by the use of a constant \Nii$/$\Ha\ ratio throughout each galaxy to correct for the neighbouring \Nii\ lines, and by neglecting both changes in the stellar-continuum spectral shape and the presence of \Ha\ photospheric absorption during the BB continuum subtraction process. Below we list some of the physical processes responsible for these variations.

\medskip

\noindent\textit{Diffuse Ionized Gas\ ---}
The relative contribution of DIG and \Hii\ regions to the total \Ha\ flux changes as a function of position inside galaxies. In regions lacking active star formation and dominated by old stellar populations (e.g.\ the central bulges of some galaxies), the diffuse \Ha\ emission comes largely from ionization by hot low-mass evolved stars \citep[e.g.][]{Flores-Fajardo_2011,Zhang_2017}, while across the whole disks of star-forming galaxies leaking ionizing radiation from \Hii\ regions also ionizes the diffuse ISM \citep[e.g.][]{Belfiore_2022}. DIG emission is characterized by a significantly higher \Nii /\Ha\ ratio than is observed in \Hii\ regions \citep[e.g.][]{Haffner_2009, Blanc_2009}. Hence, the central zones of many galaxies, as well as their inter-arm regions, tend to present higher \Nii$/$\Ha\ ratio than the star-forming regions. 

\medskip

\noindent\textit{Metallicity gradients\ ---}
The \Nii$/$\Ha\ ratio is commonly used as metallicity indicator \citep{Denicolo_2002, Pettini_2004}. Since the metallicity typically decreases towards large radii in local disk galaxies \citep{Searle_1971,Sanchez_2014,Kaplan_2016,Kreckel_2019}, the \Nii$/$\Ha\ line ratio should also decrease as a function of radius.

\medskip

\noindent\textit{Changes in stellar populations\ ---}
The ratio between the BB and the NB continuum changes depending on the properties of the stellar populations that give rise to the continuum spectrum. Not only the colour of the stellar continuum can change, but also the depth of the \Ha\ photospheric absorption line (which has a much larger relative impact in the NB continuum flux than in the BB) changes as a function of stellar population parameters (mainly age). So the scaling factor one should use when subtracting the BB derived continuum from the NB (Section~\ref{subsec:continuum_sub}) is not constant across the galaxy. This can introduce biases, which will be noticeable in galaxy centres with strong bulges, where old and red stellar populations dominate over the younger and bluer populations seen throughout the disk, and between spiral arm and inter-arm regions \citep[e.g.][]{Garner_2024}. 

\medskip

In order to quantify the impact of these biases, we rely on the PHANGS-MUSE \Ha\ line maps (hereafter MUSE \Ha) produced by the PHANGS Data Analysis Pipeline \citep[DAP, \S 5 of][]{Emsellem_2022}. First, we compare the \Ha\ flux radial profiles from \pha\ and PHANGS-MUSE for the 19 galaxies that are included in both surveys. The comparison helps us to understand how narrow-band derived \Ha\ fluxes differ from the spectroscopically measured line fluxes as a function of radius, and how the \pha\ radial profiles extend out in radius, compared to the PHANGS-MUSE observations which are limited to the central parts of the galaxies.

We also explore the comparison focusing on regions of high \ew , which are typically dominated by \Hii \ regions and young stellar populations. To select these regions, we use \Ha\ emission line \ew\ maps created from the PHANGS-MUSE data and use them to define high-EW pixel masks. In these regions, we expect the effects of DIG contamination and changes in stellar populations to be much smaller than across the whole galaxies. While the \pha\ data support a range of science cases, the fidelity of the recovered \Ha\ emission-line fluxes is particularly important for \Hii\ region studies, making the flux in these regions particularly relevant to be tested and optimized.

After having identified the \Ha\ flux differences, we apply empirical corrections to minimize the effects of biases in the \Nii\ decontamination recipe (Appendix~\ref{app:f_Nii} and Section~\ref{subsec:N2_cont}) and the effects of \Ha\ photospheric absorption, not modelled in the stellar-continuum subtraction (Section~\ref{subsec:continuum_sub}). For both corrections we make use of PHANGS-MUSE data from \citet{Emsellem_2022} to derive recipes that are applicable to the full \pha\ sample and other future \Ha\ narrow-band surveys in general.

\begin{figure*}[h!]
    \centering
    \includegraphics[width=\textwidth]{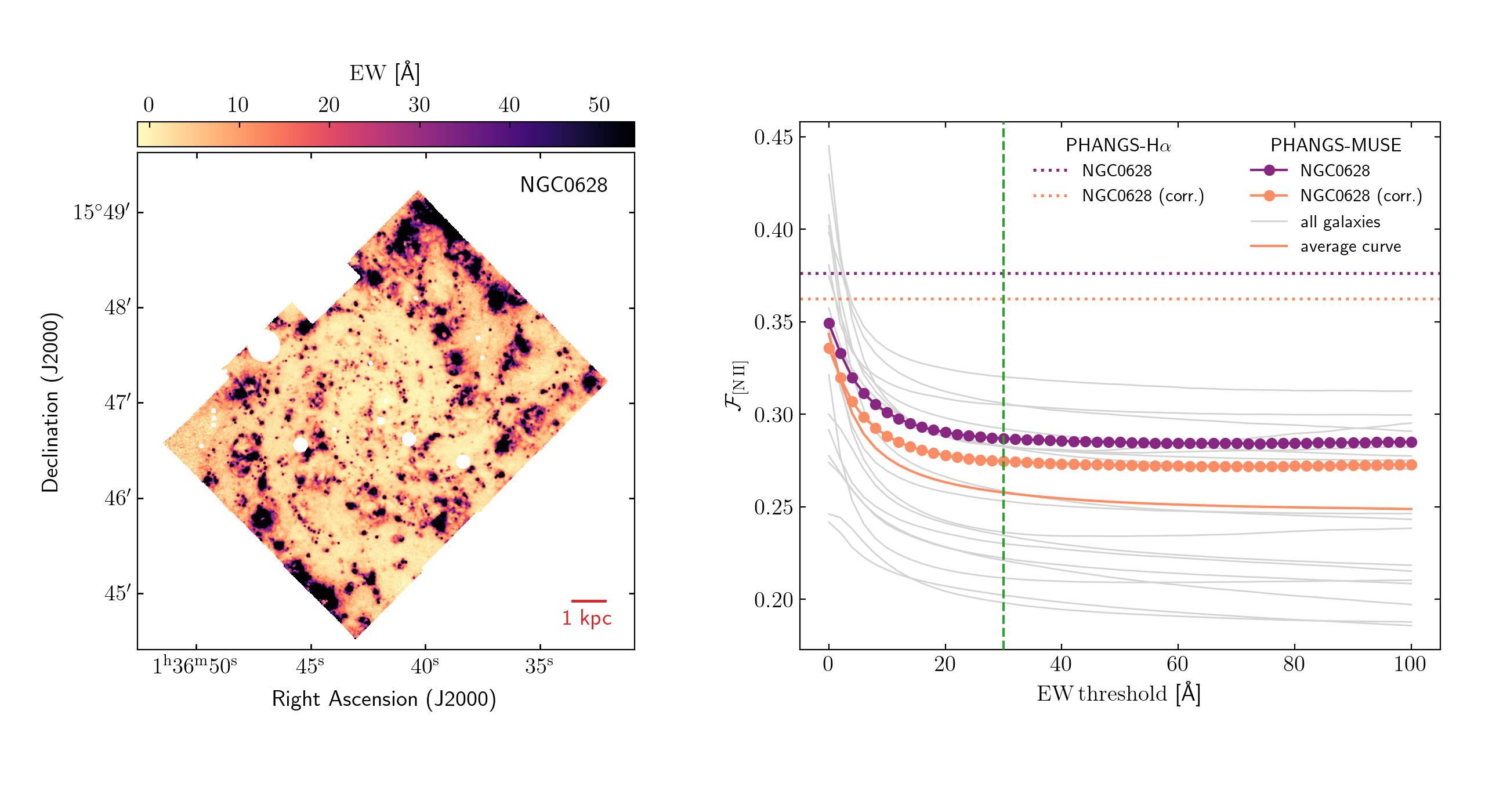}
    \caption{\textit{Left}: \ew\ map of NGC0628 computed from the PHANGS-MUSE observation of the galaxy. Masked foreground stars are visible as white circles in the image. \textit{Right}: variation of the \fNii\ computed from the PHANGS-MUSE \Nii$\lambda$6548, \Nii$\lambda$6583 and \Ha$\lambda$6562 emission line maps for different \ew\ thresholds. The curves for the \fNii\ uncorrected (purple) and corrected (orange) for the WFI NB filter transmission are shown. The level where the curves flatten, or better the $d\mathcal{F}/d\mathrm{EW}$ derivative tends asymptotically to zero, determines the \ew $=30$ cutoff (green vertical dashed line). As can be seen from the curves computed for the other 18 PHANGS-MUSE galaxies, plotted in light-gray, the same behaviour of nearly constant \fNii\ (i.e. with variations of $d\mathcal{F}/d\mathrm{EW}$ smaller than a few tenths of a percent) above \ew $=30$ and its sharp increase for EW values representing the transition from the \Hii - to the DIG-dominated regime are shared across the sample. The mean curve of the full sample of 19 galaxies is shown as a solid orange curve. The horizontal dotted lines represent the \fNii\ calculated for the \pha\ data of NGC 0628 from Equation~\ref{eq:NII_frac_th} (purple) and Equation~\ref{eq:f_Nii_transm} (orange) which are respectively without and with the correction for the NB filter.}
    \label{fig:ew_map}
\end{figure*}

\subsection{\Ha\ emission equivalent width cut}
\label{subsec:ew_cut}

We adopt an emission line \ew\ threshold to define the areas within galaxies that are dominated by \Hii\ regions. In DIG-dominated regions, enhanced low-ionization lines are observed relative to \Ha , resulting in a larger \Nii /\Ha\ fraction than in \Hii -dominated regions. A high \Ha\ surface brightness threshold has been widely adopted in the literature to separate \Hii\ regions from DIG \citep[e.g.\! 10$^{39}$ erg s$^{-1}$ kpc$^{-2}$,][]{Zhang_2017}, as it is directly related to the density of the ionized gas. Other studies have proposed a selection based on the \Ha\ equivalent width \citep[e.g.][]{Cid_2011}. As an example, \citet{Lacerda_2018} show that the equivalent width computed from spatially resolved CALIFA data correlates with the location of the different ionized regions in the ``Baldwin, Phillips \& Terlevich" \citep[BPT; ][]{Baldwin_1981} diagram, suggesting an \ew\ threshold of 14~\AA\ to isolate pure \Hii\ regions. This EW threshold is likely to be dependent on the resolution of the data, and not directly applicable to other datasets. We therefore follow the procedure described below to identify an adequate threshold for our dataset.

We measure \ew\ maps directly from the PHANGS-MUSE spectral datacubes, for all  galaxies in common with the \pha\ sample. An example of such an \ew\ map is shown in the left panel of Figure~\ref{fig:ew_map} for NGC 0628\footnote{We adopt a definition of EW that has positive values for emission lines and negative values for absorption lines}.

In parallel with these \ew\ maps, we also compute spaxel-by-spaxel maps of \fNii\ from the PHANGS-MUSE \Nii$\lambda$6548, \Nii$\lambda$6583, and \Ha$\lambda$6562 spectral emission line maps \citep[PHANGS-MUSE DAP pipeline products,][]{Emsellem_2022} using two approaches: (1) a straightforward ratio between the \Nii\ doublet flux and the total flux of the three lines and (2) the same procedure, but with each line weighted by the effective transmission of the narrow-band filters (see Equation~\ref{eq:NII_frac_corr}).

The right plot of Figure~\ref{fig:ew_map} shows, for the case of \mbox{NGC 0628}, the mean \fNii\ values of all pixels selected within the high \ew\ mask as a function of the \ew\ threshold adopted (from 0 to 100 \AA). Both the intrinsic \fNii\ values (purple) and those corrected for the effective filter transmission (orange) are shown. The orange horizontal dotted line represents the \fNii\ value used for NGC~0628 as part of the standard procedure described in Section \ref{subsec:N2_cont}. This is the value calculated from Equation~\ref{eq:f_Nii_transm} and listed in Table~\ref{tab:results} for all the galaxies. The value uncorrected for the filter transmission (purple horizontal dotted line) is also shown.

Figure~\ref{fig:ew_map} shows two important things. First, the standard recipe overestimates the \Nii\ contamination to the narrow-band flux in bright \Hii\ regions by factors of $\sim$ 10\%--50\% depending on the target ($\sim$ 30\% for NGC 0628). This bias motivates the corrections discussed below in Sections \ref{subsec:correct_nii} and \ref{subsec:fnii_vs_mass}. Second, and more relevant here, is the fact that the level of \Nii\ contamination in the narrow-band fluxes is relatively constant across high \ew\ regions, and it increases sharply below a certain threshold corresponding to the transition from the \Hii\ region- to the DIG-dominated regime. The first derivative of the \fNii\ vs. EW threshold function becomes smaller than a few tenths of a percent, indicating that the function approaches its asymptotic regime, at \ew $\simeq 30$ \AA\ for all galaxies. We therefore adopt the \ew $=30$ \AA\ threshold as the high-EW mask cut to select the areas dominated by \Hii\ region emission.

\subsection{H$\alpha$ flux profiles}
\label{subsec:compare_profile}

Radial surface brightness profiles are computed for all three \pha\ products (BB, NB, and \Ha SUB) to assess the calibration and the assumptions underlying the creation of our \Ha\ line maps. These profiles are derived from concentric 1 kpc-wide annuli deprojected onto the galaxy disk using the inclinations and position angles provided in Table~\ref{tab:sample}. For each annulus $A$, the surface brightness is calculated as $S_A = \sum_{i\in A} p_i/N$, i.e.\ total flux divided by the $N$ unmasked\footnote{The pixel mask includes the pipeline bad pixel mask with the addition of two sets of masks: one for the foreground stars that are automatically found from Gaia DR2 catalogue and one manual mask of the stellar ghosts.} pixels within the annulus. The errors are propagated from the associated standard deviation map as $\sigma_{S_A} = (\sum_{i\in A} (\sigma_i/N)^2)^{1/2}$, where $\sigma_i$ is the error of the pixel $p_i$.

For each \pha\ image, radial profiles are computed over the full field, and also after masking regions falling outside the PHANGS-MUSE data footprint, enabling a direct comparison between the two datasets. Following this approach, radial profiles are computed consistently for both MUSE \Ha\ and \Ha SUB images after reprojecting them onto the same WCS coordinates and applying an identical mask to exclude bad pixels and foreground stars in both datasets. Likewise, \pha\ BB and NB image profiles are computed and compared with reconstructed PHANGS-MUSE BB and NB images derived from the convolution of the MUSE datacubes with the WFI/DirectCCD BB and NB filters (hereafter referred to as MUSE BB and MUSE NB), as described in \S 4.2.4 of \citet{Emsellem_2022}. 
The \Ha SUB and MUSE \Ha\ line maps are converted to flux units per arcsec$^2$, whereas BB, NB, and their MUSE BB and MUSE NB counterparts are given in flux density per unit wavelength per arcsec$^2$. In addition, \Ha SUB images are corrected for MW foreground extinction in the same manner as in \citet{Emsellem_2022} to match their extinction-corrected MUSE \Ha\footnote{In this analysis we do not correct the BB and NB images for MW foreground extinction, because the MUSE BB and MUSE NB images we compare them to, are reconstructed directly from the PHANGS-MUSE datacubes, which are not corrected for this effect, unlike the PHANGS-MUSE emission line maps that are produced by the PHANGS-MUSE DAP.}.

\begin{table*}[h!]
\caption{\label{tab:flux_offsets}Flux residual offsets between PHANGS-MUSE and \pha}
\small
\centering
\begin{tabular}{@{\extracolsep{3pt}}cd{3.1}d{3.1}d{3.1}d{1.4}d{2.1}d{3.1}d{2.1}d{3.1}d{1.4}d{2.1}d{1.1}d{3.1}}
\hline\hline\\[-6pt]
 & \multicolumn{3}{c}{\pha\ Offsets} & \multicolumn{5}{c}{Per-Galaxy Corrections} & \multicolumn{4}{c}{Survey-Wide Corrections} \\[2pt]
\cline{2-4} \cline{5-9} \cline{10-13}\\[-6pt]
Name & \multicolumn{1}{c}{BB} & \multicolumn{1}{c}{NB} & \multicolumn{1}{c}{\Ha SUB} & \multicolumn{1}{c}{\fNii} & \multicolumn{1}{c}{$\Delta_{\mathcal{F}_{[\mathrm{N\,II}]}}$} & \multicolumn{1}{c}{\ewph} & \multicolumn{1}{c}{$\Delta_{\mathrm{EW_{ph}}}$} & \multicolumn{1}{c}{\Ha SUB} & \multicolumn{1}{c}{\fNii\ fit} & \multicolumn{1}{c}{$\Delta_{\mathcal{F}_{[\mathrm{N\,II}]\,\mathrm{fit}}}$} & \multicolumn{1}{c}{$\Delta_{\mathrm{med(EW_{ph})}}$ \tablefootmark{\scriptsize{a}}} & \multicolumn{1}{c}{\Ha SUB} \\[2pt]
    {} & \multicolumn{1}{c}{(\%)} & \multicolumn{1}{c}{(\%)} & \multicolumn{1}{c}{(\%)} & \multicolumn{1}{c}{} & \multicolumn{1}{c}{(\%)} & \multicolumn{1}{c}{(\AA )} & \multicolumn{1}{c}{(\%)} & \multicolumn{1}{c}{(\%)} & \multicolumn{1}{c}{} & \multicolumn{1}{c}{(\%)} & \multicolumn{1}{c}{(\%)} & \multicolumn{1}{c}{(\%)}\\[6pt]
    \multicolumn{1}{c}{(1)} & \multicolumn{1}{c}{(2)} & \multicolumn{1}{c}{(3)} & \multicolumn{1}{c}{(4)} & \multicolumn{1}{c}{(5)} & \multicolumn{1}{c}{(6)} & \multicolumn{1}{c}{(7)} & \multicolumn{1}{c}{(8)} & \multicolumn{1}{c}{(9)} & \multicolumn{1}{c}{(10)} & \multicolumn{1}{c}{(11)} & \multicolumn{1}{c}{(12)} & \multicolumn{1}{c}{(13)}\\
\hline\\[-8pt]
IC5332 & -3.6 & 0.4 & -5.1 & 0.230 & 0.9 & -3.5 & 4.7 & 0.6 & 0.232 & 0.7 & 4.0 & -0.3 \\
NGC0628 & -1.4 & -0.1 & -16.4 & 0.275 & 11.5 & -2.9 & 4.2 & -0.6 & 0.290 & 9.5 & 4.4 & -2.5 \\
NGC1087 & -5.8 & -0.6 & -11.4 & 0.222 & 3.8 & -3.4 & 5.0 & -2.6 & 0.227 & 3.3 & 4.5 & -3.6 \\
NGC1300 & -2.8 & -2.5 & -20.1 & 0.253 & 12.8 & -3.2 & 6.0 & -1.2 & 0.264 & 11.5 & 5.7 & -2.9 \\
NGC1365 & -5.0 & -0.4 & -15.3 & 0.259 & 16.2 & -3.0 & 4.0 & 4.8 & 0.264 & 15.5 & 4.0 & 4.1 \\
NGC1385 & -3.4 & -3.3 & -15.9 & 0.235 & 3.4 & -3.1 & 4.5 & -8.0 & 0.232 & 3.7 & 4.4 & -7.9 \\
NGC1433 & -3.4 & -6.2 & -28.3 & 0.305 & 13.3 & -3.1 & 5.2 & -9.9 & 0.299 & 14.0 & 5.0 & -9.3 \\
NGC1512 & -1.9 & -2.6 & -20.2 & 0.320 & 11.8 & -3.0 & 5.3 & -3.1 & 0.302 & 14.2 & 5.3 & -0.6 \\
NGC1566 & -3.5 & -1.9 & -18.3 & 0.283 & 12.6 & -2.9 & 3.8 & -1.9 & 0.273 & 13.9 & 4.0 & -0.5 \\
NGC1672 & -2.1 & -1.6 & -18.2 & 0.292 & 12.8 & -2.9 & 4.5 & -0.9 & 0.284 & 13.9 & 4.6 & 0.3 \\
NGC2835 & -2.3 & -0.2 & -13.2 & 0.221 & 9.6 & -3.1 & 4.6 & 1.0 & 0.262 & 4.5 & 4.5 & -4.1 \\
NGC3351 & -2.7 & -8.2 & -28.6 & 0.282 & 9.1 & -3.0 & 5.7 & -13.9 & 0.289 & 8.3 & 5.7 & -14.6 \\
NGC3627 & -1.5 & -0.1 & -23.0 & 0.283 & 18.4 & -2.7 & 4.0 & -0.6 & 0.307 & 15.2 & 4.5 & -3.4 \\
NGC4254 & -2.5 & -9.8 & -33.5 & 0.198 & 5.7 & -3.7 & 5.1 & -22.7 & 0.198 & 5.8 & 4.2 & -23.6 \\
NGC4303 & -1.9 & -4.9 & -19.2 & 0.282 & 7.5 & -2.8 & 3.9 & -7.7 & 0.260 & 10.4 & 4.3 & -4.6 \\
NGC4321 & -3.1 & -4.3 & -27.1 & 0.236 & 15.0 & -2.8 & 5.1 & -7.0 & 0.265 & 11.7 & 5.4 & -10.0 \\
NGC4535 & -4.7 & -14.9 & -41.0 & 0.212 & 7.8 & -3.0 & 4.8 & -28.5 & 0.225 & 6.6 & 4.8 & -29.6 \\
NGC5068 & -3.1 & -0.8 & -3.8 & 0.202 & -1.4 & -3.2 & 3.9 & -1.3 & 0.202 & -1.4 & 3.6 & -1.5 \\
NGC7496 & 7.8 & 11.1 & 16.3 & 0.306 & 4.1 & -3.2 & 3.8 & 24.2 & 0.291 & 6.7 & 3.6 & 26.6 \\
\hline\\[-8pt]
median & -2.8 & -1.9 & -18.3 &   & 9.6 & -3.0 & 4.6 & -1.9 &   & 9.5 & 4.5 & -3.4 \\
\hline
\end{tabular}
\tablefoot{Col. (1): Galaxy name. Columns (2)--(4): Residual flux offsets between PHANGS–MUSE and \pha . Offsets are computed as the mean of residual profiles: (BB~$-$~MUSE~BB)/MUSE~BB, (NB~$-$~MUSE~NB)/MUSE~NB, and (\Ha SUB~$-$~MUSE~\Ha)/MUSE~\Ha.
Columns (5)--(8): Per-galaxy corrections using measured \fNii\ (col.~5) and \ewph\ (col.~7), with corresponding correction offsets (cols.~6 and 8), and resulting corrected \Ha\ SUB values (col.~9).
Columns (10)--(13): Survey-wide corrections using fitted \fNii\ values (col.~10) and a fixed median \ewph , with corresponding offsets (cols.~11 and 12) and the final corrected \Ha\ SUB values (col.~13). All flux offsets are expressed as percent differences. Equivalent widths are in \AA. BB offsets (col.~2) are applied as a baseline correction before computing all other offsets.\\
\tablefoottext{a}{This offset is computed using the median \ewph\ across all galaxies of $-3.0$ \AA .}
}
\end{table*}

Analysing these radial profiles we find few percent differences between the \pha\ BB and MUSE BB fluxes across all galaxies (see column 2 in Table~\ref{tab:flux_offsets}). In principle, this was not expected, since the \pha\ BB data was used as a reference to set the absolute flux scale of the PHANGS-MUSE cubes \citep{Emsellem_2022}. Further investigation showed that this small offset can be attributed to differences in the background subtraction algorithm used in the \pha\ internal data release 1.0 (which was used to calibrate the PHANGS-MUSE data), and the \pha\ data release 2.3 images presented here, which were processed as described in Section \ref{subsec:bkg_subtr}. These residual BB offsets are used to scale the PHANGS-MUSE datacubes and line maps for all the subsequent comparison analysis presented in this section, in order to ensure an adequate comparison in a common absolute flux scale. For each galaxy, a correction factor is applied to the PHANGS-MUSE data, and this is accounted for when computing the residuals presented below and reported in columns 3 through 13 of Table~\ref{tab:flux_offsets}. After the correction, a residual median offset of $-1.9$\% is found between the \pha\ NB and MUSE NB images (column 3 in Table~\ref{tab:flux_offsets}).

\begin{figure*}
    \centering
    \includegraphics[width=\textwidth]{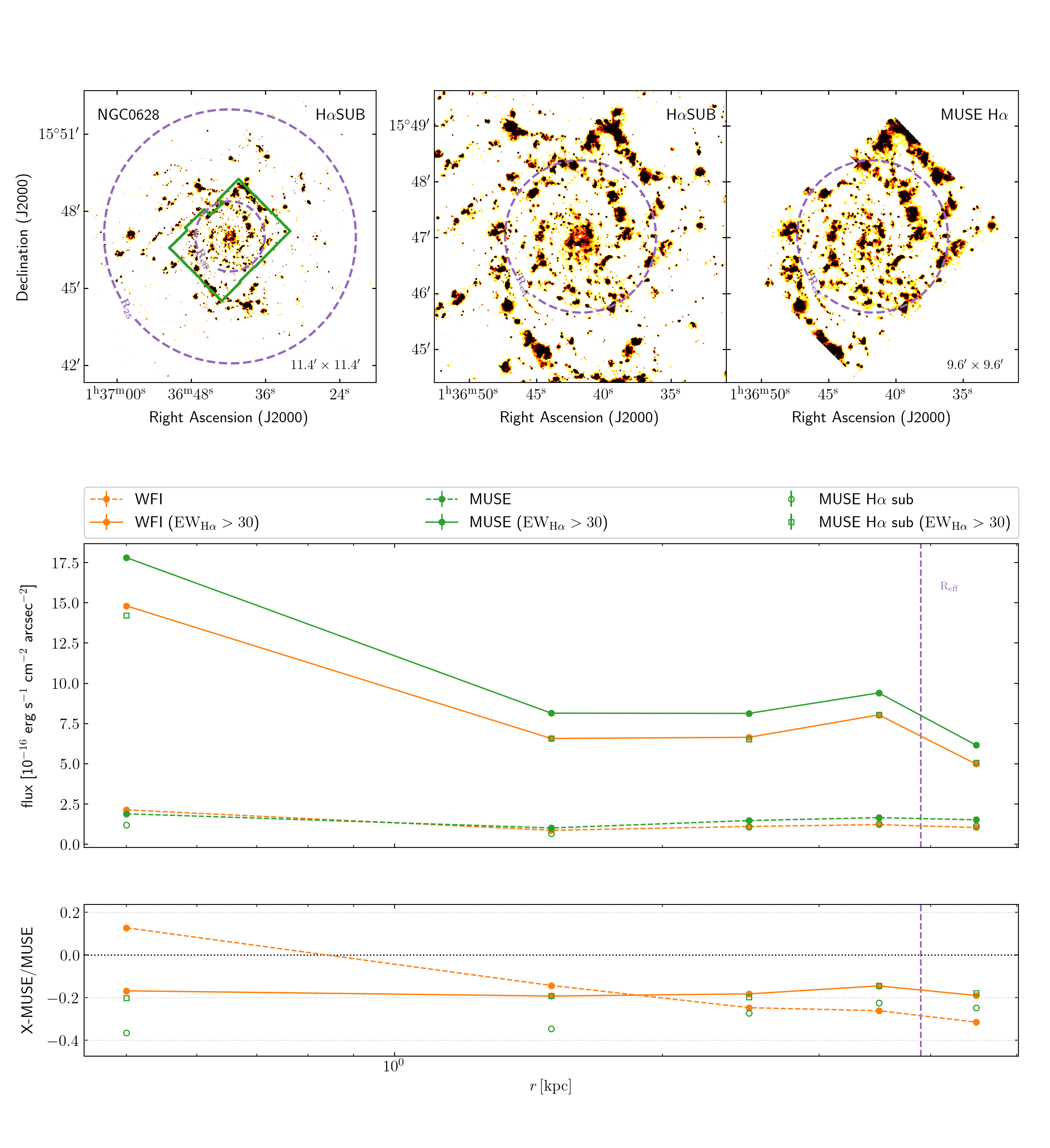}
    \caption{Top-left: cutout of \Ha SUB image up to a radius larger than the $B$-band 25th isophotal (dashed line). In green is shown the MUSE footprint. Top-right: a zoom of \Ha SUB and MUSE \Ha\ for a better view of the inner structures, within the effective radius of the galaxy (dashed line). Two surface brightness profile comparisons are shown in the central main plot. The dashed orange and green lines represent the profiles computed by using all the unmasked pixels within each annulus, respectively for WFI \Ha SUB and MUSE \Ha . On the other hand, the continuous lines show the profiles computed for pixels in each annulus with EW higher than 30 \AA , where the same colour code as for the dashed lines is used. In the bottom, the residual WFI-MUSE profiles, normalized by MUSE data, are shown with the continuous line and dashed line for the computations with all pixels and high-EW pixels, respectively. As sanity check, green open circle and open squares represent the profiles for the images computed with the same recipes adopted for \Ha SUB but employing MUSE BB and MUSE NB instead of BB and NB images, respectively for the profiles computed for all the pixels and high-EW pixels within the annuli.}
    \label{fig:Ha_profile}
\end{figure*}

An example of the resulting \Ha\ profiles for NGC~0628 is shown in Figure~\ref{fig:Ha_profile}. In this figure, a cutout of the WFI \Ha SUB image around the area of interest for \Ha\ emission is shown (top-left), with the MUSE footprint used to compute the comparison profiles overlaid in green. A pair of WFI/MUSE \Ha\ images (top-right) is shown in an enlarged view to better highlight the structures within the area of interest. The projected galaxy effective radius $R_\mathrm{eff}$ and the 25$^\mathrm{th}$ $B$-band isophotal radius $R_{25}$ are plotted in dashed lines.   

The middle panel of Figure~\ref{fig:Ha_profile} presents the radial surface brightness profile from the WFI \Ha SUB  (filled orange circles) and MUSE \Ha\ (filled green circles) emission line flux maps. Two versions of each profile are presented: (1) the dashed lines represent profiles computed by averaging the flux across all pixels in each annulus; (2) the solid lines correspond to profiles derived using only pixels within the high-EW mask defined in Section~\ref{subsec:ew_cut}, i.e.\ with \ew\ above $30$ \AA. In both cases, the pixels outside the MUSE footprint are excluded from the computation to allow comparing the profiles. 


We also compute flux profiles for all the 19 galaxies by applying the same continuum-subtraction, \Nii\ decontamination, and effective filter transmission correction recipes, described in Section~\ref{sec:final_calib}, to the MUSE NB and MUSE BB reconstructed images. The resulting images, referred to as MUSE \Ha SUB, are used to compute profiles like those shown in Figure~\ref{fig:Ha_profile} for NGC 0628, where open green circles correspond to integrals over all pixels, and open green squares include only high emission EW pixels.
\Ha SUB differences with its spectroscopic counterpart MUSE \Ha SUB (see below the discussion on the NB/BB scaling validity in the central bulge), together with their strong agreement in the high-EW regime, highlights the need for an EW cut.

The bottom panel of Figure~\ref{fig:Ha_profile} presents the residuals between the WFI and MUSE \Ha SUB images relative to the MUSE \Ha\ profiles. In this panel we only compare ``high EW region'' profiles with each other, and ``all pixels'' profiles to each other. Several trends can be observed for NGC 0628, which highlight the main effects we see across the full sample.

When including all pixels, many galaxies show a strong radial trend in the residuals between the \pha\ \Ha SUB images and the MUSE line maps, similar to the one seen in NGC 0628 (orange dashed line in the bottom panel of Figure \ref{fig:Ha_profile}). The narrow-band \Ha SUB maps overestimate the \Ha\ flux in the central regions, and underestimate it at large radii. The scale of these systematic deviations is large (up to 40\% relative offsets between galaxy centres and their outer regions). The effect is clearly visible as a flux excess in the top-right maps of Figure~\ref{fig:Ha_profile}. These trends correlate with the presence of bulges in the centres of these galaxies, and are thought to arise from the use of a constant factor to scale the BB and NB images for continuum subtraction, which does not account for differences in the stellar population spectral slope. For continuum subtraction, the BB and NB images are scaled using foreground MW field stars. The NB-BB colour of these stars is not necessarily representative of the stellar populations across the inner regions and disks of our extra-galactic targets. In low emission EW regions this systematic error in the estimation of the stellar continuum can have a very large effect in the derived \Ha\ flux. This makes it very hard to use narrow-band imaging as a robust technique to measure nebular emission lines in low EW environments. 

Selecting \Hii\ regions dominated areas via our high-EW cut largely removes these radial trends, resulting in a more constant residual profile as shown by the orange solid line in the bottom panel of Figure~\ref{fig:Ha_profile}. While the radial trends are removed, a consistent residual flux offset of -15\% to -20\% is seen across NGC 0628. That is, the narrow-band fluxes underestimate the spectroscopically measured MUSE fluxes. For the overall sample, we see a large scatter in these offsets, although most targets show underestimated narrow-band fluxes, with residuals in the -3.8\% to -41.0\% range and only one of the 19 targets (NGC 7496) showing a positive 16\% offset (see column 4 in Table \ref{tab:flux_offsets}). The median offset across all galaxies is negative (-18.3\%). 

In the high EW regions, when we compare the MUSE \Ha\ SUB image to the MUSE \Ha\ spectroscopic line maps, we see very similar offsets, indicating that the origin of these systematic errors in the fluxes is indeed associated with the procedures used to estimate the \Ha\ line flux from the narrow and broad-band images. In the following sections we will show that these systematic offsets largely stems from systematic errors in the \Nii\ decontamination prescription, and the presence of non-negligible \Ha\ photospheric absorption underneath the \Ha\ emission lines.

\subsection{\fNii\ fraction correction}
\label{subsec:correct_nii}

In Section~\ref{subsec:N2_cont}, we adopted a constant \Nii/\Ha\ ratio per galaxy to account for the contamination of \Nii\ lines within the \Ha\ narrow-band filter. The adopted \fNii\ fraction for each galaxy is derived from empirical relations (i.e. the \cite{Kewley_2008} mass-metallicity relation and the \cite{Pettini_2004} N2 metallicity calibration), rather than from direct spectroscopic measurements. For the PHANGS-MUSE overlap sample we have direct spectral measurements of the \Nii/\Ha\ ratio, which allows us to validate the adopted methodology.

As explained in Section~\ref{subsec:ew_cut}, we use the PHANGS-MUSE DAP \Nii$\lambda$6548, \Nii$\lambda$6583 and \Ha$\lambda$6562 emission line maps to compute the \fNii\ fractions from spectroscopy, with each line flux weighted by the line's effective filter transmission (Equations~\ref{eq:eff_transmission} and \ref{eq:NII_frac_corr}). The resulting spaxel-by-spaxel \fNii\ fractions can then be used to construct 2D maps of the ratio between these fractions, computed from PHANGS-MUSE, and those adopted in the processing of the \pha\ narrow-band imaging (reported in column 11 of Table~\ref{tab:results}).

\begin{figure}[h!]
    \centering
    \includegraphics[width=0.5\textwidth]{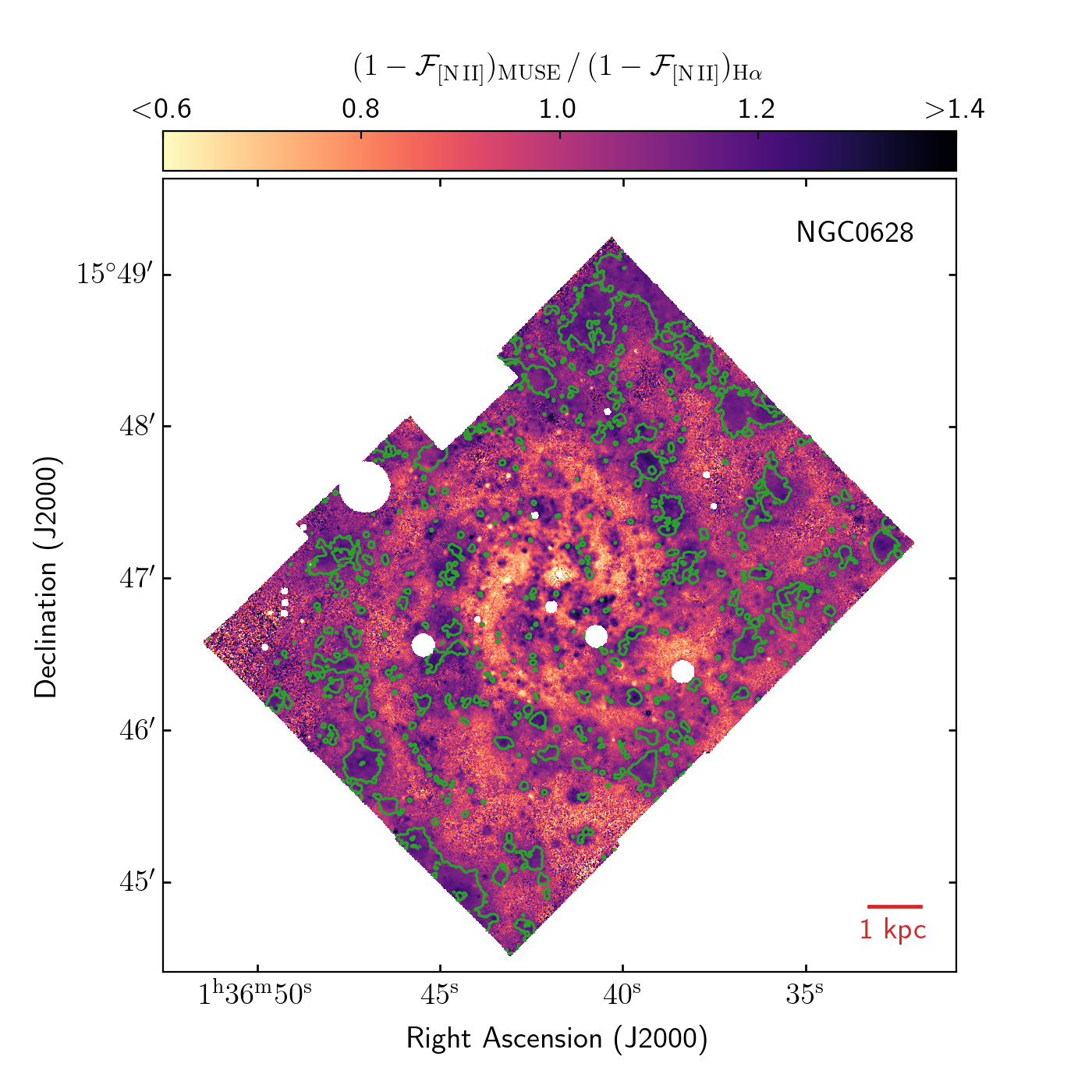}
    \caption{2D map of the scaling factor for NGC 0628 between the \fNii\ computed from PHANGS-MUSE \Nii$\lambda$6548, \Nii$\lambda$6583 and \Ha$\lambda$6562 emission line maps, each line corrected for its effective filter transmissions, and the \pha\ \fNii . Regions where the \ew $> 30$~\AA\ cut is applied are marked with green contours. The colorbar on top shows that the correction varies a lot from region to region. For the regions selected by the \ew\ threshold we observe a scaling factor in the range of $\sim$1.10--1.15.}
    \label{fig:scaling_factor_map}
\end{figure}

An example of ratio between the empirically and spectroscopically measured \Nii\ contamination factor is shown in Figure~\ref{fig:scaling_factor_map} for NGC 0628. This 2D ratio map highlights differences across the galaxy bulge, arm and inter-arm regions. As discussed in Section \ref{subsec:ew_cut}, selecting regions based on a high-EW cut reduces DIG contamination, and largely mitigates internal variations in the \Nii/\Ha\ ratio within galaxies. Regions exceeding the \ew\ threshold of $30$~\AA\ are shown within green contours in Figure~\ref{fig:scaling_factor_map}, within which the scaling factor remains fairly constant at approximately 1.10--1.15, showing minimal variations. 

This validates the idea of using a single \Nii/\Ha\ factor across all \Hii -region dominated areas within each galaxy, but shows that, in the case of NGC 0628, the empirical prescription overestimates the \Nii\ contamination in the narrow-band filter by $\sim$12\% on average. We calculate the mean \fNii\ value within the high-EW mask for all 19 galaxies, and report these values in Column (5) of Table~\ref{tab:flux_offsets}.

Applying the spectroscopically derived \fNii\ fraction correction to the narrow-band imaging, instead of the empirically derived one, decreases the scale of the residuals between the narrow-band \Ha SUB fluxes and the MUSE spectroscopic \Ha fluxes by 9.6\% (as indicated by the median in Column (6) of Table~\ref{tab:flux_offsets}). This implies that nearly half of the previously observed discrepancy with the spectroscopic data can be attributed to systematic errors in our assumptions regarding the \Nii/\Ha\ ratio.

\subsection{\fNii\ as a function of stellar mass}
\label{subsec:fnii_vs_mass}

The limitation of the spectroscopic \Nii\ correction approach presented in the previous section stems from the lack of spectral information available for the whole \pha\ sample. To address this limitation, we derive a new empirical correction by measuring the \niiha\ ratio as a function of galaxy stellar mass for the 19 PHANGS-MUSE galaxies. As in Section~\ref{subsec:correct_nii} for \fNii , the mean \niiha\ value across the spaxels with \ew\ $>30$ \AA\ is computed from each galaxy PHANGS-MUSE DAP maps, although no filter transmission correction is applied to the line fluxes in this case. This method is motivated by the fact that \Nii/\Ha\ is a metallicity indicator, therefore the \Nii/\Ha -mass relation effectively reflects the mass-metallicity relation \citep{Kewley_2008}. 


\begin{figure}[ht!]
    \centering
    \includegraphics[width=0.5\textwidth]
    {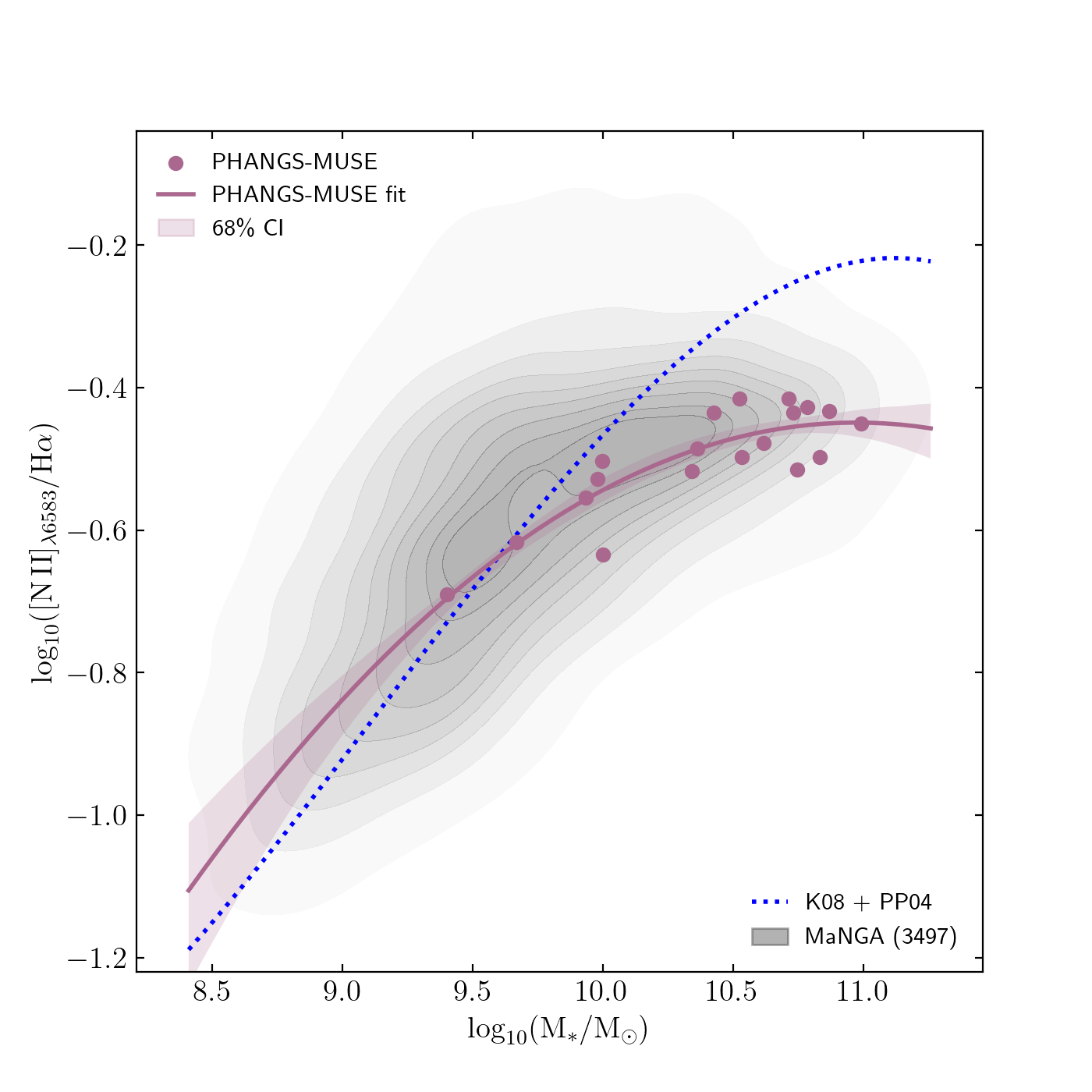}
    \caption{Mean \lognii\ as a function of \logmass\ for the 19 PHANGS-MUSE galaxies (filled circles) and a selection of 3497 MaNGA star-forming main sequence galaxies (background distribution in gray scales), computed from each survey respective emission line maps. Only the spaxels within the high-EW regime (\ew\ $>30$ \AA ) are used to calculate the mean of \lognii . The PHANGS-MUSE points trace well the MaNGA sample distribution shown in background. The best-fit line from PHANGS-MUSE data points is shown with the 68\% confidence interval, as obtained by bootstrapping the 19 points 1000 times. No corrections for the filter transmission are applied. The empirical relation as derived by combining \citet{Kewley_2008} mass-metallicity and \citet{Pettini_2004} N2-to-metallicity relations is also shown for comparison as a dotted blue line.}
    \label{fig:lognii_vs_logMstar}
\end{figure}

Figure~\ref{fig:lognii_vs_logMstar} presents the relation between \lognii\ and \logmass\ for the 19 PHANGS-MUSE galaxies, overplotted on the distribution of a large sample of MaNGA galaxies with kpc-scale resolution IFS \citep{Bundy_2015}, selected to be on the star-forming main sequence ($\log(\mathrm{sSFR}/\mathrm{yr}^{-1})>-11$) and providing reliable emission line maps from the MaNGA Data Analysis Pipeline \citep[MaNGA DAP, ][]{Westfall_2019}. For both PHANGS-MUSE and MaNGA DAP, only spaxels with the \ew $\,>30$ \AA\ mask are considered in the calculation. While MaNGA sample comparison provides a validation test to our measurements, we calculate the new empirical best-fit relation from the high-resolution PHANGS-MUSE IFS. The fitted relation can be expressed as
\begin{equation}
    \log_{10}([\mathrm{N}\,\mathrm{II}]\lambda 6583/\mathrm{H}\alpha) = -12.50 + 2.20\, x - 0.10\, x^2,
    \label{eq:nii_mass}
\end{equation}
where $x=\log_{10}(\mathrm{M_*}/\mathrm{M}_{\odot})$.

Equation~\ref{eq:nii_mass} characterizes the dependence of the \Nii/\Ha\ ratio on the stellar mass in the high-EW regime, which makes it particularly well suited for studies of \Hii\ regions. This fitted relation provides a spectroscopically based \fNii\ factor for the whole \pha\ survey, as well as for any photometric dataset, provided that the emission lines are corrected for the filter transmission as shown in Equation~\ref{eq:NII_frac_corr} in the appendix. The main advantage of this relation over the one used in Section \ref{subsec:N2_cont} is that it is independent of any mass-metallicity relation, and any N2 metallicity calibration, which are both subject to large systematic uncertainties. For example, the \cite{Kewley_2008} plus \cite{Pettini_2004} relation (blue dotted line in Figure \ref{fig:lognii_vs_logMstar}) clearly overestimates the \Nii contamination in the high stellar mass regime (where most PHANGS targets lie) and underestimates it in the low mass regime.

The \fNii\ factors derived from the fitted \Nii/\Ha -$M^*$ relation are presented in Column (10) of Table~\ref{tab:flux_offsets}. The flux profile offset between \Ha SUB and MUSE \Ha\ decreases by a median amount nearly identical to that obtained with per-galaxy corrections (compare columns (6) and (11) of Table~\ref{tab:flux_offsets}), where the \fNii\ values are computed directly from the emission line maps of each galaxy. This finding supports the use of a survey-wide correction as detailed in this section.

\subsection{Correcting for the photospheric absorption}
\label{subsec:correct_ew}

In Section~\ref{subsec:continuum_sub}, we introduced the iterative process to remove the BB contribution to the nebular emission, when scaling the BB to the NB images to achieve a correct continuum subtraction. Unfortunately, this process does not remove the underlying \Ha\ photospheric absorption whose depth is not well traced by the BB photometry.
As a consequence, using the BB as a proxy for the stellar continuum leads to an underestimation of the \Ha\ emission-line flux\footnote{As is also the case in other narrow-band imaging techniques where an off-line narrow-band filter is used to trace the continuum.}. 
Although this effect is stronger in low-EW regions dominated by older stellar populations, we focus here on the high-EW regime, where the stellar populations selected by our EW cut still span a range of ages with non-negligible Balmer absorption, making a correction necessary.

To quantify the stellar absorption correction, one can either assume a constant \ewph\ \citep[e.g.][]{Brinchmann_2004, Hopkins_2013} or measure the depth of the \Ha\ photospheric absorption by modelling the continuum with Stellar Population Synthesis (SPS) models \citep[see][among others]{GonzalezDelgado_2005, Moustakas_2006, Blanc_2009}. The latter approach has become standard in spectroscopic surveys, particularly from IFS observations where the stellar continuum can be modelled and subtracted to isolate the nebular emission in each spaxel \citep[see e.g.][]{Ho_2016,Sanchez_2016,Belfiore_2019}. 
PHANGS-MUSE precisely follows this methodology, producing spaxel-wise best-fitting linear combination of stellar population models \citep[see details in][]{Emsellem_2022}.

\begin{figure*}
    \centering
    \includegraphics[width=0.49\textwidth]{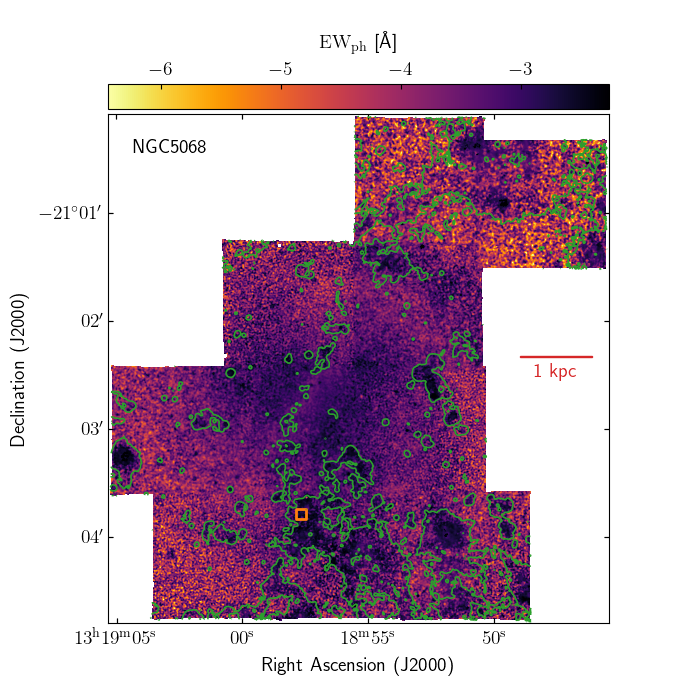}
    \includegraphics[width=0.49\textwidth]{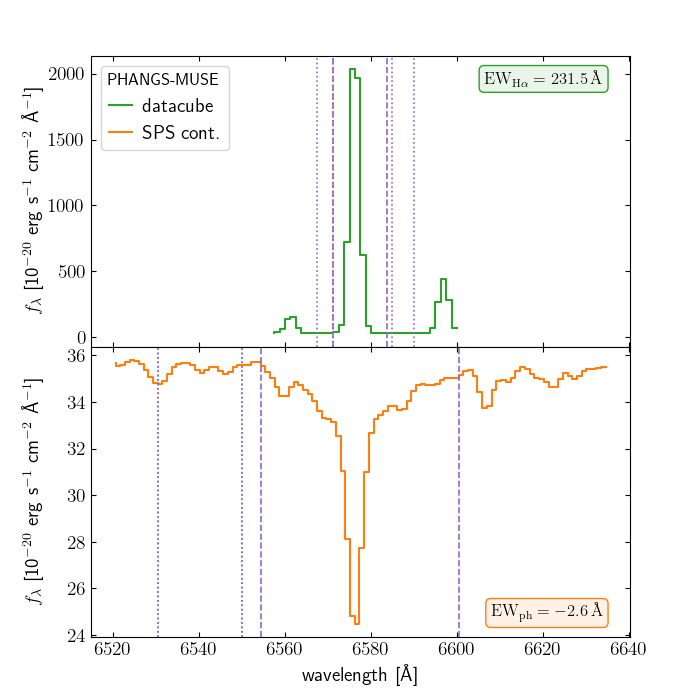}\\
    \includegraphics[width=\textwidth]{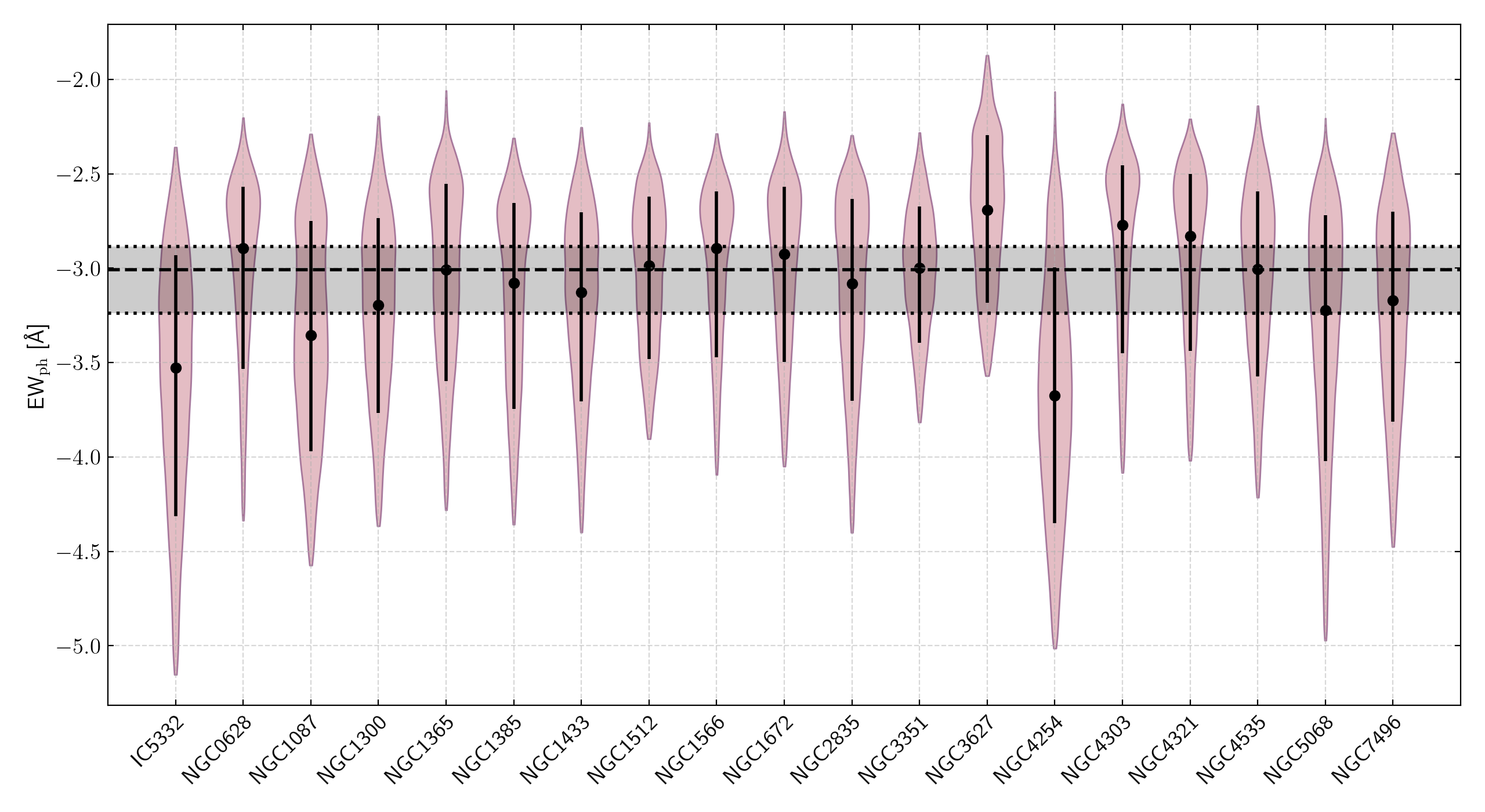}
    \caption{\emph{Top-left}: \ewph\ map of NGC 5068, where each pixel shows absorption equivalent width derived from the PHANGS-MUSE best-fitting SPS model at that position. Green contours indicate regions with \ew $>30$ \AA . The orange square marks the location of the PHANGS-MUSE datacube and SPS model spectra shown in the next panel. \emph{Top-right}: single-spaxel spectrum extracted from the PHANGS-MUSE datacube (upper panel), showing the nebular emission lines, and the corresponding best-fit stellar continuum spectrum (lower panel) where the \Ha\ absorption line is visible. The dashed and dotted lines mark the spectral windows used for line and continuum calculations, respectively. For the absorption line, a single continuum windows is used to sample a spectral region relatively free from other absorption features. The green and orange boxes report the emission and absorption EW measurements, respectively. \emph{Bottom}: plot of the \ewph\ distributions computed for each of the PHANGS-MUSE galaxies  within the \ew\ $>30$ \AA\ cut. Black dots mark the median of each distribution, while the solid lines indicate the 16\textsuperscript{th}--84\textsuperscript{th} percentile ranges. The horizontal dashed line shows the median of all galaxy medians, with dotted lines indicating the corresponding 16\textsuperscript{th} and 84\textsuperscript{th} percentiles. The \ewph\ values are negative since the EWs are all computed with the gas-centric sign convention.}
    \label{fig:EW_ph_distr}
\end{figure*}

From PHANGS-MUSE fitted continua, we derive 2D maps of \ewph\ for each galaxy, which we use to quantify and correct for the underlying \Ha\ absorption in our analysis.
The procedure and its results are illustrated in Figure~\ref{fig:EW_ph_distr}.
The top-left panel shows an example \ewph\ map for NGC 5068, where each pixel value is measured from the corresponding best-fit stellar continuum. Regions with \ew\ above our high-EW cut of 30 \AA\ are marked with green contours.
An individual spaxel PHANGS-MUSE spectrum of nebular emission lines (green) and its best-fit stellar continuum model (orange) are shown in the top-right panel, which also illustrates the adopted strategy to select the spectral windows used for line and continuum calculations in both emission and absorption line cases. For consistency with the \ew\ measurements throughout the paper, we adopt the gas-centric convention for the EW sign, which yields negative values for \ewph. Notice that different wavelength windows are used to calculate the emission and absorption EW. For the former, a narrower line window with adjacent continuum windows maximizes the S/N of the emission EW measurements. For the latter, a wider window is needed as the \Ha\ absorption feature is wider than the emission feature. The windows for the absorption EW measurements are completely encompassed by the narrow-band filter.

For each galaxy, the bottom panel of Figure~\ref{fig:EW_ph_distr} shows the resulting \ewph\ distribution obtained by applying our procedure to all the spaxels above the \ew\ threshold of 30 \AA . Represented with black dots and solid lines are the median and 16\textsuperscript{th}--84\textsuperscript{th} percentile range of each distribution, respectively. The horizontal dashed line marks the median of all galaxy medians, while the dotted lines indicate the corresponding 16\textsuperscript{th} and 84\textsuperscript{th} percentiles.

From the \ewph\ map shown in the top-left panel of Figure~\ref{fig:EW_ph_distr}, we measure absorption EWs of $\sim$ $-(2$--$7)$~\AA\ in absolute value, a range that is similarly found for the other galaxies in the sample. These values are in line with the $\sim$ $-(3$--$6)$~\AA\ range reported by \citet{GonzalezDelgado_2005}\footnote{Note that the \ewph\ sign convention is positive in \citet{GonzalezDelgado_2005}.} from simple stellar population models, supporting the use of the PHANGS-MUSE best-fitted stellar continua. 

In regions above our high-EW cut, the \ewph\ distributions shift toward lower values  since young stars dominate. Although the resulting absorptions are weaker, the measured distributions confirm that the effect of neglecting the absorption when deriving the \Ha\ fluxes should be corrected for.
For each galaxy, we adopt the median of the distribution as a representative constant \ewph , justified by the fact that the distributions span relatively narrow ranges in the high-EW regime. These values, reported in Column (7) of Table~\ref{tab:flux_offsets}, are then used to reconstruct flux correction maps by multiplying the median \ewph\ with the corresponding line-continuum map (which is part of the \ewph\ computation procedure).
This allows to build absorption-corrected flux profiles, whose effect on the residual offsets is shown in Column (8) of Table~\ref{tab:flux_offsets}. 

\begin{figure*}
    \centering
    \includegraphics[width=\textwidth]{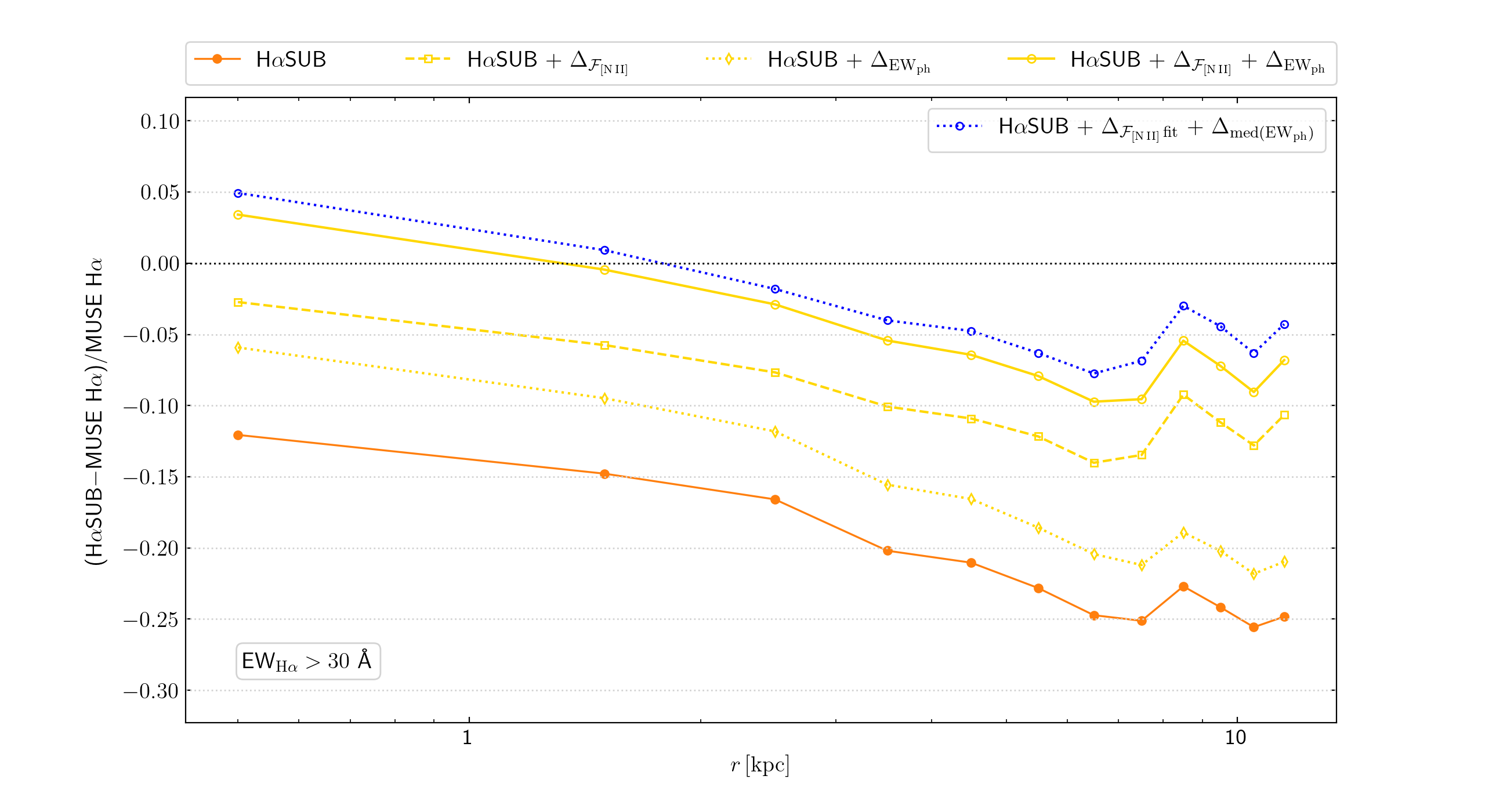}
    \caption{Radial profile residuals where in each annulus the mean value is computed across the sample of 19 \pha /PHANGS-MUSE galaxies. The residuals between the \pha\ \Ha\ emission-line equivalent fluxes (\Ha SUB) and PHANGS-MUSE \Ha\ emission-line maps (MUSE \Ha ) are taken into account and computed in the high-EW regime. The orange line represents the mean residuals of the uncorrected \Ha SUB fluxes. The yellow dashed and dotted lines show the residual offsets after the corrections for \fNii\ and \ewph , respectively. The effect of both the corrections onto the \Ha SUB fluxes is shown with the solid yellow line. Finally, a blue dotted line marks the residual offset after correcting for both \fNii\ and \ewph\ by employing the survey-wide prescriptions. MUSE \Ha\ fluxes has been corrected for the BB offset (see Table~\ref{tab:flux_offsets}).}
    \label{fig:Ha_profile_zoom}
\end{figure*}

Across the full sample, the median of the per-galaxy medians is $-3.01^{+0.12}_{-0.23}$ \AA\ with symmetric uncertainties given by the 16\textsuperscript{th}--84\textsuperscript{th} percentiles. This estimate is fairly consistent with values reported in literature (\citealp[e.g.\,$-2.8\,\pm\,0.4$\,\AA ,][]{Moustakas_2006}), although previous works average over all spaxels rather than restricting to high-EW regions as we do here.

A survey-wide correction is required for \pha\ galaxies without spectroscopic absorption estimates from PHANGS-MUSE. For these cases, we adopt the median of all the medians \ewph\ as representative value and construct flux correction maps in the same way as for the per-galaxy case. The improved offsets obtained with this approach are reported in Column (12) of Table~\ref{tab:flux_offsets} for the 19 PHANGS-MUSE galaxies for which the \Ha\ line continuum maps needed for the flux corrections are already derived.

As for the case of the \fNii\ corrections, nearly identical offset improvements are observed when the per-galaxy and the survey-wide \ewph\ corrections are applied, being 4.6\% for the first case and 4.5\% in the second (see the bottom row of columns (8) and (12)). Overall, the \fNii\ corrections account for $\sim$10\% and those for the \ewph\ account for $\sim$5\% of the flux offset between the "true" spectroscopic \Ha\ emission lines, as measured by PHANGS-MUSE, and our emission-line equivalent fluxes as derived in Section~\ref{sec:final_calib}.
These improvements are clearly reflected in the residual offsets listed in Columns (9) and (13) of Table~\ref{tab:flux_offsets}, showing a marked reduction relative to the initial offsets in Column (4).

The effect of the flux corrections is also shown in Figure~\ref{fig:Ha_profile_zoom}, which presents the stacked radial residual profiles across our sample of 19 galaxies. These profiles, computed as the mean residuals between \Ha SUB and MUSE \Ha\ fluxes in elliptical annuli and within the high-EW cut, are shown before (orange) and after the corrections for \fNii\ and \ewph . The per-galaxy corrections derived directly from the PHANGS-MUSE data are shown in yellow, and the combined effect of the survey-wide correction (i.e. from the best-fit \Nii/\Ha -$M^*$ relation and the median \ewph\ across the sample) is shown in blue. The comparison highlights the progressive improvement obtained by applying the individual corrections, either separately or in combination, yielding a reduction from the initial flux systematics of $\sim$20\% to a level of $\sim$2\%. After these corrections, we still observe residuals having radial dependence consistent with the expected effect of a metallicity gradient on the \Nii\ ratio. Further improvements could be achieved considering the effect of the metallicity gradient in future work.

Taken together, these results allow us to confidently apply the survey-wide correction prescriptions to galaxies lacking spectroscopically determined \fNii\ and \ewph , and provide a framework that can be extended to other narrow-band imaging surveys. These corrections will be applied to the next release of \pha\ data.

\section{Conclusions}
\label{sec:conclusions}

We have presented \pha , a survey of nearby star-forming galaxies that provides \Ha\ emission-line maps from narrow-band imaging. This dataset substantially expands the \Ha\ coverage of the PHANGS sample from 19 PHANGS-MUSE \citep{Emsellem_2022} and PHANGS-HST-H$\alpha$ \citep{Chandar_2025} targets to 65 galaxies, enabling broader statistical studies across the parent sample. The galaxy images, obtained with a NB filter centred on the \Ha\ line and a BB filter to trace the stellar continuum, were collected during a three-year observational campaign with the \MPG\ and \dP\ telescopes.

The dedicated \pha\ data reduction pipeline, although tailored to these observations, provides a robust procedure to calibrate astrometry and photometry on a per-exposure basis using the accurate positions of Gaia foreground stars and existing colour transformations from Gaia passbands to our filters. In particular, we fit a relation between stellar colours in the Gaia and \pha\ systems (the observed NB$-$BB colour) to calibrate the NB against the BB exposures.
These procedures yield accurate astrometric and photometric solutions that have been used to anchor the PHANGS-MUSE dataset \citep{Emsellem_2022}.

We employ state of the art techniques to derive \Ha\ emission-line flux maps from the narrow-band imaging. These include correcting for the nebular line contribution to the broad-band flux prior to continuum subtraction, and removing the \Nii $\lambda\lambda$6548,6583 doublet contamination using galaxy-specific factors informed by modelled \niiha\ ratios and the effects of the filter transmission. The resulting \Ha\ flux maps provide homogeneous measurements of the ionized-gas emission across the full sample, although subject to some intrinsic limitations of narrow-band imaging, associated with the treatment of stellar colour and \Ha\ photospheric absorption during continuum subtraction, and the lack of spectroscopic information needed to accurately remove the \Nii\ contamination.

These data have already been used in multiple PHANGS studies, ranging from the use of BB images to calibrate MUSE observations (for NGC 1672 in \citealp{Ho_2019,Kreckel_2019} and all 19 PHANGS-MUSE galaxies in \citealp{Emsellem_2022}) to analyses of the CO and \Ha\ emission spatial distributions across different scales \citep{Schinnerer_2019,Pan_2022}, the measurement of CO-\Ha\ azimuthal offsets \citep{Querejeta_2025}, and other investigations of the molecular gas and star formation lifecycle based on CO-\Ha\ displacements \citep{Chevance_2020,Chevance_2022,Kim_2022,Romanelli_2025}. The \pha\ maps have also been employed for tracing galactic structures in morphological studies of nearby galaxies\citep{Querejeta_2021, Stuber_2023} and for deriving SFR surface density maps based on \Ha\ emission \citep{Querejeta_2021,Querejeta_2024,Sun_2023}.

Besides presenting the survey, we characterize the magnitude of the systematic errors that affect the recovery of \Ha\ emission line fluxes from narrow-band imaging, we identify the main sources of systematic uncertainty, and we develop empirical corrections that can be apply to further improve the accuracy of the dataset. By comparing the \Ha\ surface brightness profiles from \pha\ and PHANGS-MUSE observations, we find that standard state-of-the-art techniques typically used in \Ha\ narrow-band imaging processing translate into flux measurements that are subject to systematic biases at the $\sim$20\% level, for bright \Hii\ regions, which we trace by using the \Ha\ emission EW.

We show that, in \Hii\ regions, this level of systematic uncertainty can be reduced by an order of magnitude, down to the $\sim$2\% level, after applying corrections based on spectroscopically informed, galaxy-specific \fNii\ factors, and properly accounting for the effects of stellar \Ha\ photospheric absorption. We present empirical recipes for these corrections that can be applied to the full \pha\ sample, and to data from other surveys, independently of the availability of integral-field spectroscopic data. Our results focus on \Hii\ region, which dominate the emission in star forming areas of galaxies. These corrections are not applicable to low surface brightness, and low emission line EW areas (e.g. bulges dominated by old stellar populations, and DIG dominated areas) where the robust recovery of \Ha\ emission line fluxes from narrow-band imaging is even more challenging.

The PHANGS collaboration is currently extending the \pha\ survey with new WFI observations at \MPG\ telescope, and with additional data from ESO archive. This ongoing effort aims to significantly expand the sample far beyond the current number of 65 galaxies, increasing the survey statistical power and providing astrometric and photometric calibrations for other PHANGS observations. We plan to incorporate the correction strategy developed in this work to future releases of the \pha\ survey dataset, in order to set a new standard for minimizing the systematic uncertainty in the recovery of \Hii\ region \Ha\ fluxes from narrow-band imaging of nearby galaxies. The forthcoming release will provide uniformly processed \Ha\ maps, fully consistent with the procedures established in this work.

\begin{acknowledgements}
MC gratefully acknowledges funding from the DFG through an Emmy Noether Research Group (grant number CH2137/1-1).
COOL Research DAO \citep{Chevance_2025} is a Decentralized Autonomous Organization supporting research in astrophysics aimed at uncovering our cosmic origins.
HAP acknowledges support from the National Science and Technology Council of Taiwan under grant 113-2112-M-032-014-MY3.
MQ acknowledges support from the Spanish grant PID2022-138560NB-I00, funded by MCIN/AEI/10.13039/501100011033/FEDER, EU.
This paper makes use of observations collected at the European Southern Observatory under ESO programme(s) 098.A-9005(A), 0101.A-9014(A), and 0102.A-9002(A), obtained with the Wide Field Imager (WFI) on the MPG/ESO 2.2-meter telescope at La Silla Observatory, as well as data gathered with the 2.5-metre du Pont telescope at Las Campanas Observatory, Chile.
\end{acknowledgements}


\bibliographystyle{aa}
\bibliography{narrow_bib}

\begin{thebibliography}{93}
\expandafter\ifx\csname natexlab\endcsname\relax\def\natexlab#1{#1}\fi

\bibitem[{{Anand} {et~al.}(2021){Anand}, {Lee}, {Van Dyk}, {Leroy}, {Rosolowsky}, {Schinnerer}, {Larson}, {Kourkchi}, {Kreckel}, {Scheuermann}, {Rizzi}, {Thilker}, {Tully}, {Bigiel}, {Blanc}, {Boquien}, {Chandar}, {Dale}, {Emsellem}, {Deger}, {Glover}, {Grasha}, {Groves}, {S. Klessen}, {Kruijssen}, {Querejeta}, {S{\'a}nchez-Bl{\'a}zquez}, {Schruba}, {Turner}, {Ubeda}, {Williams}, \& {Whitmore}}]{Gagandeep_2021}
{Anand}, G.~S., {Lee}, J.~C., {Van Dyk}, S.~D., {et~al.} 2021, \mnras, 501, 3621

\bibitem[{{Baade} {et~al.}(1999){Baade}, {Meisenheimer}, {Iwert}, {Alonso}, {Augusteijn}, {Beletic}, {Bellemann}, {Benesch}, {B{\"o}hm}, {B{\"o}hnhardt}, {Brewer}, {Deiries}, {Delabre}, {Donaldson}, {Dupuy}, {Franke}, {Gerdes}, {Gilliotte}, {Grimm}, {Haddad}, {Hess}, {Ihle}, {Klein}, {Lenzen}, {Lizon}, {Mancini}, {M{\"u}nch}, {Pizarro}, {Prado}, {Rahmer}, {Reyes}, {Richardson}, {Robledo}, {Sanchez}, {Silber}, {Sinclaire}, {Wackermann}, \& {Zaggia}}]{Baade_1999}
{Baade}, D., {Meisenheimer}, K., {Iwert}, O., {et~al.} 1999, The Messenger, 95, 15

\bibitem[{{Baldwin} {et~al.}(1981){Baldwin}, {Phillips}, \& {Terlevich}}]{Baldwin_1981}
{Baldwin}, J.~A., {Phillips}, M.~M., \& {Terlevich}, R. 1981, \pasp, 93, 5

\bibitem[{{Belfiore} {et~al.}(2023){Belfiore}, {Leroy}, {Williams}, {Barnes}, {Bigiel}, {Boquien}, {Cao}, {Chastenet}, {Congiu}, {Dale}, {Egorov}, {Eibensteiner}, {Emsellem}, {Glover}, {Groves}, {Hassani}, {Klessen}, {Kreckel}, {Neumann}, {Neumann}, {Querejeta}, {Rosolowsky}, {Sanchez-Blazquez}, {Sandstrom}, {Schinnerer}, {Sun}, {Sutter}, \& {Watkins}}]{Belfiore_2023}
{Belfiore}, F., {Leroy}, A.~K., {Williams}, T.~G., {et~al.} 2023, \aap, 678, A129

\bibitem[{{Belfiore} {et~al.}(2022){Belfiore}, {Santoro}, {Groves}, {Schinnerer}, {Kreckel}, {Glover}, {Klessen}, {Emsellem}, {Blanc}, {Congiu}, {Barnes}, {Boquien}, {Chevance}, {Dale}, {Diederik Kruijssen}, {Leroy}, {Pan}, {Pessa}, {Schruba}, \& {Williams}}]{Belfiore_2022}
{Belfiore}, F., {Santoro}, F., {Groves}, B., {et~al.} 2022, \aap, 659, A26

\bibitem[{{Belfiore} {et~al.}(2019){Belfiore}, {Westfall}, {Schaefer}, {Cappellari}, {Ji}, {Bershady}, {Tremonti}, {Law}, {Yan}, {Bundy}, {Shetty}, {Drory}, {Thomas}, {Emsellem}, \& {S{\'a}nchez}}]{Belfiore_2019}
{Belfiore}, F., {Westfall}, K.~B., {Schaefer}, A., {et~al.} 2019, \aj, 158, 160

\bibitem[{{Bertin}(2006)}]{Bertin_2006}
{Bertin}, E. 2006, in Astronomical Society of the Pacific Conference Series, Vol. 351, Astronomical Data Analysis Software and Systems XV, ed. C.~{Gabriel}, C.~{Arviset}, D.~{Ponz}, \& S.~{Enrique}, 112

\bibitem[{{Bertin} \& {Arnouts}(1996)}]{Bertin_1996}
{Bertin}, E. \& {Arnouts}, S. 1996, \aaps, 117, 393

\bibitem[{{Bertin} {et~al.}(2002){Bertin}, {Mellier}, {Radovich}, {Missonnier}, {Didelon}, \& {Morin}}]{Bertin_2002}
{Bertin}, E., {Mellier}, Y., {Radovich}, M., {et~al.} 2002, in Astronomical Society of the Pacific Conference Series, Vol. 281, Astronomical Data Analysis Software and Systems XI, ed. D.~A. {Bohlender}, D.~{Durand}, \& T.~H. {Handley}, 228

\bibitem[{{Bessell} \& {Murphy}(2012)}]{Bessell&Murphy_2012}
{Bessell}, M. \& {Murphy}, S. 2012, \pasp, 124, 140

\bibitem[{{Blanc} {et~al.}(2009){Blanc}, {Heiderman}, {Gebhardt}, {Evans}, \& {Adams}}]{Blanc_2009}
{Blanc}, G.~A., {Heiderman}, A., {Gebhardt}, K., {Evans}, Neal~J., I., \& {Adams}, J. 2009, \apj, 704, 842

\bibitem[{{Blanc} {et~al.}(2013){Blanc}, {Weinzirl}, {Song}, {Heiderman}, {Gebhardt}, {Jogee}, {Evans}, {van den Bosch}, {Luo}, {Drory}, {Fabricius}, {Fisher}, {Hao}, {Kaplan}, {Marinova}, {Vutisalchavakul}, \& {Yoachim}}]{Blanc_2013}
{Blanc}, G.~A., {Weinzirl}, T., {Song}, M., {et~al.} 2013, \aj, 145, 138

\bibitem[{Bradley {et~al.}(2021)Bradley, Sipőcz, Robitaille, Tollerud, Vinícius, Deil, Barbary, Wilson, Busko, Donath, Günther, Cara, krachyon, Conseil, Bostroem, Droettboom, Bray, Lim, Bratholm, Barentsen, Craig, Rathi, Pascual, Perren, Georgiev, de~Val-Borro, Kerzendorf, Bach, Quint, \& Souchereau}]{Bradley_2021}
Bradley, L., Sipőcz, B., Robitaille, T., {et~al.} 2021, astropy/photutils: 1.2.0

\bibitem[{{Brinchmann} {et~al.}(2004){Brinchmann}, {Charlot}, {White}, {Tremonti}, {Kauffmann}, {Heckman}, \& {Brinkmann}}]{Brinchmann_2004}
{Brinchmann}, J., {Charlot}, S., {White}, S.~D.~M., {et~al.} 2004, \mnras, 351, 1151

\bibitem[{{Bundy} {et~al.}(2015){Bundy}, {Bershady}, {Law}, {Yan}, {Drory}, {MacDonald}, {Wake}, {Cherinka}, {S{\'a}nchez-Gallego}, {Weijmans}, {Thomas}, {Tremonti}, {Masters}, {Coccato}, {Diamond-Stanic}, {Arag{\'o}n-Salamanca}, {Avila-Reese}, {Badenes}, {Falc{\'o}n-Barroso}, {Belfiore}, {Bizyaev}, {Blanc}, {Bland-Hawthorn}, {Blanton}, {Brownstein}, {Byler}, {Cappellari}, {Conroy}, {Dutton}, {Emsellem}, {Etherington}, {Frinchaboy}, {Fu}, {Gunn}, {Harding}, {Johnston}, {Kauffmann}, {Kinemuchi}, {Klaene}, {Knapen}, {Leauthaud}, {Li}, {Lin}, {Maiolino}, {Malanushenko}, {Malanushenko}, {Mao}, {Maraston}, {McDermid}, {Merrifield}, {Nichol}, {Oravetz}, {Pan}, {Parejko}, {Sanchez}, {Schlegel}, {Simmons}, {Steele}, {Steinmetz}, {Thanjavur}, {Thompson}, {Tinker}, {van den Bosch}, {Westfall}, {Wilkinson}, {Wright}, {Xiao}, \& {Zhang}}]{Bundy_2015}
{Bundy}, K., {Bershady}, M.~A., {Law}, D.~R., {et~al.} 2015, \apj, 798, 7

\bibitem[{{Chandar} {et~al.}(2025){Chandar}, {Barnes}, {Thilker}, {Caputo}, {Floyd}, {Leroy}, {{\'U}beda}, {Lee}, {Boquien}, {Maschmann}, {Belfiore}, {Kreckel}, {Glover}, {Klessen}, {Groves}, {Dale}, {Schinnerer}, {Emsellem}, {Rosolowsky}, {Bigiel}, {Blanc}, {Chevance}, {Congiu}, {Egorov}, {Faesi}, {Grasha}, {Hannon}, {Larson}, {Lopez}, {Mok}, {Neumann}, {Ostriker}, {Razza}, {S{\'a}nchez-Bl{\'a}zquez}, {Santoro}, {Schruba}, {Sun}, {Usero}, {Watkins}, {Whitmore}, \& {Williams}}]{Chandar_2025}
{Chandar}, R., {Barnes}, A.~T., {Thilker}, D.~A., {et~al.} 2025, \aj, 169, 150

\bibitem[{{Chevance} {et~al.}(2020){Chevance}, {Kruijssen}, {Hygate}, {Schruba}, {Longmore}, {Groves}, {Henshaw}, {Herrera}, {Hughes}, {Jeffreson}, {Lang}, {Leroy}, {Meidt}, {Pety}, {Razza}, {Rosolowsky}, {Schinnerer}, {Bigiel}, {Blanc}, {Emsellem}, {Faesi}, {Glover}, {Haydon}, {Ho}, {Kreckel}, {Lee}, {Liu}, {Querejeta}, {Saito}, {Sun}, {Usero}, \& {Utomo}}]{Chevance_2020}
{Chevance}, M., {Kruijssen}, J.~M.~D., {Hygate}, A. P.~S., {et~al.} 2020, \mnras, 493, 2872

\bibitem[{{Chevance} {et~al.}(2022){Chevance}, {Kruijssen}, {Krumholz}, {Groves}, {Keller}, {Hughes}, {Glover}, {Henshaw}, {Herrera}, {Kim}, {Leroy}, {Pety}, {Razza}, {Rosolowsky}, {Schinnerer}, {Schruba}, {Barnes}, {Bigiel}, {Blanc}, {Dale}, {Emsellem}, {Faesi}, {Grasha}, {Klessen}, {Kreckel}, {Liu}, {Longmore}, {Meidt}, {Querejeta}, {Saito}, {Sun}, \& {Usero}}]{Chevance_2022}
{Chevance}, M., {Kruijssen}, J.~M.~D., {Krumholz}, M.~R., {et~al.} 2022, \mnras, 509, 272

\bibitem[{{Chevance} {et~al.}(2025){Chevance}, {Kruijssen}, \& {Longmore}}]{Chevance_2025}
{Chevance}, M., {Kruijssen}, J.~M.~D., \& {Longmore}, S.~N. 2025, arXiv e-prints, arXiv:2501.13160

\bibitem[{{Cid Fernandes} {et~al.}(2011){Cid Fernandes}, {Stasi{\'n}ska}, {Mateus}, \& {Vale Asari}}]{Cid_2011}
{Cid Fernandes}, R., {Stasi{\'n}ska}, G., {Mateus}, A., \& {Vale Asari}, N. 2011, \mnras, 413, 1687

\bibitem[{{Condon}(1934)}]{Condon_1934}
{Condon}, E.~U. 1934, \apj, 79, 217

\bibitem[{{Congiu} {et~al.}(2025){Congiu}, {Scheuermann}, {Kreckel}, {Leroy}, {Emsellem}, {Belfiore}, {Hartke}, {Anand}, {Egorov}, {Groves}, {Kravtsov}, {Thilker}, {Tovo}, {Bigiel}, {Blanc}, {Bolatto}, {Cronin}, {Dale}, {McClain}, {M{\'e}ndez-Delgado}, {Oakes}, {Klessen}, {Schinnerer}, \& {Williams}}]{Congiu_2025}
{Congiu}, E., {Scheuermann}, F., {Kreckel}, K., {et~al.} 2025, \aap, 700, A125

\bibitem[{Craig {et~al.}(2016)Craig, Crawford, Seifert, Robitaille, Sipocz, Walawender, Vinícius, Ninan, Droettboom, Tollerud, Bray, walkerna22, stottsco, Günther, Rol, Bradley, Deil, Barbary, Horton, and. Nathan, Gasdia, Kerzendorf, Nelson, Singh, Streicher, Karr, Holte, Gomez, \& Weaver}]{Craig_2016}
Craig, M., Crawford, S., Seifert, M., {et~al.} 2016, astropy/ccdproc: 1.2.0

\bibitem[{{Croom} {et~al.}(2012){Croom}, {Lawrence}, {Bland-Hawthorn}, {Bryant}, {Fogarty}, {Richards}, {Goodwin}, {Farrell}, {Miziarski}, {Heald}, {Jones}, {Lee}, {Colless}, {Brough}, {Hopkins}, {Bauer}, {Birchall}, {Ellis}, {Horton}, {Leon-Saval}, {Lewis}, {L{\'o}pez-S{\'a}nchez}, {Min}, {Trinh}, \& {Trowland}}]{Croom_2012}
{Croom}, S.~M., {Lawrence}, J.~S., {Bland-Hawthorn}, J., {et~al.} 2012, \mnras, 421, 872

\bibitem[{{Denicol{\'o}} {et~al.}(2002){Denicol{\'o}}, {Terlevich}, \& {Terlevich}}]{Denicolo_2002}
{Denicol{\'o}}, G., {Terlevich}, R., \& {Terlevich}, E. 2002, \mnras, 330, 69

\bibitem[{{Drew} {et~al.}(2005){Drew}, {Greimel}, {Irwin}, {Aungwerojwit}, {Barlow}, {Corradi}, {Drake}, {G{\"a}nsicke}, {Groot}, {Hales}, {Hopewell}, {Irwin}, {Knigge}, {Leisy}, {Lennon}, {Mampaso}, {Masheder}, {Matsuura}, {Morales-Rueda}, {Morris}, {Parker}, {Phillipps}, {Rodriguez-Gil}, {Roelofs}, {Skillen}, {Sokoloski}, {Steeghs}, {Unruh}, {Viironen}, {Vink}, {Walton}, {Witham}, {Wright}, {Zijlstra}, \& {Zurita}}]{Drew_2005}
{Drew}, J.~E., {Greimel}, R., {Irwin}, M.~J., {et~al.} 2005, \mnras, 362, 753

\bibitem[{{Emsellem} {et~al.}(2022){Emsellem}, {Schinnerer}, {Santoro}, {Belfiore}, {Pessa}, {McElroy}, {Blanc}, {Congiu}, {Groves}, {Ho}, {Kreckel}, {Razza}, {Sanchez-Blazquez}, {Egorov}, {Faesi}, {Klessen}, {Leroy}, {Meidt}, {Querejeta}, {Rosolowsky}, {Scheuermann}, {Anand}, {Barnes}, {Be{\v{s}}li{\'c}}, {Bigiel}, {Boquien}, {Cao}, {Chevance}, {Dale}, {Eibensteiner}, {Glover}, {Grasha}, {Henshaw}, {Hughes}, {Koch}, {Kruijssen}, {Lee}, {Liu}, {Pan}, {Pety}, {Saito}, {Sandstrom}, {Schruba}, {Sun}, {Thilker}, {Usero}, {Watkins}, \& {Williams}}]{Emsellem_2022}
{Emsellem}, E., {Schinnerer}, E., {Santoro}, F., {et~al.} 2022, \aap, 659, A191

\bibitem[{{Evans} {et~al.}(2018){Evans}, {Riello}, {De Angeli}, {Carrasco}, {Montegriffo}, {Fabricius}, {Jordi}, {Palaversa}, {Diener}, {Busso}, {Cacciari}, {van Leeuwen}, {Burgess}, {Davidson}, {Harrison}, {Hodgkin}, {Pancino}, {Richards}, {Altavilla}, {Balaguer-N{\'u}{\~n}ez}, {Barstow}, {Bellazzini}, {Brown}, {Castellani}, {Cocozza}, {De Luise}, {Delgado}, {Ducourant}, {Galleti}, {Gilmore}, {Giuffrida}, {Holl}, {Kewley}, {Koposov}, {Marinoni}, {Marrese}, {Osborne}, {Piersimoni}, {Portell}, {Pulone}, {Ragaini}, {Sanna}, {Terrett}, {Walton}, {Wevers}, \& {Wyrzykowski}}]{Evans_2018}
{Evans}, D.~W., {Riello}, M., {De Angeli}, F., {et~al.} 2018, \aap, 616, A4

\bibitem[{{Flores-Fajardo} {et~al.}(2011){Flores-Fajardo}, {Morisset}, {Stasi{\'n}ska}, \& {Binette}}]{Flores-Fajardo_2011}
{Flores-Fajardo}, N., {Morisset}, C., {Stasi{\'n}ska}, G., \& {Binette}, L. 2011, \mnras, 415, 2182

\bibitem[{{Gaia Collaboration} {et~al.}(2018){Gaia Collaboration}, {Brown}, {Vallenari}, {Prusti}, {de Bruijne}, {Babusiaux}, {Bailer-Jones}, {Biermann}, {Evans}, {Eyer}, {Jansen}, {Jordi}, {Klioner}, {Lammers}, {Lindegren}, {Luri}, {Mignard}, {Panem}, {Pourbaix}, {Randich}, {Sartoretti}, {Siddiqui}, {Soubiran}, {van Leeuwen}, {Walton}, {Arenou}, {Bastian}, {Cropper}, {Drimmel}, {Katz}, {Lattanzi}, {Bakker}, {Cacciari}, {Casta{\~n}eda}, {Chaoul}, {Cheek}, {De Angeli}, {Fabricius}, {Guerra}, {Holl}, {Masana}, {Messineo}, {Mowlavi}, {Nienartowicz}, {Panuzzo}, {Portell}, {Riello}, {Seabroke}, {Tanga}, {Th{\'e}venin}, {Gracia-Abril}, {Comoretto}, {Garcia-Reinaldos}, {Teyssier}, {Altmann}, {Andrae}, {Audard}, {Bellas-Velidis}, {Benson}, {Berthier}, {Blomme}, {Burgess}, {Busso}, {Carry}, {Cellino}, {Clementini}, {Clotet}, {Creevey}, {Davidson}, {De Ridder}, {Delchambre}, {Dell'Oro}, {Ducourant}, {Fern{\'a}ndez-Hern{\'a}ndez}, {Fouesneau}, {Fr{\'e}mat}, {Galluccio}, {Garc{\'\i}a-Torres},
  {Gonz{\'a}lez-N{\'u}{\~n}ez}, {Gonz{\'a}lez-Vidal}, {Gosset}, {Guy}, {Halbwachs}, {Hambly}, {Harrison}, {Hern{\'a}ndez}, {Hestroffer}, {Hodgkin}, {Hutton}, {Jasniewicz}, {Jean-Antoine-Piccolo}, {Jordan}, {Korn}, {Krone-Martins}, {Lanzafame}, {Lebzelter}, {L{\"o}ffler}, {Manteiga}, {Marrese}, {Mart{\'\i}n-Fleitas}, {Moitinho}, {Mora}, {Muinonen}, {Osinde}, {Pancino}, {Pauwels}, {Petit}, {Recio-Blanco}, {Richards}, {Rimoldini}, {Robin}, {Sarro}, {Siopis}, {Smith}, {Sozzetti}, {S{\"u}veges}, {Torra}, {van Reeven}, {Abbas}, {Abreu Aramburu}, {Accart}, {Aerts}, {Altavilla}, {{\'A}lvarez}, {Alvarez}, {Alves}, {Anderson}, {Andrei}, {Anglada Varela}, {Antiche}, {Antoja}, {Arcay}, {Astraatmadja}, {Bach}, {Baker}, {Balaguer-N{\'u}{\~n}ez}, {Balm}, {Barache}, {Barata}, {Barbato}, {Barblan}, {Barklem}, {Barrado}, {Barros}, {Barstow}, {Bartholom{\'e} Mu{\~n}oz}, {Bassilana}, {Becciani}, {Bellazzini}, {Berihuete}, {Bertone}, {Bianchi}, {Bienaym{\'e}}, {Blanco-Cuaresma}, {Boch}, {Boeche}, {Bombrun}, {Borrachero},
  {Bossini}, {Bouquillon}, {Bourda}, {Bragaglia}, {Bramante}, {Breddels}, {Bressan}, {Brouillet}, {Br{\"u}semeister}, {Brugaletta}, {Bucciarelli}, {Burlacu}, {Busonero}, {Butkevich}, {Buzzi}, {Caffau}, {Cancelliere}, {Cannizzaro}, {Cantat-Gaudin}, {Carballo}, {Carlucci}, {Carrasco}, {Casamiquela}, {Castellani}, {Castro-Ginard}, {Charlot}, {Chemin}, {Chiavassa}, {Cocozza}, {Costigan}, {Cowell}, {Crifo}, {Crosta}, {Crowley}, {Cuypers}, {Dafonte}, {Damerdji}, {Dapergolas}, {David}, {David}, {de Laverny}, {De Luise}, {De March}, {de Martino}, {de Souza}, {de Torres}, {Debosscher}, {del Pozo}, {Delbo}, {Delgado}, {Delgado}, {Di Matteo}, {Diakite}, {Diener}, {Distefano}, {Dolding}, {Drazinos}, {Dur{\'a}n}, {Edvardsson}, {Enke}, {Eriksson}, {Esquej}, {Eynard Bontemps}, {Fabre}, {Fabrizio}, {Faigler}, {Falc{\~a}o}, {Farr{\`a}s Casas}, {Federici}, {Fedorets}, {Fernique}, {Figueras}, {Filippi}, {Findeisen}, {Fonti}, {Fraile}, {Fraser}, {Fr{\'e}zouls}, {Gai}, {Galleti}, {Garabato}, {Garc{\'\i}a-Sedano}, {Garofalo},
  {Garralda}, {Gavel}, {Gavras}, {Gerssen}, {Geyer}, {Giacobbe}, {Gilmore}, {Girona}, {Giuffrida}, {Glass}, {Gomes}, {Granvik}, {Gueguen}, {Guerrier}, {Guiraud}, {Guti{\'e}rrez-S{\'a}nchez}, {Haigron}, {Hatzidimitriou}, {Hauser}, {Haywood}, {Heiter}, {Helmi}, {Heu}, {Hilger}, {Hobbs}, {Hofmann}, {Holland}, {Huckle}, {Hypki}, {Icardi}, {Jan{\ss}en}, {Jevardat de Fombelle}, {Jonker}, {Juh{\'a}sz}, {Julbe}, {Karampelas}, {Kewley}, {Klar}, {Kochoska}, {Kohley}, {Kolenberg}, {Kontizas}, {Kontizas}, {Koposov}, {Kordopatis}, {Kostrzewa-Rutkowska}, {Koubsky}, {Lambert}, {Lanza}, {Lasne}, {Lavigne}, {Le Fustec}, {Le Poncin-Lafitte}, {Lebreton}, {Leccia}, {Leclerc}, {Lecoeur-Taibi}, {Lenhardt}, {Leroux}, {Liao}, {Licata}, {Lindstr{\o}m}, {Lister}, {Livanou}, {Lobel}, {L{\'o}pez}, {Managau}, {Mann}, {Mantelet}, {Marchal}, {Marchant}, {Marconi}, {Marinoni}, {Marschalk{\'o}}, {Marshall}, {Martino}, {Marton}, {Mary}, {Massari}, {Matijevi{\v{c}}}, {Mazeh}, {McMillan}, {Messina}, {Michalik}, {Millar}, {Molina}, {Molinaro},
  {Moln{\'a}r}, {Montegriffo}, {Mor}, {Morbidelli}, {Morel}, {Morris}, {Mulone}, {Muraveva}, {Musella}, {Nelemans}, {Nicastro}, {Noval}, {O'Mullane}, {Ord{\'e}novic}, {Ord{\'o}{\~n}ez-Blanco}, {Osborne}, {Pagani}, {Pagano}, {Pailler}, {Palacin}, {Palaversa}, {Panahi}, {Pawlak}, {Piersimoni}, {Pineau}, {Plachy}, {Plum}, {Poggio}, {Poujoulet}, {Pr{\v{s}}a}, {Pulone}, {Racero}, {Ragaini}, {Rambaux}, {Ramos-Lerate}, {Regibo}, {Reyl{\'e}}, {Riclet}, {Ripepi}, {Riva}, {Rivard}, {Rixon}, {Roegiers}, {Roelens}, {Romero-G{\'o}mez}, {Rowell}, {Royer}, {Ruiz-Dern}, {Sadowski}, {Sagrist{\`a} Sell{\'e}s}, {Sahlmann}, {Salgado}, {Salguero}, {Sanna}, {Santana-Ros}, {Sarasso}, {Savietto}, {Schultheis}, {Sciacca}, {Segol}, {Segovia}, {S{\'e}gransan}, {Shih}, {Siltala}, {Silva}, {Smart}, {Smith}, {Solano}, {Solitro}, {Sordo}, {Soria Nieto}, {Souchay}, {Spagna}, {Spoto}, {Stampa}, {Steele}, {Steidelm{\"u}ller}, {Stephenson}, {Stoev}, {Suess}, {Surdej}, {Szabados}, {Szegedi-Elek}, {Tapiador}, {Taris}, {Tauran}, {Taylor},
  {Teixeira}, {Terrett}, {Teyssandier}, {Thuillot}, {Titarenko}, {Torra Clotet}, {Turon}, {Ulla}, {Utrilla}, {Uzzi}, {Vaillant}, {Valentini}, {Valette}, {van Elteren}, {Van Hemelryck}, {van Leeuwen}, {Vaschetto}, {Vecchiato}, {Veljanoski}, {Viala}, {Vicente}, {Vogt}, {von Essen}, {Voss}, {Votruba}, {Voutsinas}, {Walmsley}, {Weiler}, {Wertz}, {Wevers}, {Wyrzykowski}, {Yoldas}, {{\v{Z}}erjal}, {Ziaeepour}, {Zorec}, {Zschocke}, {Zucker}, {Zurbach}, \& {Zwitter}}]{Gaia_DR2}
{Gaia Collaboration}, {Brown}, A.~G.~A., {Vallenari}, A., {et~al.} 2018, \aap, 616, A1

\bibitem[{{Gaia Collaboration} {et~al.}(2016{\natexlab{a}}){Gaia Collaboration}, {Brown}, {Vallenari}, {Prusti}, {de Bruijne}, {Mignard}, {Drimmel}, {Babusiaux}, {Bailer-Jones}, {Bastian}, {Biermann}, {Evans}, {Eyer}, {Jansen}, {Jordi}, {Katz}, {Klioner}, {Lammers}, {Lindegren}, {Luri}, {O'Mullane}, {Panem}, {Pourbaix}, {Randich}, {Sartoretti}, {Siddiqui}, {Soubiran}, {Valette}, {van Leeuwen}, {Walton}, {Aerts}, {Arenou}, {Cropper}, {H{\o}g}, {Lattanzi}, {Grebel}, {Holland}, {Huc}, {Passot}, {Perryman}, {Bramante}, {Cacciari}, {Casta{\~n}eda}, {Chaoul}, {Cheek}, {De Angeli}, {Fabricius}, {Guerra}, {Hern{\'a}ndez}, {Jean-Antoine-Piccolo}, {Masana}, {Messineo}, {Mowlavi}, {Nienartowicz}, {Ord{\'o}{\~n}ez-Blanco}, {Panuzzo}, {Portell}, {Richards}, {Riello}, {Seabroke}, {Tanga}, {Th{\'e}venin}, {Torra}, {Els}, {Gracia-Abril}, {Comoretto}, {Garcia-Reinaldos}, {Lock}, {Mercier}, {Altmann}, {Andrae}, {Astraatmadja}, {Bellas-Velidis}, {Benson}, {Berthier}, {Blomme}, {Busso}, {Carry}, {Cellino}, {Clementini},
  {Cowell}, {Creevey}, {Cuypers}, {Davidson}, {De Ridder}, {de Torres}, {Delchambre}, {Dell'Oro}, {Ducourant}, {Fr{\'e}mat}, {Garc{\'\i}a-Torres}, {Gosset}, {Halbwachs}, {Hambly}, {Harrison}, {Hauser}, {Hestroffer}, {Hodgkin}, {Huckle}, {Hutton}, {Jasniewicz}, {Jordan}, {Kontizas}, {Korn}, {Lanzafame}, {Manteiga}, {Moitinho}, {Muinonen}, {Osinde}, {Pancino}, {Pauwels}, {Petit}, {Recio-Blanco}, {Robin}, {Sarro}, {Siopis}, {Smith}, {Smith}, {Sozzetti}, {Thuillot}, {van Reeven}, {Viala}, {Abbas}, {Abreu Aramburu}, {Accart}, {Aguado}, {Allan}, {Allasia}, {Altavilla}, {{\'A}lvarez}, {Alves}, {Anderson}, {Andrei}, {Anglada Varela}, {Antiche}, {Antoja}, {Ant{\'o}n}, {Arcay}, {Bach}, {Baker}, {Balaguer-N{\'u}{\~n}ez}, {Barache}, {Barata}, {Barbier}, {Barblan}, {Barrado y Navascu{\'e}s}, {Barros}, {Barstow}, {Becciani}, {Bellazzini}, {Bello Garc{\'\i}a}, {Belokurov}, {Bendjoya}, {Berihuete}, {Bianchi}, {Bienaym{\'e}}, {Billebaud}, {Blagorodnova}, {Blanco-Cuaresma}, {Boch}, {Bombrun}, {Borrachero}, {Bouquillon},
  {Bourda}, {Bouy}, {Bragaglia}, {Breddels}, {Brouillet}, {Br{\"u}semeister}, {Bucciarelli}, {Burgess}, {Burgon}, {Burlacu}, {Busonero}, {Buzzi}, {Caffau}, {Cambras}, {Campbell}, {Cancelliere}, {Cantat-Gaudin}, {Carlucci}, {Carrasco}, {Castellani}, {Charlot}, {Charnas}, {Chiavassa}, {Clotet}, {Cocozza}, {Collins}, {Costigan}, {Crifo}, {Cross}, {Crosta}, {Crowley}, {Dafonte}, {Damerdji}, {Dapergolas}, {David}, {David}, {De Cat}, {de Felice}, {de Laverny}, {De Luise}, {De March}, {de Martino}, {de Souza}, {Debosscher}, {del Pozo}, {Delbo}, {Delgado}, {Delgado}, {Di Matteo}, {Diakite}, {Distefano}, {Dolding}, {Dos Anjos}, {Drazinos}, {Duran}, {Dzigan}, {Edvardsson}, {Enke}, {Evans}, {Eynard Bontemps}, {Fabre}, {Fabrizio}, {Faigler}, {Falc{\~a}o}, {Farr{\`a}s Casas}, {Federici}, {Fedorets}, {Fern{\'a}ndez-Hern{\'a}ndez}, {Fernique}, {Fienga}, {Figueras}, {Filippi}, {Findeisen}, {Fonti}, {Fouesneau}, {Fraile}, {Fraser}, {Fuchs}, {Gai}, {Galleti}, {Galluccio}, {Garabato}, {Garc{\'\i}a-Sedano}, {Garofalo},
  {Garralda}, {Gavras}, {Gerssen}, {Geyer}, {Gilmore}, {Girona}, {Giuffrida}, {Gomes}, {Gonz{\'a}lez-Marcos}, {Gonz{\'a}lez-N{\'u}{\~n}ez}, {Gonz{\'a}lez-Vidal}, {Granvik}, {Guerrier}, {Guillout}, {Guiraud}, {G{\'u}rpide}, {Guti{\'e}rrez-S{\'a}nchez}, {Guy}, {Haigron}, {Hatzidimitriou}, {Haywood}, {Heiter}, {Helmi}, {Hobbs}, {Hofmann}, {Holl}, {Holland}, {Hunt}, {Hypki}, {Icardi}, {Irwin}, {Jevardat de Fombelle}, {Jofr{\'e}}, {Jonker}, {Jorissen}, {Julbe}, {Karampelas}, {Kochoska}, {Kohley}, {Kolenberg}, {Kontizas}, {Koposov}, {Kordopatis}, {Koubsky}, {Krone-Martins}, {Kudryashova}, {Kull}, {Bachchan}, {Lacoste-Seris}, {Lanza}, {Lavigne}, {Le Poncin-Lafitte}, {Lebreton}, {Lebzelter}, {Leccia}, {Leclerc}, {Lecoeur-Taibi}, {Lemaitre}, {Lenhardt}, {Leroux}, {Liao}, {Licata}, {Lindstr{\o}m}, {Lister}, {Livanou}, {Lobel}, {L{\"o}ffler}, {L{\'o}pez}, {Lorenz}, {MacDonald}, {Magalh{\~a}es Fernandes}, {Managau}, {Mann}, {Mantelet}, {Marchal}, {Marchant}, {Marconi}, {Marinoni}, {Marrese}, {Marschalk{\'o}}, {Marshall},
  {Mart{\'\i}n-Fleitas}, {Martino}, {Mary}, {Matijevi{\v{c}}}, {Mazeh}, {McMillan}, {Messina}, {Michalik}, {Millar}, {Miranda}, {Molina}, {Molinaro}, {Molinaro}, {Moln{\'a}r}, {Moniez}, {Montegriffo}, {Mor}, {Mora}, {Morbidelli}, {Morel}, {Morgenthaler}, {Morris}, {Mulone}, {Muraveva}, {Musella}, {Narbonne}, {Nelemans}, {Nicastro}, {Noval}, {Ord{\'e}novic}, {Ordieres-Mer{\'e}}, {Osborne}, {Pagani}, {Pagano}, {Pailler}, {Palacin}, {Palaversa}, {Parsons}, {Pecoraro}, {Pedrosa}, {Pentik{\"a}inen}, {Pichon}, {Piersimoni}, {Pineau}, {Plachy}, {Plum}, {Poujoulet}, {Pr{\v{s}}a}, {Pulone}, {Ragaini}, {Rago}, {Rambaux}, {Ramos-Lerate}, {Ranalli}, {Rauw}, {Read}, {Regibo}, {Reyl{\'e}}, {Ribeiro}, {Rimoldini}, {Ripepi}, {Riva}, {Rixon}, {Roelens}, {Romero-G{\'o}mez}, {Rowell}, {Royer}, {Ruiz-Dern}, {Sadowski}, {Sagrist{\`a} Sell{\'e}s}, {Sahlmann}, {Salgado}, {Salguero}, {Sarasso}, {Savietto}, {Schultheis}, {Sciacca}, {Segol}, {Segovia}, {Segransan}, {Shih}, {Smareglia}, {Smart}, {Solano}, {Solitro}, {Sordo}, {Soria
  Nieto}, {Souchay}, {Spagna}, {Spoto}, {Stampa}, {Steele}, {Steidelm{\"u}ller}, {Stephenson}, {Stoev}, {Suess}, {S{\"u}veges}, {Surdej}, {Szabados}, {Szegedi-Elek}, {Tapiador}, {Taris}, {Tauran}, {Taylor}, {Teixeira}, {Terrett}, {Tingley}, {Trager}, {Turon}, {Ulla}, {Utrilla}, {Valentini}, {van Elteren}, {Van Hemelryck}, {van Leeuwen}, {Varadi}, {Vecchiato}, {Veljanoski}, {Via}, {Vicente}, {Vogt}, {Voss}, {Votruba}, {Voutsinas}, {Walmsley}, {Weiler}, {Weingrill}, {Wevers}, {Wyrzykowski}, {Yoldas}, {{\v{Z}}erjal}, {Zucker}, {Zurbach}, {Zwitter}, {Alecu}, {Allen}, {Allende Prieto}, {Amorim}, {Anglada-Escud{\'e}}, {Arsenijevic}, {Azaz}, {Balm}, {Beck}, {Bernstein}, {Bigot}, {Bijaoui}, {Blasco}, {Bonfigli}, {Bono}, {Boudreault}, {Bressan}, {Brown}, {Brunet}, {Bunclark}, {Buonanno}, {Butkevich}, {Carret}, {Carrion}, {Chemin}, {Ch{\'e}reau}, {Corcione}, {Darmigny}, {de Boer}, {de Teodoro}, {de Zeeuw}, {Delle Luche}, {Domingues}, {Dubath}, {Fodor}, {Fr{\'e}zouls}, {Fries}, {Fustes}, {Fyfe}, {Gallardo}, {Gallegos},
  {Gardiol}, {Gebran}, {Gomboc}, {G{\'o}mez}, {Grux}, {Gueguen}, {Heyrovsky}, {Hoar}, {Iannicola}, {Isasi Parache}, {Janotto}, {Joliet}, {Jonckheere}, {Keil}, {Kim}, {Klagyivik}, {Klar}, {Knude}, {Kochukhov}, {Kolka}, {Kos}, {Kutka}, {Lainey}, {LeBouquin}, {Liu}, {Loreggia}, {Makarov}, {Marseille}, {Martayan}, {Martinez-Rubi}, {Massart}, {Meynadier}, {Mignot}, {Munari}, {Nguyen}, {Nordlander}, {Ocvirk}, {O'Flaherty}, {Olias Sanz}, {Ortiz}, {Osorio}, {Oszkiewicz}, {Ouzounis}, {Palmer}, {Park}, {Pasquato}, {Peltzer}, {Peralta}, {P{\'e}turaud}, {Pieniluoma}, {Pigozzi}, {Poels}, {Prat}, {Prod'homme}, {Raison}, {Rebordao}, {Risquez}, {Rocca-Volmerange}, {Rosen}, {Ruiz-Fuertes}, {Russo}, {Sembay}, {Serraller Vizcaino}, {Short}, {Siebert}, {Silva}, {Sinachopoulos}, {Slezak}, {Soffel}, {Sosnowska}, {Strai{\v{z}}ys}, {ter Linden}, {Terrell}, {Theil}, {Tiede}, {Troisi}, {Tsalmantza}, {Tur}, {Vaccari}, {Vachier}, {Valles}, {Van Hamme}, {Veltz}, {Virtanen}, {Wallut}, {Wichmann}, {Wilkinson}, {Ziaeepour}, \&
  {Zschocke}}]{Gaia_DR1}
{Gaia Collaboration}, {Brown}, A.~G.~A., {Vallenari}, A., {et~al.} 2016{\natexlab{a}}, \aap, 595, A2

\bibitem[{{Gaia Collaboration} {et~al.}(2016{\natexlab{b}}){Gaia Collaboration}, {Prusti}, {de Bruijne}, {Brown}, {Vallenari}, {Babusiaux}, {Bailer-Jones}, {Bastian}, {Biermann}, {Evans}, {Eyer}, {Jansen}, {Jordi}, {Klioner}, {Lammers}, {Lindegren}, {Luri}, {Mignard}, {Milligan}, {Panem}, {Poinsignon}, {Pourbaix}, {Randich}, {Sarri}, {Sartoretti}, {Siddiqui}, {Soubiran}, {Valette}, {van Leeuwen}, {Walton}, {Aerts}, {Arenou}, {Cropper}, {Drimmel}, {H{\o}g}, {Katz}, {Lattanzi}, {O'Mullane}, {Grebel}, {Holland}, {Huc}, {Passot}, {Bramante}, {Cacciari}, {Casta{\~n}eda}, {Chaoul}, {Cheek}, {De Angeli}, {Fabricius}, {Guerra}, {Hern{\'a}ndez}, {Jean-Antoine-Piccolo}, {Masana}, {Messineo}, {Mowlavi}, {Nienartowicz}, {Ord{\'o}{\~n}ez-Blanco}, {Panuzzo}, {Portell}, {Richards}, {Riello}, {Seabroke}, {Tanga}, {Th{\'e}venin}, {Torra}, {Els}, {Gracia-Abril}, {Comoretto}, {Garcia-Reinaldos}, {Lock}, {Mercier}, {Altmann}, {Andrae}, {Astraatmadja}, {Bellas-Velidis}, {Benson}, {Berthier}, {Blomme}, {Busso}, {Carry}, {Cellino},
  {Clementini}, {Cowell}, {Creevey}, {Cuypers}, {Davidson}, {De Ridder}, {de Torres}, {Delchambre}, {Dell'Oro}, {Ducourant}, {Fr{\'e}mat}, {Garc{\'\i}a-Torres}, {Gosset}, {Halbwachs}, {Hambly}, {Harrison}, {Hauser}, {Hestroffer}, {Hodgkin}, {Huckle}, {Hutton}, {Jasniewicz}, {Jordan}, {Kontizas}, {Korn}, {Lanzafame}, {Manteiga}, {Moitinho}, {Muinonen}, {Osinde}, {Pancino}, {Pauwels}, {Petit}, {Recio-Blanco}, {Robin}, {Sarro}, {Siopis}, {Smith}, {Smith}, {Sozzetti}, {Thuillot}, {van Reeven}, {Viala}, {Abbas}, {Abreu Aramburu}, {Accart}, {Aguado}, {Allan}, {Allasia}, {Altavilla}, {{\'A}lvarez}, {Alves}, {Anderson}, {Andrei}, {Anglada Varela}, {Antiche}, {Antoja}, {Ant{\'o}n}, {Arcay}, {Atzei}, {Ayache}, {Bach}, {Baker}, {Balaguer-N{\'u}{\~n}ez}, {Barache}, {Barata}, {Barbier}, {Barblan}, {Baroni}, {Barrado y Navascu{\'e}s}, {Barros}, {Barstow}, {Becciani}, {Bellazzini}, {Bellei}, {Bello Garc{\'\i}a}, {Belokurov}, {Bendjoya}, {Berihuete}, {Bianchi}, {Bienaym{\'e}}, {Billebaud}, {Blagorodnova}, {Blanco-Cuaresma},
  {Boch}, {Bombrun}, {Borrachero}, {Bouquillon}, {Bourda}, {Bouy}, {Bragaglia}, {Breddels}, {Brouillet}, {Br{\"u}semeister}, {Bucciarelli}, {Budnik}, {Burgess}, {Burgon}, {Burlacu}, {Busonero}, {Buzzi}, {Caffau}, {Cambras}, {Campbell}, {Cancelliere}, {Cantat-Gaudin}, {Carlucci}, {Carrasco}, {Castellani}, {Charlot}, {Charnas}, {Charvet}, {Chassat}, {Chiavassa}, {Clotet}, {Cocozza}, {Collins}, {Collins}, {Costigan}, {Crifo}, {Cross}, {Crosta}, {Crowley}, {Dafonte}, {Damerdji}, {Dapergolas}, {David}, {David}, {De Cat}, {de Felice}, {de Laverny}, {De Luise}, {De March}, {de Martino}, {de Souza}, {Debosscher}, {del Pozo}, {Delbo}, {Delgado}, {Delgado}, {di Marco}, {Di Matteo}, {Diakite}, {Distefano}, {Dolding}, {Dos Anjos}, {Drazinos}, {Dur{\'a}n}, {Dzigan}, {Ecale}, {Edvardsson}, {Enke}, {Erdmann}, {Escolar}, {Espina}, {Evans}, {Eynard Bontemps}, {Fabre}, {Fabrizio}, {Faigler}, {Falc{\~a}o}, {Farr{\`a}s Casas}, {Faye}, {Federici}, {Fedorets}, {Fern{\'a}ndez-Hern{\'a}ndez}, {Fernique}, {Fienga}, {Figueras},
  {Filippi}, {Findeisen}, {Fonti}, {Fouesneau}, {Fraile}, {Fraser}, {Fuchs}, {Furnell}, {Gai}, {Galleti}, {Galluccio}, {Garabato}, {Garc{\'\i}a-Sedano}, {Gar{\'e}}, {Garofalo}, {Garralda}, {Gavras}, {Gerssen}, {Geyer}, {Gilmore}, {Girona}, {Giuffrida}, {Gomes}, {Gonz{\'a}lez-Marcos}, {Gonz{\'a}lez-N{\'u}{\~n}ez}, {Gonz{\'a}lez-Vidal}, {Granvik}, {Guerrier}, {Guillout}, {Guiraud}, {G{\'u}rpide}, {Guti{\'e}rrez-S{\'a}nchez}, {Guy}, {Haigron}, {Hatzidimitriou}, {Haywood}, {Heiter}, {Helmi}, {Hobbs}, {Hofmann}, {Holl}, {Holland }, {Hunt}, {Hypki}, {Icardi}, {Irwin}, {Jevardat de Fombelle}, {Jofr{\'e}}, {Jonker}, {Jorissen}, {Julbe}, {Karampelas}, {Kochoska}, {Kohley}, {Kolenberg}, {Kontizas}, {Koposov}, {Kordopatis}, {Koubsky}, {Kowalczyk}, {Krone-Martins}, {Kudryashova}, {Kull}, {Bachchan}, {Lacoste-Seris}, {Lanza}, {Lavigne}, {Le Poncin-Lafitte}, {Lebreton}, {Lebzelter}, {Leccia}, {Leclerc}, {Lecoeur-Taibi}, {Lemaitre}, {Lenhardt}, {Leroux}, {Liao}, {Licata}, {Lindstr{\o}m}, {Lister}, {Livanou}, {Lobel},
  {L{\"o}ffler}, {L{\'o}pez}, {Lopez-Lozano}, {Lorenz}, {Loureiro}, {MacDonald}, {Magalh{\~a}es Fernandes}, {Managau}, {Mann}, {Mantelet}, {Marchal}, {Marchant}, {Marconi}, {Marie}, {Marinoni}, {Marrese}, {Marschalk{\'o}}, {Marshall}, {Mart{\'\i}n-Fleitas}, {Martino}, {Mary}, {Matijevi{\v{c}}}, {Mazeh}, {McMillan}, {Messina}, {Mestre}, {Michalik}, {Millar}, {Miranda}, {Molina}, {Molinaro}, {Molinaro}, {Moln{\'a}r}, {Moniez}, {Montegriffo}, {Monteiro}, {Mor}, {Mora}, {Morbidelli}, {Morel}, {Morgenthaler}, {Morley}, {Morris}, {Mulone}, {Muraveva}, {Musella}, {Narbonne}, {Nelemans}, {Nicastro}, {Noval}, {Ord{\'e}novic}, {Ordieres-Mer{\'e}}, {Osborne}, {Pagani}, {Pagano}, {Pailler}, {Palacin}, {Palaversa}, {Parsons}, {Paulsen}, {Pecoraro}, {Pedrosa}, {Pentik{\"a}inen}, {Pereira}, {Pichon}, {Piersimoni}, {Pineau}, {Plachy}, {Plum}, {Poujoulet}, {Pr{\v{s}}a}, {Pulone}, {Ragaini}, {Rago}, {Rambaux}, {Ramos-Lerate}, {Ranalli}, {Rauw}, {Read}, {Regibo}, {Renk}, {Reyl{\'e}}, {Ribeiro}, {Rimoldini}, {Ripepi}, {Riva},
  {Rixon}, {Roelens}, {Romero-G{\'o}mez}, {Rowell}, {Royer}, {Rudolph}, {Ruiz-Dern}, {Sadowski}, {Sagrist{\`a} Sell{\'e}s}, {Sahlmann}, {Salgado}, {Salguero}, {Sarasso}, {Savietto}, {Schnorhk}, {Schultheis}, {Sciacca}, {Segol}, {Segovia}, {Segransan}, {Serpell}, {Shih}, {Smareglia}, {Smart}, {Smith}, {Solano}, {Solitro}, {Sordo}, {Soria Nieto}, {Souchay}, {Spagna}, {Spoto}, {Stampa}, {Steele}, {Steidelm{\"u}ller}, {Stephenson}, {Stoev}, {Suess}, {S{\"u}veges}, {Surdej}, {Szabados}, {Szegedi-Elek}, {Tapiador}, {Taris}, {Tauran}, {Taylor}, {Teixeira}, {Terrett}, {Tingley}, {Trager}, {Turon}, {Ulla}, {Utrilla}, {Valentini}, {van Elteren}, {Van Hemelryck}, {van Leeuwen}, {Varadi}, {Vecchiato}, {Veljanoski}, {Via}, {Vicente}, {Vogt}, {Voss}, {Votruba}, {Voutsinas}, {Walmsley}, {Weiler}, {Weingrill}, {Werner}, {Wevers}, {Whitehead}, {Wyrzykowski}, {Yoldas}, {{\v{Z}}erjal}, {Zucker}, {Zurbach}, {Zwitter}, {Alecu}, {Allen}, {Allende Prieto}, {Amorim}, {Anglada-Escud{\'e}}, {Arsenijevic}, {Azaz}, {Balm}, {Beck},
  {Bernstein}, {Bigot}, {Bijaoui}, {Blasco}, {Bonfigli}, {Bono}, {Boudreault}, {Bressan}, {Brown}, {Brunet}, {Bunclark}, {Buonanno}, {Butkevich}, {Carret}, {Carrion}, {Chemin}, {Ch{\'e}reau}, {Corcione}, {Darmigny}, {de Boer}, {de Teodoro}, {de Zeeuw}, {Delle Luche}, {Domingues}, {Dubath}, {Fodor}, {Fr{\'e}zouls}, {Fries}, {Fustes}, {Fyfe}, {Gallardo}, {Gallegos}, {Gardiol}, {Gebran}, {Gomboc}, {G{\'o}mez}, {Grux}, {Gueguen}, {Heyrovsky}, {Hoar}, {Iannicola}, {Isasi Parache}, {Janotto}, {Joliet}, {Jonckheere}, {Keil}, {Kim}, {Klagyivik}, {Klar}, {Knude}, {Kochukhov}, {Kolka}, {Kos}, {Kutka}, {Lainey}, {LeBouquin}, {Liu}, {Loreggia}, {Makarov}, {Marseille}, {Martayan}, {Martinez-Rubi}, {Massart}, {Meynadier}, {Mignot}, {Munari}, {Nguyen}, {Nordlander}, {Ocvirk}, {O'Flaherty}, {Olias Sanz}, {Ortiz}, {Osorio}, {Oszkiewicz}, {Ouzounis}, {Palmer}, {Park}, {Pasquato}, {Peltzer}, {Peralta}, {P{\'e}turaud}, {Pieniluoma}, {Pigozzi}, {Poels}, {Prat}, {Prod'homme}, {Raison}, {Rebordao}, {Risquez}, {Rocca-Volmerange},
  {Rosen}, {Ruiz-Fuertes}, {Russo}, {Sembay}, {Serraller Vizcaino}, {Short}, {Siebert}, {Silva}, {Sinachopoulos}, {Slezak}, {Soffel}, {Sosnowska}, {Strai{\v{z}}ys}, {ter Linden}, {Terrell}, {Theil}, {Tiede}, {Troisi}, {Tsalmantza}, {Tur}, {Vaccari}, {Vachier}, {Valles}, {Van Hamme}, {Veltz}, {Virtanen}, {Wallut}, {Wichmann}, {Wilkinson}, {Ziaeepour}, \& {Zschocke}}]{Gaia_mission}
{Gaia Collaboration}, {Prusti}, T., {de Bruijne}, J.~H.~J., {et~al.} 2016{\natexlab{b}}, \aap, 595, A1

\bibitem[{{Galavis} {et~al.}(1997){Galavis}, {Mendoza}, \& {Zeippen}}]{Galavis_1997}
{Galavis}, M.~E., {Mendoza}, C., \& {Zeippen}, C.~J. 1997, \aaps, 123, 159

\bibitem[{{Garner} {et~al.}(2024){Garner}, {Mihos}, {Harding}, \& {Garner}}]{Garner_2024}
{Garner}, R., {Mihos}, J.~C., {Harding}, P., \& {Garner}, C.~R. 2024, \apj, 961, 217

\bibitem[{{Gonz{\'a}lez Delgado} {et~al.}(2005){Gonz{\'a}lez Delgado}, {Cervi{\~n}o}, {Martins}, {Leitherer}, \& {Hauschildt}}]{GonzalezDelgado_2005}
{Gonz{\'a}lez Delgado}, R.~M., {Cervi{\~n}o}, M., {Martins}, L.~P., {Leitherer}, C., \& {Hauschildt}, P.~H. 2005, \mnras, 357, 945

\bibitem[{{Groves} {et~al.}(2012){Groves}, {Brinchmann}, \& {Walcher}}]{Groves_2012}
{Groves}, B., {Brinchmann}, J., \& {Walcher}, C.~J. 2012, \mnras, 419, 1402

\bibitem[{{Haffner} {et~al.}(2009){Haffner}, {Dettmar}, {Beckman}, {Wood}, {Slavin}, {Giammanco}, {Madsen}, {Zurita}, \& {Reynolds}}]{Haffner_2009}
{Haffner}, L.~M., {Dettmar}, R.~J., {Beckman}, J.~E., {et~al.} 2009, Reviews of Modern Physics, 81, 969

\bibitem[{{Ho} {et~al.}(2019){Ho}, {Kreckel}, {Meidt}, {Groves}, {Blanc}, {Bigiel}, {Dale}, {Emsellem}, {Glover}, {Grasha}, {Kewley}, {Kruijssen}, {Lang}, {McElroy}, {Kudritzki}, {Sanchez-Blazquez}, {Sandstrom}, {Santoro}, {Schinnerer}, \& {Schruba}}]{Ho_2019}
{Ho}, I.~T., {Kreckel}, K., {Meidt}, S.~E., {et~al.} 2019, \apjl, 885, L31

\bibitem[{{Ho} {et~al.}(2016){Ho}, {Medling}, {Groves}, {Rich}, {Rupke}, {Hampton}, {Kewley}, {Bland-Hawthorn}, {Croom}, {Richards}, {Schaefer}, {Sharp}, \& {Sweet}}]{Ho_2016}
{Ho}, I.~T., {Medling}, A.~M., {Groves}, B., {et~al.} 2016, \apss, 361, 280

\bibitem[{{Hopkins} {et~al.}(2013){Hopkins}, {Driver}, {Brough}, {Owers}, {Bauer}, {Gunawardhana}, {Cluver}, {Colless}, {Foster}, {Lara-L{\'o}pez}, {Roseboom}, {Sharp}, {Steele}, {Thomas}, {Baldry}, {Brown}, {Liske}, {Norberg}, {Robotham}, {Bamford}, {Bland-Hawthorn}, {Drinkwater}, {Loveday}, {Meyer}, {Peacock}, {Tuffs}, {Agius}, {Alpaslan}, {Andrae}, {Cameron}, {Cole}, {Ching}, {Christodoulou}, {Conselice}, {Croom}, {Cross}, {De Propris}, {Delhaize}, {Dunne}, {Eales}, {Ellis}, {Frenk}, {Graham}, {Grootes}, {H{\"a}u{\ss}ler}, {Heymans}, {Hill}, {Hoyle}, {Hudson}, {Jarvis}, {Johansson}, {Jones}, {van Kampen}, {Kelvin}, {Kuijken}, {L{\'o}pez-S{\'a}nchez}, {Maddox}, {Madore}, {Maraston}, {McNaught-Roberts}, {Nichol}, {Oliver}, {Parkinson}, {Penny}, {Phillipps}, {Pimbblet}, {Ponman}, {Popescu}, {Prescott}, {Proctor}, {Sadler}, {Sansom}, {Seibert}, {Staveley-Smith}, {Sutherland}, {Taylor}, {Van Waerbeke}, {V{\'a}zquez-Mata}, {Warren}, {Wijesinghe}, {Wild}, \& {Wilkins}}]{Hopkins_2013}
{Hopkins}, A.~M., {Driver}, S.~P., {Brough}, S., {et~al.} 2013, \mnras, 430, 2047

\bibitem[{{Hunter} \& {Elmegreen}(2004)}]{Hunter&Elmegreen_2004}
{Hunter}, D.~A. \& {Elmegreen}, B.~G. 2004, \aj, 128, 2170

\bibitem[{{James} {et~al.}(2004){James}, {Shane}, {Beckman}, {Cardwell}, {Collins}, {Etherton}, {de Jong}, {Fathi}, {Knapen}, {Peletier}, {Percival}, {Pollacco}, {Seigar}, {Stedman}, \& {Steele}}]{James_2004}
{James}, P.~A., {Shane}, N.~S., {Beckman}, J.~E., {et~al.} 2004, \aap, 414, 23

\bibitem[{{Janesick}(2001)}]{Janesick_2001}
{Janesick}, J.~R. 2001, {Scientific charge-coupled devices}

\bibitem[{{Kaplan} {et~al.}(2016){Kaplan}, {Jogee}, {Kewley}, {Blanc}, {Weinzirl}, {Song}, {Drory}, {Luo}, \& {van den Bosch}}]{Kaplan_2016}
{Kaplan}, K.~F., {Jogee}, S., {Kewley}, L., {et~al.} 2016, \mnras, 462, 1642

\bibitem[{{Kennicutt}(1998)}]{Kennicutt_1998}
{Kennicutt}, Robert~C., J. 1998, \araa, 36, 189

\bibitem[{{Kennicutt} {et~al.}(2003){Kennicutt}, {Armus}, {Bendo}, {Calzetti}, {Dale}, {Draine}, {Engelbracht}, {Gordon}, {Grauer}, {Helou}, {Hollenbach}, {Jarrett}, {Kewley}, {Leitherer}, {Li}, {Malhotra}, {Regan}, {Rieke}, {Rieke}, {Roussel}, {Smith}, {Thornley}, \& {Walter}}]{Kennicutt_2003}
{Kennicutt}, Robert~C., J., {Armus}, L., {Bendo}, G., {et~al.} 2003, \pasp, 115, 928

\bibitem[{{Kennicutt} {et~al.}(2008){Kennicutt}, {Lee}, {Funes}, {J.}, {Sakai}, \& {Akiyama}}]{Kennicutt_2008}
{Kennicutt}, Robert~C., J., {Lee}, J.~C., {Funes}, J.~G., {et~al.} 2008, \apjs, 178, 247

\bibitem[{{Kewley} \& {Ellison}(2008)}]{Kewley_2008}
{Kewley}, L.~J. \& {Ellison}, S.~L. 2008, \apj, 681, 1183

\bibitem[{{Kim} {et~al.}(2022){Kim}, {Chevance}, {Kruijssen}, {Leroy}, {Schruba}, {Barnes}, {Bigiel}, {Blanc}, {Cao}, {Congiu}, {Dale}, {Faesi}, {Glover}, {Grasha}, {Groves}, {Hughes}, {Klessen}, {Kreckel}, {McElroy}, {Pan}, {Pety}, {Querejeta}, {Razza}, {Rosolowsky}, {Saito}, {Schinnerer}, {Sun}, {Tomi{\v{c}}i{\'c}}, {Usero}, \& {Williams}}]{Kim_2022}
{Kim}, J., {Chevance}, M., {Kruijssen}, J.~M.~D., {et~al.} 2022, \mnras, 516, 3006

\bibitem[{{Kreckel} {et~al.}(2019){Kreckel}, {Ho}, {Blanc}, {Groves}, {Santoro}, {Schinnerer}, {Bigiel}, {Chevance}, {Congiu}, {Emsellem}, {Faesi}, {Glover}, {Grasha}, {Kruijssen}, {Lang}, {Leroy}, {Meidt}, {McElroy}, {Pety}, {Rosolowsky}, {Saito}, {Sandstrom}, {Sanchez-Blazquez}, \& {Schruba}}]{Kreckel_2019}
{Kreckel}, K., {Ho}, I.~T., {Blanc}, G.~A., {et~al.} 2019, \apj, 887, 80

\bibitem[{{Krisciunas} {et~al.}(2017){Krisciunas}, {Contreras}, {Burns}, {Phillips}, {Stritzinger}, {Morrell}, {Hamuy}, {Anais}, {Boldt}, {Busta}, {Campillay}, {Castell{\'o}n}, {Folatelli}, {Freedman}, {Gonz{\'a}lez}, {Hsiao}, {Krzeminski}, {Persson}, {Roth}, {Salgado}, {Ser{\'o}n}, {Suntzeff}, {Torres}, {Filippenko}, {Li}, {Madore}, {DePoy}, {Marshall}, {Rheault}, \& {Villanueva}}]{Krisciunas_2017}
{Krisciunas}, K., {Contreras}, C., {Burns}, C.~R., {et~al.} 2017, \aj, 154, 211

\bibitem[{{Lacerda} {et~al.}(2018){Lacerda}, {Cid Fernandes}, {Couto}, {Stasi{\'n}ska}, {Garc{\'\i}a-Benito}, {Vale Asari}, {P{\'e}rez}, {Gonz{\'a}lez Delgado}, {S{\'a}nchez}, \& {de Amorim}}]{Lacerda_2018}
{Lacerda}, E.~A.~D., {Cid Fernandes}, R., {Couto}, G.~S., {et~al.} 2018, \mnras, 474, 3727

\bibitem[{{Landolt}(1992)}]{Landolt_1992}
{Landolt}, A.~U. 1992, \aj, 104, 340

\bibitem[{{Lang} {et~al.}(2020){Lang}, {Meidt}, {Rosolowsky}, {Nofech}, {Schinnerer}, {Leroy}, {Emsellem}, {Pessa}, {Glover}, {Groves}, {Hughes}, {Kruijssen}, {Querejeta}, {Schruba}, {Bigiel}, {Blanc}, {Chevance}, {Colombo}, {Faesi}, {Henshaw}, {Herrera}, {Liu}, {Pety}, {Puschnig}, {Saito}, {Sun}, \& {Usero}}]{Lang_2020}
{Lang}, P., {Meidt}, S.~E., {Rosolowsky}, E., {et~al.} 2020, \apj, 897, 122

\bibitem[{{Lee} {et~al.}(2007){Lee}, {Kennicutt}, {Funes}, {Sakai}, \& {Akiyama}}]{Lee_2007}
{Lee}, J.~C., {Kennicutt}, R.~C., {Funes}, S.~J., J.~G., {Sakai}, S., \& {Akiyama}, S. 2007, \apjl, 671, L113

\bibitem[{{Lee} {et~al.}(2023){Lee}, {Sandstrom}, {Leroy}, {Thilker}, {Schinnerer}, {Rosolowsky}, {Larson}, {Egorov}, {Williams}, {Schmidt}, {Emsellem}, {Anand}, {Barnes}, {Belfiore}, {Be{\v{s}}li{\'c}}, {Bigiel}, {Blanc}, {Bolatto}, {Boquien}, {den Brok}, {Cao}, {Chandar}, {Chastenet}, {Chevance}, {Chiang}, {Congiu}, {Dale}, {Deger}, {Eibensteiner}, {Faesi}, {Glover}, {Grasha}, {Groves}, {Hassani}, {Henny}, {Henshaw}, {Hoyer}, {Hughes}, {Jeffreson}, {Jim{\'e}nez-Donaire}, {Kim}, {Kim}, {Klessen}, {Koch}, {Kreckel}, {Kruijssen}, {Li}, {Liu}, {Lopez}, {Maschmann}, {Chen}, {Meidt}, {Murphy}, {Neumann}, {Neumayer}, {Pan}, {Pessa}, {Pety}, {Querejeta}, {Pinna}, {Rodr{\'\i}guez}, {Saito}, {S{\'a}nchez-Bl{\'a}zquez}, {Santoro}, {Sardone}, {Smith}, {Sormani}, {Scheuermann}, {Stuber}, {Sutter}, {Sun}, {Teng}, {Tre{\ss}}, {Usero}, {Watkins}, {Whitmore}, \& {Razza}}]{Lee_2023}
{Lee}, J.~C., {Sandstrom}, K.~M., {Leroy}, A.~K., {et~al.} 2023, \apjl, 944, L17

\bibitem[{{Lee} {et~al.}(2022){Lee}, {Whitmore}, {Thilker}, {Deger}, {Larson}, {Ubeda}, {Anand}, {Boquien}, {Chandar}, {Dale}, {Emsellem}, {Leroy}, {Rosolowsky}, {Schinnerer}, {Schmidt}, {Lilly}, {Turner}, {Van Dyk}, {White}, {Barnes}, {Belfiore}, {Bigiel}, {Blanc}, {Cao}, {Chevance}, {Congiu}, {Egorov}, {Glover}, {Grasha}, {Groves}, {Henshaw}, {Hughes}, {Klessen}, {Koch}, {Kreckel}, {Kruijssen}, {Liu}, {Lopez}, {Mayker}, {Meidt}, {Murphy}, {Pan}, {Pety}, {Querejeta}, {Razza}, {Saito}, {S{\'a}nchez-Bl{\'a}zquez}, {Santoro}, {Sardone}, {Scheuermann}, {Schruba}, {Sun}, {Usero}, {Watkins}, \& {Williams}}]{Lee_2022}
{Lee}, J.~C., {Whitmore}, B.~C., {Thilker}, D.~A., {et~al.} 2022, \apjs, 258, 10

\bibitem[{{Leroy} {et~al.}(2019){Leroy}, {Sandstrom}, {Lang}, {Lewis}, {Salim}, {Behrens}, {Chastenet}, {Chiang}, {Gallagher}, {Kessler}, \& {Utomo}}]{Leroy_2019}
{Leroy}, A.~K., {Sandstrom}, K.~M., {Lang}, D., {et~al.} 2019, \apjs, 244, 24

\bibitem[{{Leroy} {et~al.}(2021){Leroy}, {Schinnerer}, {Hughes}, {Rosolowsky}, {Pety}, {Schruba}, {Usero}, {Blanc}, {Chevance}, {Emsellem}, {Faesi}, {Herrera}, {Liu}, {Meidt}, {Querejeta}, {Saito}, {Sandstrom}, {Sun}, {Williams}, {Anand}, {Barnes}, {Behrens}, {Belfiore}, {Benincasa}, {Be{\v{s}}li{\'c}}, {Bigiel}, {Bolatto}, {den Brok}, {Cao}, {Chandar}, {Chastenet}, {Chiang}, {Congiu}, {Dale}, {Deger}, {Eibensteiner}, {Egorov}, {Garc{\'\i}a-Rodr{\'\i}guez}, {Glover}, {Grasha}, {Henshaw}, {Ho}, {Kepley}, {Kim}, {Klessen}, {Kreckel}, {Koch}, {Kruijssen}, {Larson}, {Lee}, {Lopez}, {Machado}, {Mayker}, {McElroy}, {Murphy}, {Ostriker}, {Pan}, {Pessa}, {Puschnig}, {Razza}, {S{\'a}nchez-Bl{\'a}zquez}, {Santoro}, {Sardone}, {Scheuermann}, {Sliwa}, {Sormani}, {Stuber}, {Thilker}, {Turner}, {Utomo}, {Watkins}, \& {Whitmore}}]{Leroy_2021}
{Leroy}, A.~K., {Schinnerer}, E., {Hughes}, A., {et~al.} 2021, \apjs, 257, 43

\bibitem[{{Leroy} {et~al.}(2025){Leroy}, {Sun}, {Meidt}, {Agertz}, {Chiang}, {Gensior}, {Glover}, {Gnedin}, {Hughes}, {Schinnerer}, {Barnes}, {Bigiel}, {Bolatto}, {Colombo}, {den Brok}, {Chevance}, {Chown}, {Eibensteiner}, {Gleis}, {Grasha}, {Henshaw}, {Klessen}, {Koch}, {Oakes}, {Pan}, {Querejeta}, {Rosolowsky}, {Saito}, {Sandstrom}, {Sarbadhicary}, {Teng}, {Usero}, {Utomo}, \& {Williams}}]{Leroy_2025}
{Leroy}, A.~K., {Sun}, J., {Meidt}, S., {et~al.} 2025, \apj, 985, 14

\bibitem[{{Makarov} {et~al.}(2014){Makarov}, {Prugniel}, {Terekhova}, {Courtois}, \& {Vauglin}}]{Makarov_2014}
{Makarov}, D., {Prugniel}, P., {Terekhova}, N., {Courtois}, H., \& {Vauglin}, I. 2014, \aap, 570, A13

\bibitem[{{Martin} {et~al.}(2005){Martin}, {Fanson}, {Schiminovich}, {Morrissey}, {Friedman}, {Barlow}, {Conrow}, {Grange}, {Jelinsky}, {Milliard}, {Siegmund}, {Bianchi}, {Byun}, {Donas}, {Forster}, {Heckman}, {Lee}, {Madore}, {Malina}, {Neff}, {Rich}, {Small}, {Surber}, {Szalay}, {Welsh}, \& {Wyder}}]{Martin_2005}
{Martin}, D.~C., {Fanson}, J., {Schiminovich}, D., {et~al.} 2005, \apjl, 619, L1

\bibitem[{{Maschmann} {et~al.}(2024){Maschmann}, {Lee}, {Thilker}, {Whitmore}, {Deger}, {Boquien}, {Chandar}, {Dale}, {Wofford}, {Hannon}, {Larson}, {Leroy}, {Schinnerer}, {Rosolowsky}, {{\'U}beda}, {Barnes}, {Emsellem}, {Grasha}, {Groves}, {Indebetouw}, {Kim}, {Klessen}, {Kreckel}, {Levy}, {Pinna}, {Rodr{\'\i}guez}, {Tian}, \& {Williams}}]{Maschmann_2024}
{Maschmann}, D., {Lee}, J.~C., {Thilker}, D.~A., {et~al.} 2024, \apjs, 273, 14

\bibitem[{{McCully} {et~al.}(2018){McCully}, {Crawford}, {Kovacs}, {Tollerud}, {Betts}, {Bradley}, {Craig}, {Turner}, {Streicher}, {Sipocz}, {Robitaille}, \& {Deil}}]{McCully_2018}
{McCully}, C., {Crawford}, S., {Kovacs}, G., {et~al.} 2018, {Astropy/Astroscrappy: V1.0.5 Zenodo Release}

\bibitem[{{Meurer} {et~al.}(2006){Meurer}, {Hanish}, {Ferguson}, {Knezek}, {Kilborn}, {Putman}, {Smith}, {Koribalski}, {Meyer}, {Oey}, {Ryan-Weber}, {Zwaan}, {Heckman}, {Kennicutt}, {Lee}, {Webster}, {Bland-Hawthorn}, {Dopita}, {Freeman}, {Doyle}, {Drinkwater}, {Staveley-Smith}, \& {Werk}}]{Meurer_2006}
{Meurer}, G.~R., {Hanish}, D.~J., {Ferguson}, H.~C., {et~al.} 2006, \apjs, 165, 307

\bibitem[{{Mor{\'e}}(1978)}]{LMA_1978}
{Mor{\'e}}, J.~J. 1978, in Lecture Notes in Mathematics, Berlin Springer Verlag, Vol. 630, 105--116

\bibitem[{{Moustakas} \& {Kennicutt}(2006)}]{Moustakas_2006}
{Moustakas}, J. \& {Kennicutt}, Jr., R.~C. 2006, \apjs, 164, 81

\bibitem[{{Pan} {et~al.}(2022){Pan}, {Schinnerer}, {Hughes}, {Leroy}, {Groves}, {Barnes}, {Belfiore}, {Bigiel}, {Blanc}, {Cao}, {Chevance}, {Congiu}, {Dale}, {Eibensteiner}, {Emsellem}, {Faesi}, {Glover}, {Grasha}, {Herrera}, {Ho}, {Klessen}, {Kruijssen}, {Lang}, {Liu}, {McElroy}, {Meidt}, {Murphy}, {Pety}, {Querejeta}, {Razza}, {Rosolowsky}, {Saito}, {Santoro}, {Schruba}, {Sun}, {Tomi{\v{c}}i{\'c}}, {Usero}, {Utomo}, \& {Williams}}]{Pan_2022}
{Pan}, H.-A., {Schinnerer}, E., {Hughes}, A., {et~al.} 2022, \apj, 927, 9

\bibitem[{{Pancino} {et~al.}(2021){Pancino}, {Sanna}, {Altavilla}, {Marinoni}, {Rainer}, {Cocozza}, {Ragaini}, {Galleti}, {Bellazzini}, {Bragaglia}, {Tessicini}, {Voss}, {Carrasco}, {Jordi}, {Harrison}, {De Angeli}, {Evans}, \& {Fanari}}]{Pancino_2021}
{Pancino}, E., {Sanna}, N., {Altavilla}, G., {et~al.} 2021, \mnras, 503, 3660

\bibitem[{{Pettini} \& {Pagel}(2004)}]{Pettini_2004}
{Pettini}, M. \& {Pagel}, B. E.~J. 2004, \mnras, 348, L59

\bibitem[{{Querejeta} {et~al.}(2024){Querejeta}, {Leroy}, {Meidt}, {Schinnerer}, {Belfiore}, {Emsellem}, {Klessen}, {Sun}, {Sormani}, {Be{\v{s}}li{\'c}}, {Cao}, {Chevance}, {Colombo}, {Dale}, {Garc{\'\i}a-Burillo}, {Glover}, {Grasha}, {Groves}, {Koch}, {Neumann}, {Pan}, {Pessa}, {Pety}, {Pinna}, {Ramambason}, {Razza}, {Romanelli}, {Rosolowsky}, {Ruiz-Garc{\'\i}a}, {S{\'a}nchez-Bl{\'a}zquez}, {Smith}, {Stuber}, {Ubeda}, {Usero}, \& {Williams}}]{Querejeta_2024}
{Querejeta}, M., {Leroy}, A.~K., {Meidt}, S.~E., {et~al.} 2024, \aap, 687, A293

\bibitem[{{Querejeta} {et~al.}(2025){Querejeta}, {Meidt}, {Cao}, {Colombo}, {Emsellem}, {Garc{\'\i}a-Burillo}, {Klessen}, {Koch}, {Leroy}, {Ruiz-Garc{\'\i}a}, {Schinnerer}, {Smith}, {Stuber}, {Thorp}, {Williams}, {Boquien}, {Dale}, {Faesi}, {Gleis}, {Grasha}, {Hughes}, {Jim{\'e}nez-Donaire}, {Kreckel}, {Liu}, {Neumann}, {Pan}, {Pinna}, {Razza}, {Saito}, {Sun}, \& {Usero}}]{Querejeta_2025}
{Querejeta}, M., {Meidt}, S.~E., {Cao}, Y., {et~al.} 2025, \aap, 701, A183

\bibitem[{{Querejeta} {et~al.}(2021){Querejeta}, {Schinnerer}, {Meidt}, {Sun}, {Leroy}, {Emsellem}, {Klessen}, {Mu{\~n}oz-Mateos}, {Salo}, {Laurikainen}, {Be{\v{s}}li{\'c}}, {Blanc}, {Chevance}, {Dale}, {Eibensteiner}, {Faesi}, {Garc{\'\i}a-Rodr{\'\i}guez}, {Glover}, {Grasha}, {Henshaw}, {Herrera}, {Hughes}, {Kreckel}, {Kruijssen}, {Liu}, {Murphy}, {Pan}, {Pety}, {Razza}, {Rosolowsky}, {Saito}, {Schruba}, {Usero}, {Watkins}, \& {Williams}}]{Querejeta_2021}
{Querejeta}, M., {Schinnerer}, E., {Meidt}, S., {et~al.} 2021, \aap, 656, A133

\bibitem[{{Romanelli} {et~al.}(2025){Romanelli}, {Chevance}, {Kruijssen}, {Ramambason}, {Querejeta}, {Boquien}, {Dale}, {den Brok}, {Glover}, {Grasha}, {Hughes}, {Kim}, {Longmore}, {Meidt}, {Mendez-Delgado}, {Neumann}, {Pety}, {Schinnerer}, {Smith}, {Sun}, \& {Williams}}]{Romanelli_2025}
{Romanelli}, A., {Chevance}, M., {Kruijssen}, J.~M.~D., {et~al.} 2025, \aap, 698, A296

\bibitem[{{Rossa} \& {Dettmar}(2003)}]{Rossa&Dettmar_2003}
{Rossa}, J. \& {Dettmar}, R.~J. 2003, \aap, 406, 505

\bibitem[{{Salo} {et~al.}(2015){Salo}, {Laurikainen}, {Laine}, {Comer{\'o}n}, {Gadotti}, {Buta}, {Sheth}, {Zaritsky}, {Ho}, {Knapen}, {Athanassoula}, {Bosma}, {Laine}, {Cisternas}, {Kim}, {Mu{\~n}oz-Mateos}, {Regan}, {Hinz}, {Gil de Paz}, {Menendez-Delmestre}, {Mizusawa}, {Erroz-Ferrer}, {Meidt}, \& {Querejeta}}]{Salo_2015}
{Salo}, H., {Laurikainen}, E., {Laine}, J., {et~al.} 2015, \apjs, 219, 4

\bibitem[{{S{\'a}nchez} {et~al.}(2012){S{\'a}nchez}, {Kennicutt}, {Gil de Paz}, {van de Ven}, {V{\'\i}lchez}, {Wisotzki}, {Walcher}, {Mast}, {Aguerri}, {Albiol-P{\'e}rez}, {Alonso-Herrero}, {Alves}, {Bakos}, {Bart{\'a}kov{\'a}}, {Bland-Hawthorn}, {Boselli}, {Bomans}, {Castillo-Morales}, {Cortijo-Ferrero}, {de Lorenzo-C{\'a}ceres}, {Del Olmo}, {Dettmar}, {D{\'\i}az}, {Ellis}, {Falc{\'o}n-Barroso}, {Flores}, {Gallazzi}, {Garc{\'\i}a-Lorenzo}, {Gonz{\'a}lez Delgado}, {Gruel}, {Haines}, {Hao}, {Husemann}, {Igl{\'e}sias-P{\'a}ramo}, {Jahnke}, {Johnson}, {Jungwiert}, {Kalinova}, {Kehrig}, {Kupko}, {L{\'o}pez-S{\'a}nchez}, {Lyubenova}, {Marino}, {M{\'a}rmol-Queralt{\'o}}, {M{\'a}rquez}, {Masegosa}, {Meidt}, {Mendez-Abreu}, {Monreal-Ibero}, {Montijo}, {Mour{\~a}o}, {Palacios-Navarro}, {Papaderos}, {Pasquali}, {Peletier}, {P{\'e}rez}, {P{\'e}rez}, {Quirrenbach}, {Rela{\~n}o}, {Rosales-Ortega}, {Roth}, {Ruiz-Lara}, {S{\'a}nchez-Bl{\'a}zquez}, {Sengupta}, {Singh}, {Stanishev}, {Trager}, {Vazdekis}, {Viironen}, {Wild},
  {Zibetti}, \& {Ziegler}}]{Sanchez_2012}
{S{\'a}nchez}, S.~F., {Kennicutt}, R.~C., {Gil de Paz}, A., {et~al.} 2012, \aap, 538, A8

\bibitem[{{S{\'a}nchez} {et~al.}(2016){S{\'a}nchez}, {P{\'e}rez}, {S{\'a}nchez-Bl{\'a}zquez}, {Gonz{\'a}lez}, {Ros{\'a}les-Ortega}, {Cano-D{\'\i}az}, {L{\'o}pez-Cob{\'a}}, {Marino}, {Gil de Paz}, {Moll{\'a}}, {L{\'o}pez-S{\'a}nchez}, {Ascasibar}, \& {Barrera-Ballesteros}}]{Sanchez_2016}
{S{\'a}nchez}, S.~F., {P{\'e}rez}, E., {S{\'a}nchez-Bl{\'a}zquez}, P., {et~al.} 2016, \rmxaa, 52, 21

\bibitem[{{S{\'a}nchez} {et~al.}(2014){S{\'a}nchez}, {Rosales-Ortega}, {Iglesias-P{\'a}ramo}, {Moll{\'a}}, {Barrera-Ballesteros}, {Marino}, {P{\'e}rez}, {S{\'a}nchez-Blazquez}, {Gonz{\'a}lez Delgado}, {Cid Fernandes}, {de Lorenzo-C{\'a}ceres}, {Mendez-Abreu}, {Galbany}, {Falcon-Barroso}, {Miralles-Caballero}, {Husemann}, {Garc{\'\i}a-Benito}, {Mast}, {Walcher}, {Gil de Paz}, {Garc{\'\i}a-Lorenzo}, {Jungwiert}, {V{\'\i}lchez}, {J{\'\i}lkov{\'a}}, {Lyubenova}, {Cortijo-Ferrero}, {D{\'\i}az}, {Wisotzki}, {M{\'a}rquez}, {Bland-Hawthorn}, {Ellis}, {van de Ven}, {Jahnke}, {Papaderos}, {Gomes}, {Mendoza}, \& {L{\'o}pez-S{\'a}nchez}}]{Sanchez_2014}
{S{\'a}nchez}, S.~F., {Rosales-Ortega}, F.~F., {Iglesias-P{\'a}ramo}, J., {et~al.} 2014, \aap, 563, A49

\bibitem[{{Schinnerer} {et~al.}(2019){Schinnerer}, {Hughes}, {Leroy}, {Groves}, {Blanc}, {Kreckel}, {Bigiel}, {Chevance}, {Dale}, {Emsellem}, {Faesi}, {Glover}, {Grasha}, {Henshaw}, {Hygate}, {Kruijssen}, {Meidt}, {Pety}, {Querejeta}, {Rosolowsky}, {Saito}, {Schruba}, {Sun}, \& {Utomo}}]{Schinnerer_2019}
{Schinnerer}, E., {Hughes}, A., {Leroy}, A., {et~al.} 2019, \apj, 887, 49

\bibitem[{{Searle}(1971)}]{Searle_1971}
{Searle}, L. 1971, \apj, 168, 327

\bibitem[{{Stritzinger} {et~al.}(2011){Stritzinger}, {Phillips}, {Boldt}, {Burns}, {Campillay}, {Contreras}, {Gonzalez}, {Folatelli}, {Morrell}, {Krzeminski}, {Roth}, {Salgado}, {DePoy}, {Hamuy}, {Freedman}, {Madore}, {Marshall}, {Persson}, {Rheault}, {Suntzeff}, {Villanueva}, {Li}, \& {Filippenko}}]{Stritzinger_2011}
{Stritzinger}, M.~D., {Phillips}, M.~M., {Boldt}, L.~N., {et~al.} 2011, \aj, 142, 156

\bibitem[{{Stuber} {et~al.}(2023){Stuber}, {Schinnerer}, {Williams}, {Querejeta}, {Meidt}, {Emsellem}, {Barnes}, {Klessen}, {Leroy}, {Neumann}, {Sormani}, {Bigiel}, {Chevance}, {Dale}, {Faesi}, {Glover}, {Grasha}, {Kruijssen}, {Liu}, {Pan}, {Pety}, {Pinna}, {Saito}, {Usero}, \& {Watkins}}]{Stuber_2023}
{Stuber}, S.~K., {Schinnerer}, E., {Williams}, T.~G., {et~al.} 2023, \aap, 676, A113

\bibitem[{{Sun} {et~al.}(2023){Sun}, {Leroy}, {Ostriker}, {Meidt}, {Rosolowsky}, {Schinnerer}, {Wilson}, {Utomo}, {Belfiore}, {Blanc}, {Emsellem}, {Faesi}, {Groves}, {Hughes}, {Koch}, {Kreckel}, {Liu}, {Pan}, {Pety}, {Querejeta}, {Razza}, {Saito}, {Sardone}, {Usero}, {Williams}, {Bigiel}, {Bolatto}, {Chevance}, {Dale}, {Gensior}, {Glover}, {Grasha}, {Henshaw}, {Jim{\'e}nez-Donaire}, {Klessen}, {Kruijssen}, {Murphy}, {Neumann}, {Teng}, \& {Thilker}}]{Sun_2023}
{Sun}, J., {Leroy}, A.~K., {Ostriker}, E.~C., {et~al.} 2023, \apjl, 945, L19

\bibitem[{{Tody}(1986)}]{Tody_1986}
{Tody}, D. 1986, in Society of Photo-Optical Instrumentation Engineers (SPIE) Conference Series, Vol. 627, Instrumentation in astronomy VI, ed. D.~L. {Crawford}, 733

\bibitem[{{Tokunaga} \& {Vacca}(2005)}]{Tokunaga_2005}
{Tokunaga}, A.~T. \& {Vacca}, W.~D. 2005, \pasp, 117, 421

\bibitem[{{van Dokkum}(2001)}]{vanDokkum_2001}
{van Dokkum}, P.~G. 2001, \pasp, 113, 1420

\bibitem[{{Van Sistine} {et~al.}(2016){Van Sistine}, {Salzer}, {Sugden}, {Giovanelli}, {Haynes}, {Janowiecki}, {Jaskot}, \& {Wilcots}}]{vanSistine_2016}
{Van Sistine}, A., {Salzer}, J.~J., {Sugden}, A., {et~al.} 2016, \apj, 824, 25

\bibitem[{{Vilella-Rojo} {et~al.}(2021){Vilella-Rojo}, {Logro{\~n}o-Garc{\'\i}a}, {L{\'o}pez-Sanjuan}, {Viironen}, {Varela}, {Moles}, {Cenarro}, {Crist{\'o}bal-Hornillos}, {Ederoclite}, {Hern{\'a}ndez-Monteagudo}, {Mar{\'\i}n-Franch}, {V{\'a}zquez Rami{\'o}}, {Galbany}, {Gonz{\'a}lez Delgado}, {Hern{\'a}n-Caballero}, {Lumbreras-Calle}, {S{\'a}nchez-Bl{\'a}zquez}, {Sobral}, {V{\'\i}lchez}, {Alcaniz}, {Angulo}, {Dupke}, \& {Sodr{\'e}}}]{Vilella_2021}
{Vilella-Rojo}, G., {Logro{\~n}o-Garc{\'\i}a}, R., {L{\'o}pez-Sanjuan}, C., {et~al.} 2021, \aap, 650, A68

\bibitem[{{Westfall} {et~al.}(2019){Westfall}, {Cappellari}, {Bershady}, {Bundy}, {Belfiore}, {Ji}, {Law}, {Schaefer}, {Shetty}, {Tremonti}, {Yan}, {Andrews}, {Brownstein}, {Cherinka}, {Coccato}, {Drory}, {Maraston}, {Parikh}, {S{\'a}nchez-Gallego}, {Thomas}, {Weijmans}, {Barrera-Ballesteros}, {Du}, {Goddard}, {Li}, {Masters}, {Ibarra Medel}, {S{\'a}nchez}, {Yang}, {Zheng}, \& {Zhou}}]{Westfall_2019}
{Westfall}, K.~B., {Cappellari}, M., {Bershady}, M.~A., {et~al.} 2019, \aj, 158, 231

\bibitem[{{Williams} {et~al.}(2024){Williams}, {Lee}, {Larson}, {Leroy}, {Sandstrom}, {Schinnerer}, {Thilker}, {Belfiore}, {Egorov}, {Rosolowsky}, {Sutter}, {DePasquale}, {Pagan}, {Berger}, {Anand}, {Barnes}, {Bigiel}, {Boquien}, {Cao}, {Chastenet}, {Chevance}, {Chown}, {Dale}, {Deger}, {Eibensteiner}, {Emsellem}, {Faesi}, {Glover}, {Grasha}, {Hannon}, {Hassani}, {Henshaw}, {Jim{\'e}nez-Donaire}, {Kim}, {Klessen}, {Koch}, {Li}, {Liu}, {Meidt}, {M{\'e}ndez-Delgado}, {Murphy}, {Neumann}, {Neumann}, {Neumayer}, {Oakes}, {Pathak}, {Pety}, {Pinna}, {Querejeta}, {Ramambason}, {Romanelli}, {Sormani}, {Stuber}, {Sun}, {Teng}, {Usero}, {Watkins}, \& {Weinbeck}}]{Williams_2024}
{Williams}, T.~G., {Lee}, J.~C., {Larson}, K.~L., {et~al.} 2024, \apjs, 273, 13

\bibitem[{{Wright} {et~al.}(2010){Wright}, {Eisenhardt}, {Mainzer}, {Ressler}, {Cutri}, {Jarrett}, {Kirkpatrick}, {Padgett}, {McMillan}, {Skrutskie}, {Stanford}, {Cohen}, {Walker}, {Mather}, {Leisawitz}, {Gautier}, {McLean}, {Benford}, {Lonsdale}, {Blain}, {Mendez}, {Irace}, {Duval}, {Liu}, {Royer}, {Heinrichsen}, {Howard}, {Shannon}, {Kendall}, {Walsh}, {Larsen}, {Cardon}, {Schick}, {Schwalm}, {Abid}, {Fabinsky}, {Naes}, \& {Tsai}}]{Wright_2010}
{Wright}, E.~L., {Eisenhardt}, P. R.~M., {Mainzer}, A.~K., {et~al.} 2010, \aj, 140, 1868

\bibitem[{{Zhang} {et~al.}(2017){Zhang}, {Yan}, {Bundy}, {Bershady}, {Haffner}, {Walterbos}, {Maiolino}, {Tremonti}, {Thomas}, {Drory}, {Jones}, {Belfiore}, {S{\'a}nchez}, {Diamond-Stanic}, {Bizyaev}, {Nitschelm}, {Andrews}, {Brinkmann}, {Brownstein}, {Cheung}, {Li}, {Law}, {Roman Lopes}, {Oravetz}, {Pan}, {Storchi Bergmann}, \& {Simmons}}]{Zhang_2017}
{Zhang}, K., {Yan}, R., {Bundy}, K., {et~al.} 2017, \mnras, 466, 3217

\end{thebibliography}

\begin{appendix}
\nolinenumbers

\section{Photometric calibration details}
\label{app:photocal}

Each \pha\ exposure is calibrated independently. In every exposure, stars are detected and cross-matched to Gaia DR2 stars with available passpand photometry. 
The ZP is then measured for every matched star as
\begin{equation}
    \mathrm{ZP}_\mathrm{X} = \mathrm{X}^\mathrm{Gaia} + 2.5\log (f_\mathrm{X}) + k\chi,
\label{eq:zp_formula}
\end{equation}
where X represents an arbitrary \pha\ filter (Table~\ref{tab:filters}), $f_\mathrm{X}$ is the flux measured from the aperture photometry (Section~\ref{subsec:photometry}), $\chi$ the airmass at the beginning of the exposure, and $\kappa$ the atmosphere extinction coefficient.
The term $- 2.5\log (f_\mathrm{X}) - \kappa\chi$ is then the above-atmosphere instrumental magnitude for that star detected in the exposure observed with filter X.
We use $\kappa=0.07\pm 0.01$ for the WFI \Rc\ filter\footnote{\url{http://www.eso.org/sci/facilities/lasilla/instruments/wfi/inst/zeropoints.html}} and $\kappa=0.103\pm 0.019$ taken from \citet{Krisciunas_2017} measurements at the Swope telescope with the Sloan $r$ filter. The colour term in Equation~\ref{eq:zp_formula} is negligible since its coefficient is zero for the WFI/\Rc\ and -0.016 for Swope/$r$.

The calibration term $\mathrm{X}^\mathrm{Gaia}$ represents the Gaia-predicted X magnitude computed by the \citet{Evans_2018} transformations from Gaia DR2 passbands. For \pha\ BB filters the transformations are given by the following equations:
\begin{equation}
\begin{aligned}
    R_\mathrm{c}^\mathrm{Gaia} &=  G + 0.003226 - 0.3833\cdot (G_\mathrm{BP}-G_\mathrm{RP})\\ 
    &+ 0.1345\cdot (G_\mathrm{BP}-G_\mathrm{RP})^2 + 0.163,
\end{aligned}
\label{eq:Gaia_transf_wfi}
\end{equation}
\begin{equation}
\begin{aligned}
    r^\mathrm{Gaia} &= G + 0.12879 - 0.24662\cdot (G_\mathrm{BP}-G_\mathrm{RP})\\ 
    &+ 0.027464\cdot (G_\mathrm{BP}-G_\mathrm{RP})^2\\ 
    &+ 0.049465\cdot (G_\mathrm{BP}-G_\mathrm{RP})^3.
\end{aligned}
\label{eq:Gaia_transf_ccd}
\end{equation}
The extra term of 0.163 on the right-hand side of Equation~\ref{eq:Gaia_transf_wfi}, adopted from \citet{Bessell&Murphy_2012}, converts $R_\mathrm{c}^\mathrm{Gaia}$ magnitudes from the Vega to the AB photometric system and is required since the trasformation from Gaia photometry to Johnson-Cousins \Rc\ in \citet{Evans_2018} is based on Vega-magnitude Landolt stars \citep{Landolt_1992}.
On the other hand, to obtain the Gaia-predicted NB magnitudes, we employ our derived relations between BB$-$NB and \gcol\ colours (see Section~\ref{subsec:narrow}) which then allow to find the ZP by means of Equation~\ref{eq:zp_formula}.

Once the ZPs are found for each star, a mean $\mathrm{\overline{ZP}}$ is then computed for each exposure as the inverse-variance weighted combination of all the individual ZPs with an iterative sigma clipping of all the values above $3\sigma$ to cut residual non-stellar sources. Each pixel element $p_\mathrm{x,y}$ and its error $\sigma_\mathrm{x,y}$ is then calibrated for each individual exposure as follows
\begin{equation}
\begin{aligned}
    p^\prime_{x,y}\,(\mu\mathrm{Jy}) &= p_{x,y} \cdot 10^{-0.4\cdot(\rm{\overline{ZP}}-\kappa\chi-\rm{ZP}_0)}\\
    \sigma^\prime_{x,y}\,(\mu\mathrm{Jy}) &= 10^{-0.4\cdot(\rm{\overline{ZP}}-\kappa\chi-\rm{ZP}_0)}\cdot\\ 
    &\cdot[\sigma_{x,y}^2+0.848\cdot p_{x,y}^2\cdot(\sigma_{\rm{\overline{ZP}}}^2+\chi^2\sigma_\kappa^2)]^{1/2}
\end{aligned}
\end{equation}
where $\sigma_{\mathrm{\overline{ZP}}}$ and $\sigma_\kappa$ are the errors for $\mathrm{\overline{ZP}}$ and the atmospheric extinction coefficient, respectively, and ZP$_0=23.9$ converts AB magnitudes in $\mu$Jy units.


\section{Removing the \Ha\ emission line contribution in the continuum subtraction}
\label{app:remove_em_line}

The emission-line flux and emission EW is initially calculated from the \Ha\ narrow-band continuum-subtracted images as
\begin{equation}
\begin{aligned}
    &\mathrm{F}_{\mathrm{H}\alpha + [\mathrm{N}\,\mathrm{II}]} = (f_\mathrm{NB}-k\,f_\mathrm{BB})\,\Delta_\mathrm{NB}\\
    &\mathrm{EW}_{\mathrm{H}\alpha + [\mathrm{N}\,\mathrm{II}]} = \frac{\mathrm{F}_{\mathrm{H}\alpha + [\mathrm{N}\,\mathrm{II}]}}{f_\mathrm{BB}},
\end{aligned}
\label{eq:ew_nb}
\end{equation}
where $f_\mathrm{NB}$ and $f_\mathrm{BB}$ are, respectively, the NB and BB image fluxes, $k=f_\mathrm{NB}^*/f_\mathrm{BB}^*$ is the scaling factor computed from a set of common stars in the two images and $\Delta_\mathrm{NB}$ is the NB filter effective width.

Equations~\ref{eq:ew_nb} underestimates the $\mathrm{F}_{\mathrm{H}\alpha + [\mathrm{N}\,\mathrm{II}]}$ flux and overestimates the continuum $f_\mathrm{BB}$ for the presence of the emission lines in the BB image. To remove this bias, we compute the fraction of BB flux coming from the emission lines as
\begin{equation}
    \mathcal{X} = \frac{f_\mathrm{BB}^\mathrm{\,L}}{f_\mathrm{BB}^\mathrm{\,L} + f_\mathrm{BB}^\mathrm{\,C}} = \frac{\mathrm{EW}}{\mathrm{EW} + \Delta_\mathrm{BB}},
\end{equation}
where $f_\mathrm{BB}^\mathrm{\,L}$ and $f_\mathrm{BB}^\mathrm{\,C}$ are, respectively, the line and continuum flux in the BB image with filter effective width $\Delta_\mathrm{BB}$, and where we used the equivalence EW$=f_\mathrm{BB}^\mathrm{\,L}\,\Delta_\mathrm{BB}/f_\mathrm{BB}^\mathrm{\,C}$.
A first guess of $\mathcal{X}$ fraction is derived from the \ewhan\ to then iterate in $\mathcal{X}$ to calculate the emission-line fluxes and emission EW from new continuum estimates. At each $i$\textsuperscript{\,th} iteration, we compute
\begin{equation}
\begin{aligned}
        &\mathrm{F}_{\mathrm{H}\alpha + [\mathrm{N}\,\mathrm{II}]}^{(i)} = (f_\mathrm{NB}-(1-\mathcal{X}_{i-1})\,k\,f_\mathrm{BB})\,\Delta_\mathrm{NB}\\
        &\mathrm{EW}_{\mathrm{H}\alpha + [\mathrm{N}\,\mathrm{II}]}^{(i)} = \frac{\mathrm{F}_{\mathrm{H}\alpha + [\mathrm{N}\,\mathrm{II}]}^{(i)}}{f_\mathrm{BB} - \mathrm{F}_{\mathrm{H}\alpha + [\mathrm{N}\,\mathrm{II}]}^{(i)}/\Delta_\mathrm{BB}}\\
        &\quad\longrightarrow\,\mathcal{X}_i = \frac{\mathrm{EW}_{\mathrm{H}\alpha + [\mathrm{N}\,\mathrm{II}]}^{(i)}}{\mathrm{EW}_{\mathrm{H}\alpha + [\mathrm{N}\,\mathrm{II}]}^{(i)} + \Delta_\mathrm{BB}},
\end{aligned}
\label{eq:iter_proc}
\end{equation}
thus adding the estimated $f_\mathrm{BB}^\mathrm{\,L}$ to $\mathrm{F}_{\mathrm{H}\alpha + [\mathrm{N}\,\mathrm{II}]}$ and subtracting it from $f_\mathrm{BB}$. When two consecutive continuum flux estimates differ by less than 1\%, the convergence to the corrected $\mathrm{F}_{\mathrm{H}\alpha + [\mathrm{N}\,\mathrm{II}]}$ flux is achieved.


\section{\fNii\ and filter effective transmission}
\label{app:f_Nii}

We define the \Nii\ fraction as
\begin{equation}
    \mathcal{F}^{\,0}_{[\mathrm{N\,II}]} = \frac{F_{\rm{NIIa}} + F_{\rm{NIIb}}}{F_{\rm{H}\alpha} + F_{\rm{NIIa}} + F_{\rm{NIIb}}},
    \label{eq:fnii_appendix}
\end{equation}
where $F_{\rm{H}\alpha}$, $F_{\rm{NIIa}}$ and $F_{\rm{NIIb}}$ are the fluxes for the emission lines $\rm{H}\alpha\lambda 6562$, $[\mathrm{N}\,\mathrm{II}]\lambda 6548$ and $[\mathrm{N}\,\mathrm{II}]\lambda 6583$, respectively. Using the theoretical ratio $\rm{F}_{\rm{NIIa}}/\rm{F}_{\rm{NIIb}}=0.34$ \citep{Galavis_1997}, this expression can be rewritten in terms of the ratio $F_{\rm{NIIb}}/F_{\rm{H}\alpha}$ as
\begin{equation}
    \mathcal{F}^{\,0}_{[\mathrm{N}\,\mathrm{II}]} = 1 - \frac{1}{1+1.34\,(F_{\rm{NIIb}}/F_{\rm{H}\alpha})}.
    \label{eq:NII_frac_th}
\end{equation}
With this relation, once the $F_{\rm{NIIb}}/F_{\rm{H}\alpha}$ ratio is inferred from the galaxy stellar mass through the \citet{Kewley_2008} mass-metallicity relation combined with the \citet{Pettini_2004} N2 calibration, a $\mathcal{F}^{\,0}_{[\mathrm{N}\,\mathrm{II}]}$ fraction can be determined for each galaxy. The superscript '0' indicates that this \Nii\ fraction is derived from intrinsic line fluxes, i.e. without correction for the filter transmission curve. 

For our purpose, it is important to take into account the effect of the NB filter transmission curve by defining the line flux weighted by the transmission curve as
\begin{equation}
    \mathrm{F}_\mathrm{L^\prime} = \frac{\int f_\mathrm{L^\prime}(\lambda)T_\mathrm{NB}(\lambda)\mathrm{d}\lambda}{\int T_\mathrm{NB}(\lambda)\mathrm{d}\lambda}
    \label{eq:line_flux_weighted}
\end{equation}
where L can be any of the three emission lines lying within the filter, $T_{\rm{NB}}(\lambda)$ is the normalized NB filter transmission curve, $f_\mathrm{L^\prime}(\lambda)$ is the line profile with the prime symbol (${}^\prime$) indicating that the line is shifted by the galaxy recessional velocity. This formula is valid for multiple lines in the filter bandpass, as well.

The line profile is modelled with a Gaussian shape as follows
\begin{equation}
    f_\mathrm{L^\prime}(\lambda,v_r,v_d) = e^{-(\lambda-\lambda_\mathrm{L^\prime})^2/2\sigma^2}
    \label{eq:gaussian_line}
\end{equation}
where $\lambda_\mathrm{L^\prime} = \lambda_\mathrm{L}\,(1+v_r/c)$ is the line wavelength shifted for the galaxy recessional velocity $v_r$ and $\sigma=(v_d/c)\lambda_\mathrm{L^\prime}$ is the Gaussian line dispersion computed for a velocity dispersion $v_d=15$ km s$^{-1}$. Equation~\ref{eq:gaussian_line} defines an emission line with peak amplitude of 1.0.

The \Nii\ fraction corrected for the filter transmission is then derived with the following formula:
\begin{equation}
\begin{aligned}
     &\mathcal{F}_{[\mathrm{N}\,\mathrm{II}]} = 1 - \dfrac{\mathrm{F}_{\mathrm{H}\alpha^\prime}}{\mathrm{F}_{\mathrm{H}\alpha^\prime} + \mathrm{F}_{\mathrm{NIIa}^\prime} + \mathrm{F}_{\mathrm{NIIb}^\prime}} \approx 1\\
     &- \dfrac{\sum_\lambda f_{\mathrm{H}\alpha^\prime}\,T_{\mathrm{NB}}}{\sum_\lambda [f_{\mathrm{H}\alpha^\prime} + 0.34\,(F_{\rm{NIIb}}/F_{\rm{H}\alpha})\, f_{\mathrm{NIIa}^\prime} + (F_{\rm{NIIb}}/F_{\rm{H}\alpha})\, f_{\mathrm{NIIb}^\prime}]\,T_{\mathrm{NB}}}
\end{aligned}
\label{eq:f_Nii_transm}
\end{equation}
where we made explicit the numerical form of the Equation~\ref{eq:line_flux_weighted} integrals over the NB filter wavelengths and we again made use of the theoretical $F_{\rm{NIIa}}/F_{\rm{NIIb}}=0.34$ and the $F_{\rm{NIIb}}/F_{\rm{H}\alpha}$ ratios.

As an alternative to modelling the filter transmission effect on multiple line profiles, we also define each line's effective filter transmission as
\begin{equation}
    T_{\mathrm{L}^\prime} = \frac{\int T_\mathrm{NB}(\lambda)f_\mathrm{L^\prime}(\lambda)\mathrm{d}\lambda}{\int f_\mathrm{L^\prime}(\lambda)\mathrm{d}\lambda}.
    \label{eq:eff_transmission}
\end{equation}
The effective transmission of the \Ha\ line is then computed as
\begin{equation}
    T_{\mathrm{H}\alpha} = \frac{\sum_\lambda T_\mathrm{NB}(\lambda)f_{\mathrm{H}\alpha^\prime}(\lambda)}{\sum_\lambda f_{\mathrm{H}\alpha^\prime}(\lambda)}.
    \label{eq:Ha_eff_transmission}
\end{equation}
For each galaxy, we compute \fNii\ and $T_{\mathrm{H}\alpha}$ from Equation~\ref{eq:f_Nii_transm} and Equation~\ref{eq:Ha_eff_transmission} to correct the NB fluxes for the filter transmission and the \Nii\ lines contamination.

By adopting the effective filter transmission definition from Equation~\ref{eq:eff_transmission}, we can rewrite Equation~\ref{eq:NII_frac_th} to account for the impact of the filter transmission curve as
\begin{equation}
    \mathcal{F}_{[\mathrm{N}\,\mathrm{II}]} = 1 - \frac{1}{1+(1+0.34\cdot T_{\rm{NIIa}}/T_{\rm{NIIb}})\,(F_{\rm{NIIb}}/F_{\rm{H}\alpha}\cdot T_{\rm{NIIb}}/T_{\rm{H}\alpha})},
    \label{eq:NII_frac_corr}
\end{equation}
where the same naming convention for the \Nii\ lines is used as introduced at the beginning of this appendix.
This formulation is particularly useful when simulating the effect of a photometric filter transmission on spectroscopic emission lines.


\section{Contributions}
\label{app:contributions}

The \pha\ Survey, its observational campaign and the associated paper are the outcome of a large team effort, with major direct and indirect contributions from many people in the PHANGS team. The key contributions are summarized here.

\smallskip

\noindent \textbf{Observation design, data processing, and data quality assurance:} 
The \pha\ observations, data reduction and analysis, together with data quality check, were the result of a collaborative effort made over several years by the \pha\ working group, led by I-T.~Ho until the first internal data release and by Hsi-An Pan thereafter. The working group included G.~Blanc, B.~Groves, and A.~Razza from the beginning, and later E.~Congiu and J.~Neumann.
The survey data were obtained in service mode at the \MPG , with the exception of the first observation conducted by G. Blanc, A.~Razza, and N.~Tomi\v{c}i\'{c}. E.~Schinnerer was PI for the ESO program 098.A-9005(A), with the Observation Blocks (OBs) prepared by N.~Tomi\v{c}i\'{c}, whereas I-T.~Ho was in charge as PI (and OB preparation) for the ESO programmes 0101.A-9014(A) and 0102.A-9002(A).
G.~Blanc served as PI for the data collected in observing mode at the \dP\ telescope in Las Campanas, devised the main observing strategy, and coordinated the first half of the observational campaign, with A.~Razza leading the second half. Other PHANGS team members who observed at the \dP\ were P.~Lang, R.~McElroy, C.~Faesi, I-T.~Ho, and E.~Congiu.
The data reduction pipeline was developed by A.~Razza with guidance from G.~Blanc and B.~Groves, who also wrote the final scripts used to produce the \Ha\ emission-line maps from the calibrated images (sections \ref{subsec:psf_match}, \ref{subsec:continuum_sub}, and \ref{subsec:N2_cont}). J.~Neumann and E.~Congiu collaborated on adapting the pipeline to reduce new and archival WFI observations, which, although not included in this paper, significantly broadened the capabilities of the data reduction framework.

\smallskip

\begin{figure*}[h!]
    \centering
    \includegraphics[width=\textwidth]{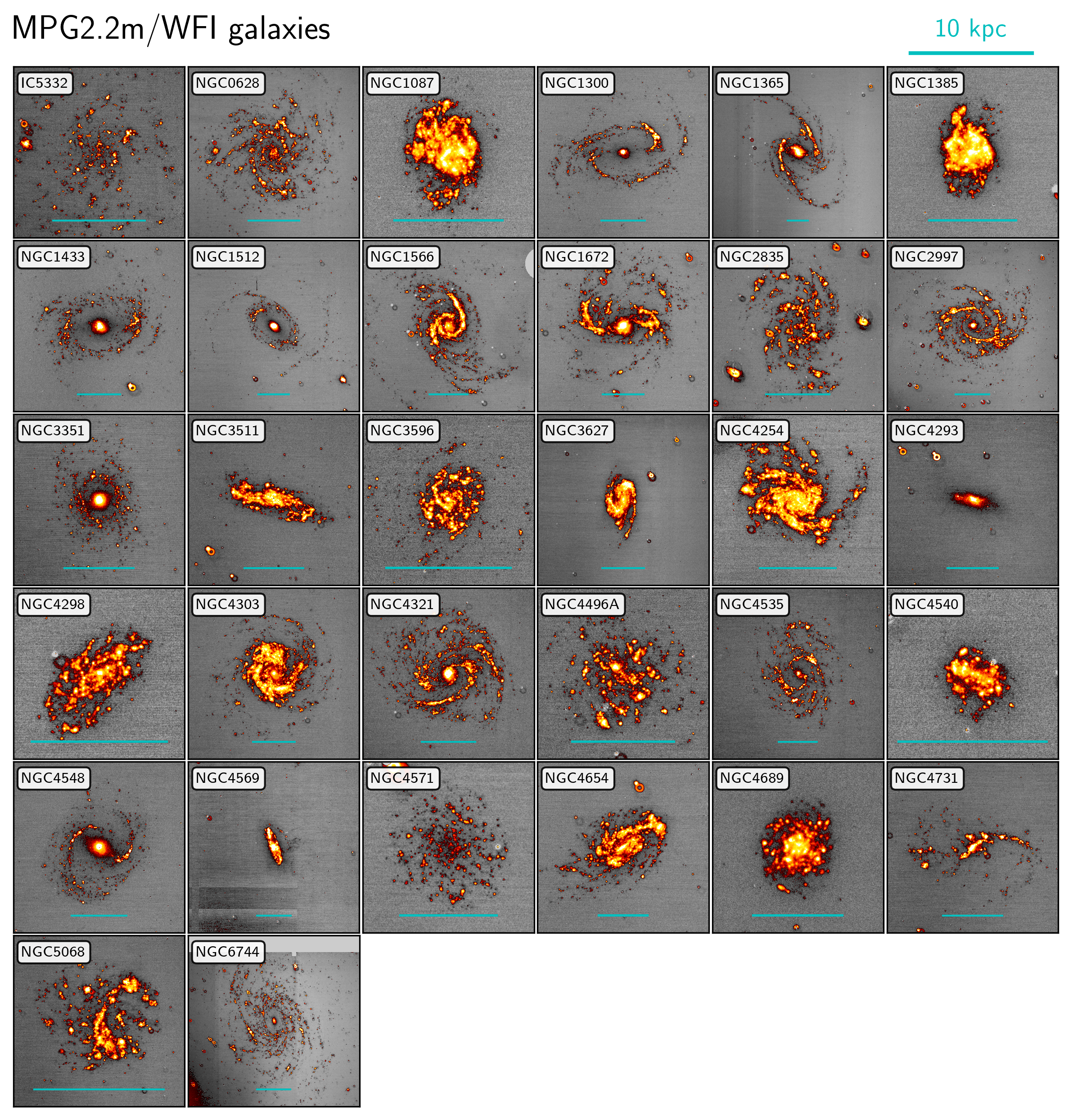}
    \caption{Collection of the final \Ha\ continuum-subtracted and \Nii -corrected (\Ha SUB) images, as produced by the \pha\ pipeline (Section~\ref{sec:reduction}) and subsequent flux corrections (Section~\ref{sec:final_calib}) from the WFI observations at the \MPG\ telescope. The cyan bars indicate the physical scale of 10 kpc.}
    \label{fig:wfi_img_collection}
\end{figure*}

\begin{figure*}[h!]
    \centering
    \includegraphics[width=\textwidth]{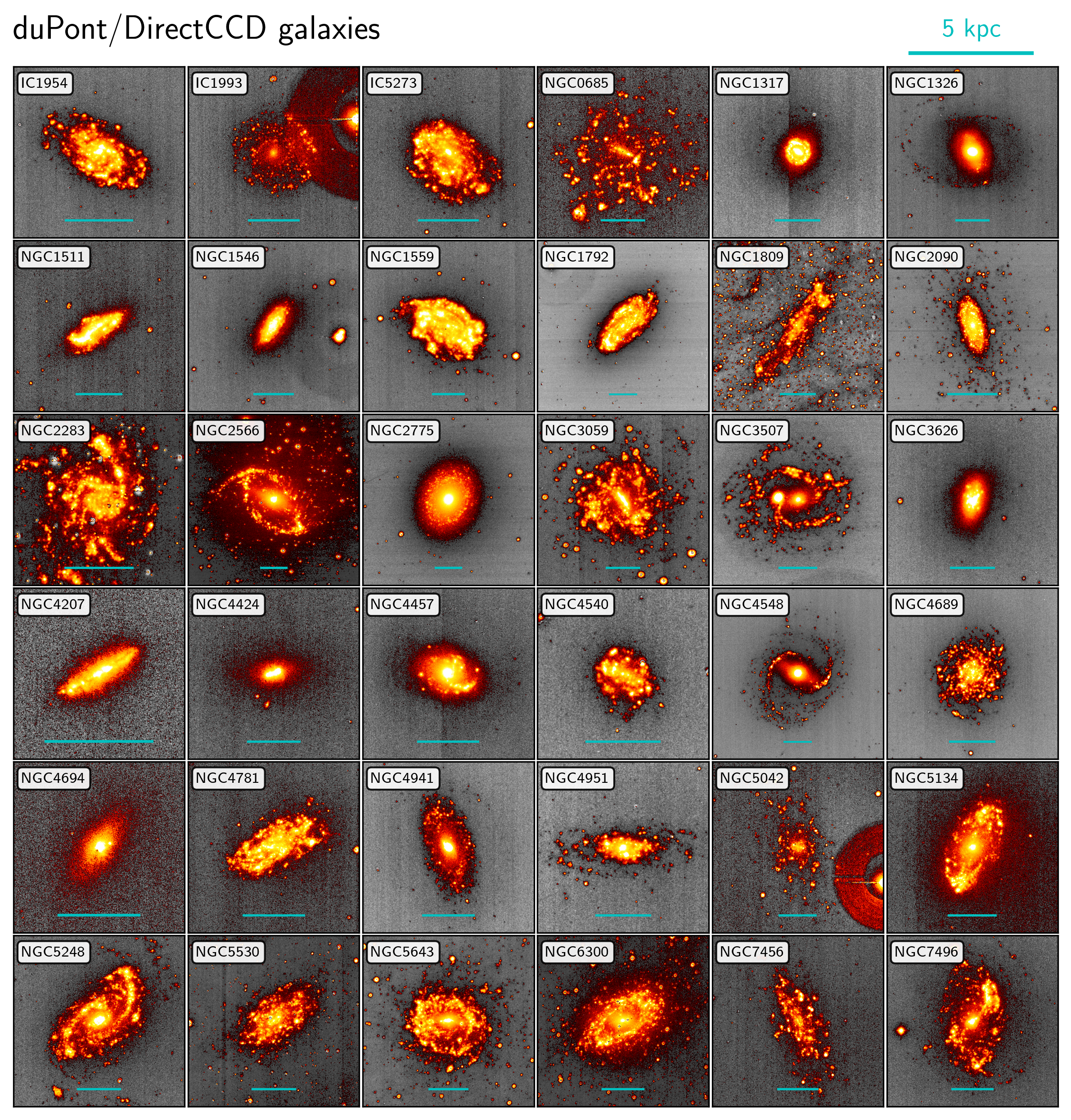}
    \caption{Collection of the final \Ha\ continuum-subtracted and \Nii -corrected (\Ha SUB) images, as produced by the \pha\ pipeline (Section~\ref{sec:reduction}) and subsequent flux corrections (Section~\ref{sec:final_calib}) from the DirectCCD observations at the \dP\ telescope. The cyan bars indicate the physical scale of 5 kpc.}
    \label{fig:ccd_img_collection}
\end{figure*}

\noindent \textbf{Paper analysis and preparation:} 
The development of the spectroscopic corrections benefited from the synergy with the PHANGS–MUSE working group, since this work relied on the calibrated data cubes and DAP emission-line maps produced by E.~Emsellem, F.~Belfiore, and I.~Pessa as part of the PHANGS–MUSE release. An early stage of the correction scheme was allowed by K.~Kreckel's emission EW maps and O.~Egorov input on how to use the PHANGS-MUSE stellar continuum. But the \pha\ team later adopted their own calculations with specific work from A.~Razza, G.~Blanc, and E.~Congiu.
Important feedback for the manuscript was provided by K.~Kreckel's comments, that helped shaping the initial draft, and by B.~Groves, J.~Neumann, E.~Congiu, and G.~Blanc whose extensive reviews brought the paper to the final stage.
Additional insights were provided by F.~Belfiore, R.~Klessen, M.~Boquien, S.~Glover, K.~Grasha, C.~Burton, and A.~Barnes (who produced Figs.~\ref{fig:wfi_img_collection} and~\ref{fig:ccd_img_collection}).
A.~Razza prepared the paper figures, tables, and text, and, together with G.~Blanc, made heavy reviews of all sections until finalization.

\smallskip

\noindent \textbf{Management of the PHANGS Collaboration:} The PHANGS Steering Committee (E.~Schinnerer, team leader 2015--; E.~Rosolowsky, team manager 2018--; A.~Leroy, team project scientist 2017--; E.~Emsellem 2017--; G.~Blanc 2017--2024) and the \href{https://sites.google.com/view/phangs/credits#h.oc5g7pb9f3az}{PHANGS Advisory Board} 
provide key input and oversight to all major collaboration decisions. Scientific exploitation of \pha\ has taken place largely in the context of the `Ionised ISM and its Relation to Star Formation' (led by F.~Belfiore 2023--, O.~Egorov 2024--, B.~Groves 2019--2024, K.~Kreckel 2020--2023, K.~Sandstrom 2019) and ‘Large Scale Dynamical Processes’ (led by S. Meidt and M. Querejeta since 2019) science working groups.


\onecolumn

\section{Collections of the \Ha\ narrow-band final images}
\label{app:imgs_coll}

\twocolumn


\onecolumn

\begin{landscape}
\section{\pha\ sample and result tables}\label{app:tables}
\begin{longtable}{cccccccccccc}
\caption{The \pha\ Sample} 
\label{tab:sample}\\
\hline\hline\\[-8pt]
Name & Telescope & Type & R.A. & Decl. & Dist. & $v_r$ & R$_{25}$ & Incl. & P.A. & St.Mass & SFR\\[4pt] 
  &   &   &   &   & $(\mathrm{Mpc})$ & $(\mathrm{km\,s^{-1}})$ & $(\mathrm{{}^{\prime\prime}})$ & $(\mathrm{{}^{\circ}})$ & $(\mathrm{{}^{\circ}})$ & $(\mathrm{10^{10}\,M_{\odot}})$ & $(\mathrm{M_{\odot}\,yr^{-1}})$\\[8pt]
(1) & (2) & (3) & (4) & (5) & (6) & (7) & (8) & (9) & 
(10) & (11) & (12) \\[2pt]
\hline\\[-8pt]
\endfirsthead
\caption{continued.}\\
\hline\\[-8pt]
Name & Telescope & Type & R.A. & Decl. & Dist. & $v_r$ & R$_{25}$ & Incl. & P.A. & St.Mass & SFR\\[4pt] 
  &   &   &   &   & $(\mathrm{Mpc})$ & $(\mathrm{km\,s^{-1}})$ & $(\mathrm{{}^{\prime\prime}})$ & $(\mathrm{{}^{\circ}})$ & $(\mathrm{{}^{\circ}})$ & $(\mathrm{10^{10}\,M_{\odot}})$ & $(\mathrm{M_{\odot}\,yr^{-1}})$\\[8pt]
(1) & (2) & (3) & (4) & (5) & (6) & (7) & (8) & (9) & 
(10) & (11) & (12) \\[2pt]
\hline\\[-8pt]
\endhead
\hline
\endfoot

IC1954 & du Pont & Sb & 03:31:31.1 & -51:54:17.5 & 12.8 & 1039 & 89 & 57.1 & 63.4 & 0.47 & 0.36 \\
IC1993 & du Pont & SABb & 03:47:04.8 & -33:42:34.8 & 18.09 & 1060 & 84 & 21.3 & 353.5 & 1.06 & 0.18 \\
IC5273 & du Pont & SBc & 22:59:26.7 & -37:42:10.2 & 14.18 & 1285 & 91 & 52.0 & 234.1 & 0.53 & 0.54 \\
IC5332\tablefootmark{\scriptsize{a}} & MPG 2.2m & SABc & 23:34:27.5 & -36:06:03.9 & 9.01 & 699 & 182 & 26.9 & 74.4 & 0.47 & 0.41 \\
NGC0628\tablefootmark{\scriptsize{a}} & MPG 2.2m & Sc & 01:36:41.7 & +15:47:01.1 & 9.84 & 650 & 296 & 8.9 & 20.7 & 2.19 & 1.75 \\
NGC0685 & du Pont & Sc & 01:47:42.8 & -52:45:43.1 & 19.94 & 1346 & 90 & 23.0 & 100.9 & 1.15 & 0.42 \\
NGC1087\tablefootmark{\scriptsize{a}} & MPG 2.2m & Sc & 02:46:25.2 & -00:29:55.4 & 15.85 & 1501 & 89 & 42.9 & 359.1 & 0.86 & 1.31 \\
NGC1300\tablefootmark{\scriptsize{a}} & MPG 2.2m & Sbc & 03:19:41.0 & -19:24:40.0 & 18.99 & 1545 & 178 & 31.8 & 278.0 & 4.14 & 1.17 \\
NGC1317 & du Pont & SABa & 03:22:44.3 & -37:06:13.6 & 19.11 & 1930 & 92 & 23.2 & 221.5 & 4.17 & 0.48 \\
NGC1326 & du Pont & S0-a & 03:23:56.4 & -36:27:52.3 & 18.34 & 1343 & 128 & 52.71 & 70.91 & 3.71 & 1.00 \\
NGC1365\tablefootmark{\scriptsize{a}} & MPG 2.2m & Sb & 03:33:36.4 & -36:08:25.5 & 19.57 & 1613 & 360 & 55.4 & 201.1 & 9.78 & 16.90 \\
NGC1385\tablefootmark{\scriptsize{a}} & MPG 2.2m & Sc & 03:37:28.6 & -24:30:04.2 & 17.22 & 1476 & 102 & 44.0 & 181.3 & 0.95 & 2.09 \\
NGC1433\tablefootmark{\scriptsize{a}} & MPG 2.2m & SBa & 03:42:01.5 & -47:13:19.0 & 18.63 & 1057 & 185 & 28.6 & 199.7 & 7.34 & 1.13 \\
NGC1511 & du Pont & Sab & 03:59:36.6 & -67:38:02.1 & 15.28 & 1331 & 110 & 72.7 & 297.0 & 0.81 & 2.27 \\
NGC1512\tablefootmark{\scriptsize{a}} & MPG 2.2m & Sa & 04:03:54.1 & -43:20:55.4 & 18.83 & 871 & 253 & 42.5 & 261.9 & 5.16 & 1.28 \\
NGC1546 & du Pont & S0-a & 04:14:36.3 & -56:03:39.2 & 17.69 & 1243 & 111 & 70.3 & 147.8 & 2.24 & 0.83 \\
NGC1559 & du Pont & SBc & 04:17:36.6 & -62:47:00.3 & 19.44 & 1275 & 125 & 65.4 & 244.5 & 2.31 & 3.76 \\
NGC1566\tablefootmark{\scriptsize{a}} & MPG 2.2m & SABb & 04:20:00.4 & -54:56:16.8 & 17.69 & 1483 & 216 & 29.5 & 214.7 & 6.09 & 4.54 \\
NGC1672\tablefootmark{\scriptsize{a}} & MPG 2.2m & Sb & 04:45:42.5 & -59:14:50.1 & 19.4 & 1318 & 184 & 42.6 & 134.3 & 5.36 & 7.60 \\
NGC1809 & du Pont & Sc & 05:02:05.0 & -69:34:04.6 & 19.95 & 1290 & 112 & 57.6 & 138.2 & 0.59 & 5.74 \\
NGC1792 & du Pont & Sbc & 05:05:14.3 & -37:58:50.0 & 16.2 & 1175 & 166 & 65.1 & 318.9 & 4.10 & 3.70 \\
NGC2090 & du Pont & Sbc & 05:47:01.9 & -34:15:02.2 & 11.75 & 898 & 134 & 64.5 & 192.46 & 1.09 & 0.41 \\
NGC2283 & du Pont & Sc & 06:45:52.8 & -18:12:38.9 & 13.68 & 821 & 82 & 43.7 & -4.1 & 0.78 & 0.52 \\
NGC2566 & du Pont & Sb & 08:18:45.6 & -25:29:58.3 & 23.44 & 1609 & 127 & 48.5 & 312.0 & 5.12 & 8.72 \\
NGC2775 & du Pont & Sab & 09:10:20.2 & +07:02:17.0 & 23.15 & 1339 & 127 & 41.2 & 156.5 & 11.76 & 0.87 \\
NGC2835\tablefootmark{\scriptsize{a}} & MPG 2.2m & Sc & 09:17:52.9 & -22:21:16.8 & 12.22 & 867 & 192 & 41.3 & 1.0 & 1.00 & 1.24 \\
NGC2997 & MPG 2.2m & SABc & 09:45:38.8 & -31:11:27.9 & 14.06 & 1076 & 307 & 33.0 & 108.1 & 5.41 & 4.37 \\
NGC3059 & du Pont & SBbc & 09:50:08.2 & -73:55:19.9 & 20.23 & 1236 & 113 & 29.4 & -14.8 & 2.39 & 2.38 \\
NGC3351\tablefootmark{\scriptsize{a}} & MPG 2.2m & Sb & 10:43:57.8 & +11:42:13.2 & 9.96 & 774 & 216 & 45.1 & 193.2 & 2.30 & 1.32 \\
NGC3511 & MPG 2.2m & SABc & 11:03:23.8 & -23:05:12.2 & 13.94 & 1096 & 181 & 75.1 & 256.8 & 1.07 & 0.81 \\
NGC3507 & du Pont & SBb & 11:03:25.4 & +18:08:07.9 & 23.55 & 969 & 87 & 21.7 & 55.8 & 2.49 & 0.99 \\
NGC3596 & MPG 2.2m & SABc & 11:15:06.2 & +14:47:13.4 & 11.3 & 1187 & 109 & 25.1 & 78.4 & 0.45 & 0.30 \\
NGC3626 & du Pont & S0-a & 11:20:03.8 & +18:21:24.6 & 20.05 & 1470 & 88 & 46.6 & 165.2 & 2.90 & 0.21 \\
NGC3627\tablefootmark{\scriptsize{a}} & MPG 2.2m & Sb & 11:20:15.0 & +12:59:29.4 & 11.32 & 715 & 308 & 57.3 & 173.1 & 6.81 & 3.84 \\
NGC4207 & du Pont & Sd & 12:15:30.4 & +09:35:05.7 & 15.78 & 606 & 45 & 64.5 & 121.9 & 0.51 & 0.19 \\
NGC4254\tablefootmark{\scriptsize{a}} & MPG 2.2m & Sc & 12:18:49.6 & +14:24:59.1 & 13.1 & 2388 & 151 & 34.4 & 68.1 & 2.66 & 3.07 \\
NGC4293 & MPG 2.2m & S0-a & 12:21:12.8 & +18:22:57.3 & 15.76 & 926 & 187 & 65.0 & 48.3 & 3.21 & 0.51 \\
NGC4298 & MPG 2.2m & Sc & 12:21:32.8 & +14:36:22.0 & 14.92 & 1138 & 76 & 59.2 & 313.9 & 1.05 & 0.46 \\
NGC4303\tablefootmark{\scriptsize{a}} & MPG 2.2m & Sbc & 12:21:54.9 & +04:28:25.5 & 16.99 & 1559 & 206 & 23.5 & 312.4 & 3.34 & 5.33 \\
NGC4321\tablefootmark{\scriptsize{a}} & MPG 2.2m & SABb & 12:22:54.9 & +15:49:20.3 & 15.21 & 1572 & 182 & 38.5 & 156.2 & 5.56 & 3.56 \\
NGC4424 & du Pont & Sa & 12:27:11.6 & +09:25:14.3 & 16.2 & 447 & 91 & 58.2 & 88.3 & 0.81 & 0.30 \\
NGC4457 & du Pont & S0-a & 12:28:59.0 & +03:34:14.2 & 15.1 & 886 & 83 & 17.4 & 78.7 & 2.60 & 0.31 \\
NGC4496A & MPG 2.2m & Scd & 12:31:39.3 & +03:56:22.6 & 14.86 & 1721 & 101 & 53.8 & 51.1 & 0.34 & 0.61 \\
NGC4535\tablefootmark{\scriptsize{a}} & MPG 2.2m & Sc & 12:34:20.3 & +08:11:52.7 & 15.77 & 1953 & 244 & 44.7 & 179.7 & 3.40 & 2.16 \\
NGC4540 & MPG 2.2m/du Pont & SABc & 12:34:50.9 & +15:33:06.2 & 15.76 & 1286 & 65 & 28.7 & 12.8 & 0.61 & 0.17 \\
NGC4548 & MPG 2.2m/du Pont & Sb & 12:35:26.5 & +14:29:46.8 & 16.22 & 482 & 166 & 38.3 & 138.0 & 4.92 & 0.52 \\
NGC4569 & MPG 2.2m & Sab & 12:36:49.8 & +13:09:46.4 & 15.76 & -225 & 273 & 70.0 & 18.0 & 6.40 & 1.32 \\
NGC4571 & MPG 2.2m & Sc & 12:36:56.4 & +14:13:02.4 & 14.9 & 342 & 106 & 32.7 & 217.5 & 1.23 & 0.29 \\
NGC4654 & MPG 2.2m & Sc & 12:43:56.6 & +13:07:36.2 & 21.98 & 1051 & 141 & 55.6 & 123.2 & 3.69 & 3.79 \\
NGC4689 & MPG 2.2m/du Pont & Sc & 12:47:45.6 & +13:45:45.8 & 15.0 & 1614 & 114 & 38.7 & 164.1 & 1.65 & 0.40 \\
NGC4694 & du Pont & S0 & 12:48:15.0 & +10:59:01.4 & 15.76 & 1168 & 59 & 60.7 & 143.3 & 0.72 & 0.16 \\
NGC4731 & MPG 2.2m & SBc & 12:51:01.2 & -06:23:34.2 & 13.28 & 1483 & 189 & 64.0 & 255.4 & 0.30 & 0.60 \\
NGC4781 & du Pont & Scd & 12:54:23.8 & -10:32:13.6 & 11.31 & 1248 & 111 & 59.0 & 290.0 & 0.44 & 0.48 \\
NGC4941 & du Pont & SABa & 13:04:13.1 & -05:33:05.5 & 15.0 & 1116 & 100 & 53.4 & 202.2 & 1.49 & 0.44 \\
NGC4951 & du Pont & SABc & 13:05:07.7 & -06:29:37.8 & 15.0 & 1176 & 94 & 70.2 & 91.2 & 0.62 & 0.35 \\
NGC5042 & du Pont & SABc & 13:15:31.0 & -23:59:02.0 & 16.78 & 1385 & 125 & 49.4 & 190.6 & 0.80 & 0.60 \\
NGC5068\tablefootmark{\scriptsize{a}} & MPG 2.2m & Sc & 13:18:54.7 & -21:02:19.5 & 5.2 & 667 & 224 & 35.7 & 342.4 & 0.25 & 0.28 \\
NGC5134 & du Pont & SABb & 13:25:18.5 & -21:08:03.1 & 19.92 & 1749 & 81 & 22.7 & 311.6 & 2.58 & 0.45 \\
NGC5248 & du Pont & SABb & 13:37:32.0 & +08:53:06.7 & 14.87 & 1163 & 122 & 47.4 & 109.2 & 2.55 & 2.29 \\
NGC5530 & du Pont & SABb & 14:18:27.3 & -43:23:17.7 & 12.27 & 1183 & 144 & 61.9 & 305.4 & 1.20 & 0.33 \\
NGC5643 & du Pont & Sc & 14:32:40.8 & -44:10:28.6 & 12.68 & 1191 & 157 & 29.9 & 318.7 & 2.17 & 2.59 \\
NGC6300 & du Pont & SBb & 17:16:59.5 & -62:49:14.0 & 11.58 & 1102 & 160 & 49.6 & 105.4 & 2.95 & 1.89 \\
NGC6744 & MPG 2.2m & Sbc & 19:09:46.1 & -63:51:27.1 & 9.39 & 832 & 470 & 52.7 & 14.0 & 5.29 & 2.41 \\
NGC7456 & du Pont & Sc & 23:02:10.3 & -39:34:09.9 & 15.7 & 1192 & 123 & 67.3 & 16.0 & 0.44 & 0.37 \\
NGC7496\tablefootmark{\scriptsize{a}} & du Pont & Sb & 23:09:47.3 & -43:25:40.3 & 18.72 & 1639 & 100 & 35.9 & 193.7 & 0.99 & 2.26\\
\end{longtable}
\tablefoot{Col.\ (1): Galaxy name. Col.\ (2): Telescope(s) used to observed the galaxy. Col.\ (3): Morphological T-type as reported in Lyon-Meudon Extragalactic Database (LEDA). Col.\ (4) and Col.\ (5): near-infrared (J2000) right ascension and declination from the Spitzer Survey of Stellar Structure in Galaxies \citep[S4G,][]{Salo_2015}. Col.\ (6): Distances in Mpc from \citet{Gagandeep_2021} and references therein. Col.\ (7): Radial velocity from \citet{Lang_2020}. Col.\ (8): R$_{25}$ in arcsecond from HyperLEDA \citep{Makarov_2014}. Col.\ (9): Inclination in degree from \citet{Lang_2020}. Col.\ (10): Position Angle in degree from \citet{Lang_2020}. Col.\ (11) and Col.\ (12): Galaxy stellar mass and SFR from \citet{Leroy_2021}.\\
\tablefoottext{a}{Galaxies overlapping with PHANGS-MUSE.}
}
\end{landscape}

\twocolumn

\onecolumn

\begin{landscape}
\begin{longtable}{@{\extracolsep{4pt}}ccd{4.0}cccd{5.0}cccd{1.3}d{1.3}c}
\caption{\pha\ Sample Results}
\label{tab:results}\\
\hline\hline\\[-8pt]
 & \multicolumn{4}{c}{Broad Band} & \multicolumn{4}{c}{Narrow Band} & \multicolumn{4}{c}{\Ha\ continuum-subtracted}\\[2pt]
\cline{2-5} \cline{6-9} \cline{10-13}\\[-8pt]
Name & Filter & \multicolumn{1}{c}{Int.Time} & $\mu_\mathrm{lim}$\tablefootmark{\scriptsize{a}} & Resolution & Filter & \multicolumn{1}{c}{Int.Time} & $\mu_\mathrm{lim}$\tablefootmark{\scriptsize{a}} & Resolution & $\Sigma^\mathrm{lim}_\mathrm{H\alpha}$ \hspace{2pt}\tablefootmark{\scriptsize{b}} & \multicolumn{1}{c}{\fNii} & \multicolumn{1}{c}{$T_{\mathrm{H}\alpha}$} & $\log_{10} (L_{\mathrm{H}\alpha})$\\[4pt]
  &  & \multicolumn{1}{c}{$(\mathrm{s})$} &  & $(\mathrm{{}^{\prime\prime}/pc})$ &  & \multicolumn{1}{c}{$(\mathrm{s})$} &  & $(\mathrm{{}^{\prime\prime}/pc})$ &  &  &  & \\[6pt]
  (1) & (2) & \multicolumn{1}{c}{(3)} & (4) & (5) & (6) & \multicolumn{1}{c}{(7)} & (8) & (9) & 
(10) & \multicolumn{1}{c}{(11)} & \multicolumn{1}{c}{(12)} & (13) \\[2pt]
\hline\\[-8pt]
\endfirsthead
\caption{continued.}\\
\hline\\[-8pt]
 & \multicolumn{4}{c}{Broad Band} & \multicolumn{4}{c}{Narrow Band} & \multicolumn{4}{c}{\Ha\ continuum-subtracted}\\[2pt]
\cline{2-5} \cline{6-9} \cline{10-13}\\[-8pt]
Name & Filter & \multicolumn{1}{c}{Int.Time} & $\mu_\mathrm{lim}$\tablefootmark{\scriptsize{a}} & Resolution & Filter & \multicolumn{1}{c}{Int.Time} & $\mu_\mathrm{lim}$\tablefootmark{\scriptsize{a}} & Resolution & $\Sigma^\mathrm{lim}_\mathrm{H\alpha}$ \hspace{2pt}\tablefootmark{\scriptsize{b}} & \multicolumn{1}{c}{\fNii} & \multicolumn{1}{c}{$T_{\mathrm{H}\alpha}$} & $\log_{10} (L_{\mathrm{H}\alpha})$\\[4pt]
  &  & \multicolumn{1}{c}{$(\mathrm{s})$} &  & $(\mathrm{{}^{\prime\prime}/pc})$ &  & \multicolumn{1}{c}{$(\mathrm{s})$} &  & $(\mathrm{{}^{\prime\prime}/pc})$ &  &  &  & \\[6pt]
  (1) & (2) & \multicolumn{1}{c}{(3)} & (4) & (5) & (6) & \multicolumn{1}{c}{(7)} & (8) & (9) & 
(10) & \multicolumn{1}{c}{(11)} & \multicolumn{1}{c}{(12)} & (13) \\[2pt]
\hline\\[-8pt]
\endhead
\hline
\endfoot

IC1954 & $r$ & 920 & 24.93 & 0.87$/$53 & \Ha 657 & 5400 & 23.74 & 1.47$/$91 & 0.44 & 0.180 & 0.805 & 40.70 \\
IC1993 & $r$ & 900 & 24.99 & 0.8$/$69 & \Ha 657 & 5400 & 23.47 & 0.81$/$71 & 0.61 & 0.236 & 0.798 & 40.97 \\
IC5273 & $r$ & 900 & 24.43 & 1.03$/$70 & \Ha 657 & 5400 & 23.61 & 1.04$/$71 & 0.50 & 0.176 & 0.715 & 40.83 \\
IC5332 & $R_\mathrm{c}$ & 959 & 25.18 & 0.67$/$29 & \Ha & 1620 & 24.00 & 0.79$/$34 & 0.48 & 0.238 & 0.992 & 40.08 \\
NGC0628 & $R_\mathrm{c}$ & 959 & 25.16 & 0.93$/$44 & \Ha & 2699 & 23.83 & 0.9$/$42 & 0.62 & 0.363 & 0.988 & 40.93 \\
NGC0685 & $r$ & 900 & 25.06 & 0.84$/$80 & \Ha 657 & 5400 & 23.95 & 0.89$/$85 & 0.46 & 0.225 & 0.691 & 40.98 \\
NGC1087 & $R_\mathrm{c}$ & 900 & 25.30 & 0.75$/$57 & \Ha & 2670 & 24.54 & 0.91$/$70 & 0.24 & 0.254 & 0.973 & 41.04 \\
NGC1300 & $R_\mathrm{c}$ & 900 & 25.59 & 0.78$/$71 & \Ha & 2670 & 24.51 & 0.83$/$76 & 0.30 & 0.356 & 0.969 & 40.82 \\
NGC1317 & $r$ & 920 & 24.14 & 0.94$/$86 & \Ha 663 & 5460 & 22.78 & 0.82$/$75 & 1.04 & 0.414 & 0.916 & 41.09 \\
NGC1326 & $r$ & 1100 & 24.43 & 0.8$/$71 & \Ha 657 & 5500 & 23.73 & 0.86$/$76 & 0.37 & 0.294 & 0.692 & 41.20 \\
NGC1365 & $R_\mathrm{c}$ & 800 & 25.66 & 0.67$/$63 & \Ha & 2670 & 23.35 & 0.62$/$58 & 0.83 & 0.377 & 0.962 & 41.58 \\
NGC1385 & $R_\mathrm{c}$ & 900 & 25.56 & 1.02$/$85 & \Ha & 2670 & 24.55 & 0.98$/$82 & 0.40 & 0.264 & 0.976 & 41.18 \\
NGC1433 & $R_\mathrm{c}$ & 1000 & 25.52 & 0.95$/$85 & \Ha & 2670 & 24.38 & 1.14$/$102 & 0.33 & 0.414 & 0.998 & 40.93 \\
NGC1511 & $r$ & 1220 & 24.07 & 1.12$/$83 & \Ha 657 & 5400 & 24.18 & 1.15$/$84 & 0.39 & 0.202 & 0.697 & 40.97 \\
NGC1512 & $R_\mathrm{c}$ & 959 & 25.58 & 0.73$/$66 & \Ha & 2699 & 24.15 & 0.92$/$83 & 0.43 & 0.408 & 0.998 & 40.62 \\
NGC1546 & $r$ & 1840 & 25.14 & 1.05$/$89 & \Ha 657 & 11300 & 24.09 & 1.46$/$125 & 0.41 & 0.272 & 0.732 & 41.02 \\
NGC1559 & $r$ & 920 & 25.66 & 1.04$/$98 & \Ha 657 & 5400 & 24.22 & 1.39$/$130 & 0.38 & 0.272 & 0.719 & 41.63 \\
NGC1566 & $R_\mathrm{c}$ & 800 & 25.28 & 0.82$/$70 & \Ha & 2670 & 24.08 & 0.79$/$67 & 0.57 & 0.379 & 0.975 & 41.34 \\
NGC1672 & $R_\mathrm{c}$ & 899 & 25.23 & 0.74$/$69 & \Ha & 2670 & 24.44 & 0.97$/$91 & 0.32 & 0.388 & 0.988 & 41.45 \\
NGC1809 & $r$ & 920 & 22.87 & 1.36$/$131 & \Ha 657 & 5400 & 22.62 & 1.3$/$126 & 1.51 & 0.183 & 0.713 & 41.08 \\
NGC1792 & $r$ & 920 & 25.92 & 1.05$/$82 & \Ha 657 & 5420 & 24.63 & 0.85$/$66 & 0.26 & 0.309 & 0.757 & 41.39 \\
NGC2090 & $r$ & 920 & 25.50 & 0.87$/$49 & \Ha 657 & 5420 & 24.61 & 0.88$/$50 & 0.20 & 0.249 & 0.848 & 40.73 \\
NGC2283 & $r$ & 920 & 23.83 & 1.06$/$70 & \Ha 657 & 6320 & 23.61 & 0.8$/$52 & 0.35 & 0.229 & 0.867 & 41.11 \\
NGC2566 & $r$ & 960 & 25.03 & 0.94$/$106 & \Ha 663 & 5460 & 24.00 & 0.79$/$89 & 0.51 & 0.447 & 0.802 & 41.55 \\
NGC2775 & $r$ & 960 & 24.86 & 1.06$/$118 & \Ha 657 & 5420 & 24.21 & 1.06$/$118 & 0.23 & 0.327 & 0.694 & 41.50 \\
NGC2835 & $R_\mathrm{c}$ & 1919 & 25.39 & 1.07$/$63 & \Ha & 2339 & 24.47 & 0.94$/$55 & 0.33 & 0.298 & 0.998 & 40.89 \\
NGC2997 & $R_\mathrm{c}$ & 959 & 25.03 & 1.29$/$88 & \Ha & 2699 & 23.50 & 1.01$/$68 & 1.49 & 0.403 & 0.998 & 41.21 \\
NGC3059 & $r$ & 920 & 25.50 & 1.26$/$123 & \Ha 657 & 5420 & 23.95 & 1.32$/$129 & 0.48 & 0.276 & 0.735 & 41.57 \\
NGC3351 & $R_\mathrm{c}$ & 580 & 24.91 & 1.12$/$54 & \Ha & 2520 & 23.49 & 1.16$/$56 & 0.76 & 0.363 & 0.995 & 40.27 \\
NGC3511 & $R_\mathrm{c}$ & 959 & 25.34 & 0.88$/$59 & \Ha & 1800 & 24.05 & 1.19$/$80 & 0.45 & 0.297 & 0.998 & 40.87 \\
NGC3507 & $r$ & 960 & 25.68 & 1.17$/$133 & \Ha 657 & 5420 & 24.50 & 1.47$/$168 & 0.29 & 0.299 & 0.827 & 41.22 \\
NGC3596 & $R_\mathrm{c}$ & 580 & 25.28 & 1.15$/$63 & \Ha & 2580 & 23.97 & 1.29$/$70 & 0.50 & 0.224 & 0.995 & 40.34 \\
NGC3626 & $r$ & 960 & 25.37 & 1.26$/$122 & \Ha 663 & 5420 & 23.43 & 1.48$/$143 & 0.76 & 0.429 & 0.743 & 41.09 \\
NGC3627 & $R_\mathrm{c}$ & 959 & 24.62 & 0.81$/$44 & \Ha & 2699 & 23.52 & 0.91$/$49 & 0.79 & 0.421 & 0.993 & 41.09 \\
NGC4207 & $r$ & 900 & 23.97 & 0.97$/$74 & \Ha 657 & 5400 & 23.04 & 0.92$/$70 & 0.70 & 0.211 & 0.909 & 40.24 \\
NGC4254 & $R_\mathrm{c}$ & 580 & 25.01 & 1.04$/$66 & \Ha & 2520 & 23.71 & 0.99$/$62 & 0.55 & 0.262 & 0.723 & 41.16 \\
NGC4293 & $R_\mathrm{c}$ & 840 & 24.68 & 0.63$/$48 & \Ha & 2699 & 24.04 & 0.79$/$60 & 0.54 & 0.382 & 1.000 & 39.92 \\
NGC4298 & $R_\mathrm{c}$ & 580 & 24.33 & 1.19$/$85 & \Ha & 2719 & 23.90 & 1.08$/$78 & 0.35 & 0.293 & 0.997 & 40.41 \\
NGC4303 & $R_\mathrm{c}$ & 420 & 25.01 & 0.81$/$66 & \Ha & 2699 & 23.96 & 0.84$/$69 & 0.52 & 0.344 & 0.968 & 41.51 \\
NGC4321 & $R_\mathrm{c}$ & 959 & 23.99 & 0.73$/$53 & \Ha & 1979 & 23.81 & 0.71$/$52 & 0.65 & 0.367 & 0.966 & 41.02 \\
NGC4424 & $r$ & 920 & 24.18 & 0.96$/$75 & \Ha 657 & 5400 & 22.93 & 1.1$/$86 & 0.97 & 0.252 & 0.943 & 40.83 \\
NGC4457 & $r$ & 920 & 24.27 & 1.25$/$91 & \Ha 657 & 5400 & 23.15 & 1.28$/$93 & 0.77 & 0.308 & 0.851 & 40.89 \\
NGC4496A & $R_\mathrm{c}$ & 580 & 25.30 & 1.1$/$79 & \Ha & 2719 & 24.35 & 1.09$/$78 & 0.34 & 0.171 & 0.944 & 40.60 \\
NGC4535 & $R_\mathrm{c}$ & 659 & 24.88 & 1.23$/$94 & \Ha & 1530 & 23.58 & 1.32$/$100 & 0.74 & 0.303 & 0.893 & 40.66 \\
NGC4540 & $R_\mathrm{c}$ & 959 & 25.35 & 1.08$/$82 & \Ha & 2699 & 24.54 & 1.03$/$78 & 0.26 & 0.242 & 0.990 & 40.20 \\
NGC4548 & $R_\mathrm{c}$ & 959 & 25.42 & 0.91$/$71 & \Ha & 2699 & 24.36 & 0.9$/$70 & 0.35 & 0.412 & 0.971 & 40.60 \\
NGC4569 & $R_\mathrm{c}$ & 580 & 24.76 & 1.25$/$95 & \Ha & 2520 & 23.44 & 1.18$/$90 & 0.77 & 0.469 & 0.725 & 40.63 \\
NGC4571 & $R_\mathrm{c}$ & 1119 & 24.95 & 1.47$/$106 & \Ha & 2719 & 24.25 & 1.22$/$88 & 0.45 & 0.323 & 0.944 & 40.53 \\
NGC4654 & $R_\mathrm{c}$ & 580 & 25.13 & 0.91$/$96 & \Ha & 2719 & 24.30 & 0.97$/$103 & 0.35 & 0.385 & 0.998 & 41.28 \\
NGC4689 & $R_\mathrm{c}$ & 959 & 24.96 & 2.69$/$195 & \Ha & 2699 & 24.31 & 2.73$/$198 & 0.30 & 0.293 & 0.962 & 40.56 \\
NGC4694 & $r$ & 920 & 24.21 & 0.99$/$75 & \Ha 657 & 5400 & 23.34 & 0.95$/$72 & 0.57 & 0.202 & 0.760 & 40.44 \\
NGC4731 & $R_\mathrm{c}$ & 580 & 25.28 & 0.8$/$51 & \Ha & 2719 & 24.12 & 1.03$/$66 & 0.41 & 0.178 & 0.975 & 40.50 \\
NGC4781 & $r$ & 920 & 23.57 & 0.88$/$48 & \Ha 657 & 5400 & 23.05 & 0.97$/$52 & 0.73 & 0.165 & 0.730 & 40.82 \\
NGC4941 & $r$ & 960 & 25.44 & 1.4$/$101 & \Ha 657 & 5420 & 24.49 & 1.31$/$95 & 0.24 & 0.255 & 0.779 & 40.78 \\
NGC4951 & $r$ & 920 & 24.20 & 1.34$/$97 & \Ha 657 & 5400 & 22.85 & 1.16$/$84 & 0.97 & 0.192 & 0.757 & 40.86 \\
NGC5042 & $r$ & 900 & 22.76 & 0.82$/$66 & \Ha 657 & 5400 & 22.65 & 1.05$/$85 & 1.15 & 0.199 & 0.675 & 40.92 \\
NGC5068 & $R_\mathrm{c}$ & 660 & 25.21 & 1.06$/$26 & \Ha & 2249 & 23.85 & 1.38$/$34 & 0.56 & 0.191 & 0.990 & 40.48 \\
NGC5134 & $r$ & 960 & 23.46 & 0.96$/$92 & \Ha 663 & 5460 & 21.88 & 0.96$/$92 & 2.40 & 0.397 & 0.857 & 41.34 \\
NGC5248 & $r$ & 920 & 23.83 & 0.93$/$67 & \Ha 657 & 5400 & 22.91 & 0.87$/$62 & 0.88 & 0.285 & 0.762 & 41.29 \\
NGC5530 & $r$ & 960 & 24.06 & 1.2$/$71 & \Ha 657 & 5400 & 23.23 & 1.24$/$73 & 0.49 & 0.236 & 0.754 & 40.85 \\
NGC5643 & $r$ & 960 & 24.09 & 1.16$/$71 & \Ha 657 & 5430 & 23.40 & 1.22$/$75 & 0.55 & 0.274 & 0.751 & 41.39 \\
NGC6300 & $r$ & 960 & 24.53 & 0.9$/$50 & \Ha 657 & 5450 & 23.82 & 0.88$/$49 & 0.59 & 0.298 & 0.784 & 41.06 \\
NGC6744 & $R_\mathrm{c}$ & 899 & 24.87 & 1.31$/$59 & \Ha & 2339 & 22.48 & 1.37$/$62 & 1.95 & 0.410 & 0.997 & 40.22 \\
NGC7456 & $r$ & 900 & 24.25 & 0.95$/$72 & \Ha 657 & 5400 & 23.28 & 1.1$/$83 & 0.63 & 0.169 & 0.751 & 40.82 \\
NGC7496 & $r$ & 960 & 24.47 & 0.79$/$71 & \Ha 663 & 5400 & 22.81 & 0.86$/$77 & 1.17 & 0.330 & 0.815 & 41.37 \\
NGC4540 & $r$ & 1820 & 25.39 & 1.28$/$97 & \Ha 657 & 5460 & 24.62 & 1.07$/$81 & 0.23 & 0.185 & 0.715 & 40.53 \\
NGC4548 & $r$ & 920 & 25.39 & 1.09$/$85 & \Ha 657 & 5460 & 24.66 & 1.17$/$91 & 0.22 & 0.370 & 0.936 & 41.07 \\
NGC4689 & $r$ & 920 & 25.17 & 1.12$/$81 & \Ha 663 & 6360 & 24.47 & 1.34$/$97 & 0.26 & 0.374 & 0.804 & 40.90
\end{longtable}
\tablefoot{Col. (1): Galaxy name. Col. (2): Broad-band filter name. Col.\. (3): Total integration time of the combined BB images. Col. (4): 3$\sigma$ surface brightness sensitivity of the BB image measured over $\sim$2 arcsec$^2$. Col. (5): Resolution in arcseconds and parsec of the BB image from the estimated PSF. Col. (6): Narrow-band filter name. Col. (7): Total integration time of the combined NB images. Col. (8): 3$\sigma$ surface brightness sensitivity of the NB image measured over $\sim$2 arcsec$^2$. Col. (9): Resolution in arcseconds and parsec of the NB image from the estimated PSF. Col. (10): 3$\sigma$ surface brightness sensitivity of the \Ha\ continuum-subtracted image measured over $\sim$2 arcsec$^2$. Col. (11): \fNii\ as determined by Equation~\ref{eq:NII_frac_formula}. Col. (12): $T_{\mathrm{H}\alpha}$ as determined in Section~\ref{subsec:N2_cont}. Col. (13): Integrated \Ha\ luminosity corrected for the Galactic foreground extinction, with no correction applied for the galaxy internal extinction.\\
\tablefoottext{a}{\,Detection limit given in mag arcsec$^{-2}$.}\\
\tablefoottext{b}{\,Detection limit given in $10^{-16}$ erg s$^{-1}$ cm$^{-2}$ arcsec$^{-2}$.}
}
\end{landscape}

\twocolumn

\end{appendix}

\end{document}